\documentclass[%
superscriptaddress,
 amsmath,amssymb,
 aps, prx, twocolumn,
longbibliography
]{revtex4-1}

\usepackage{graphicx}% Include figure files
\usepackage{dcolumn}% Align table columns on decimal point
\usepackage{bm}% bold math
\usepackage{footnote}
\usepackage{mathtools}
\usepackage{psfrag}
\usepackage{xcolor}
\usepackage{amsthm}

\usepackage[colorlinks = true,
            linkcolor = blue,
            urlcolor  = blue,
            citecolor = blue,
            anchorcolor = blue]{hyperref}
\usepackage{dsfont}
\usepackage{tabularx}
\usepackage[capitalize]{cleveref}
\newcommand{\eq}[1]{\begin{align}#1\end{align}}

\def\T{\mathcal{T}}

\definecolor{jasper}{rgb}{0.88, 0.23, 0.24}
 
\begin{document}
\title{Thermodynamics of computations with absolute irreversibility, \\ unidirectional transitions, and stochastic computation times}

\author{Gonzalo Manzano}
\affiliation{Institute for Cross-Disciplinary Physics and Complex Systems (IFISC) UIB-CSIC, Mallorca, Spain}
 
\author{G\"ulce Karde\c{s}}
\affiliation{University of Colorado, Boulder, Colorado 80309, United States}
\affiliation{Santa Fe Institute, Santa Fe, New Mexico 87501, United States} 

\author{\'Edgar Rold\'an}
\affiliation{ICTP -- The Abdus Salam International Centre for Theoretical Physics, Strada Costiera 11, 34151 Trieste, Italy}

\author{David Wolpert}
\affiliation{Santa Fe Institute, Santa Fe, New Mexico 87501, United States} 
\affiliation{ICTP -- The Abdus Salam International Centre for Theoretical Physics, Strada Costiera 11, 34151 Trieste, Italy}

             
\begin{abstract} 
Developing a thermodynamic theory of computation is a challenging task at the interface of non-equilibrium thermodynamics and computer science. In particular, this task requires dealing with difficulties such as stochastic halting times, unidirectional (possibly deterministic) transitions, and restricted initial conditions, features common in real-world computers. Here, we present a framework which tackles all such difficulties by extending the martingale theory of non-equilibrium thermodynamics to generic non-stationary Markovian processes, including those with broken detailed balance and/or absolute irreversibility.
We derive several universal fluctuation relations and second-law-like inequalities that provide both lower and upper bounds on the intrinsic dissipation (mismatch cost) associated with any periodic process --- in particular the periodic processes underlying all current digital computation. Crucially, these bounds apply even if the process has stochastic stopping times, as it does in many computational machines.
We illustrate our results with exhaustive numerical simulations of  
deterministic finite automata (DFA) processing bit strings,
one of the fundamental models of computation from theoretical computer science. We also provide universal equalities and inequalities for the acceptance probability of words of a given length by a deterministic finite automaton in terms of thermodynamic quantities, and outline connections between computer science and stochastic resetting.  Our results, while motivated from the computational context, are applicable far more broadly.
\end{abstract}

\maketitle

\section{Introduction}

\subsection{Background and motivation}
In the last three decades there has been major progress in formulating far from equilibrium systems and processes. Using stochastic thermodynamics, we can now rigorously formulate the thermodynamic behavior of systems ranging from biological molecular machines to electronic circuits, evolving arbitrarily away from equilibrium. Celebrated results of stochastic thermodynamics include fluctuation relations that generalize the second law of thermodynamics \cite{sekimoto2010stochastic,jarzynski2011equalities,Seifert12}, speed limit theorems \cite{PhysRevLett.121.070601,van2020unified,van2023thermodynamic}, thermodynamic uncertainty relations ~\cite{barato2015thermodynamic,Todd2016,Horowitz2020,utsumi2022computation}, large deviation approaches~\cite{chetrite2015nonequilibrium,hoppenau2016level}, martingale fluctuation relations for extrema and stopping times~\cite{Neri17,Chetrite2018,quantum19,gambling2021,survival22,10.21468/SciPostPhys.14.5.131,Edgar2022}, and universal bounds on various kinetic and frenetic properties~\cite{baiesi2015inflow,DiTerlizzi2018,maes2020frenesy}.

\begin{figure*}[t!]
\centering
   \includegraphics[width=0.99\linewidth]{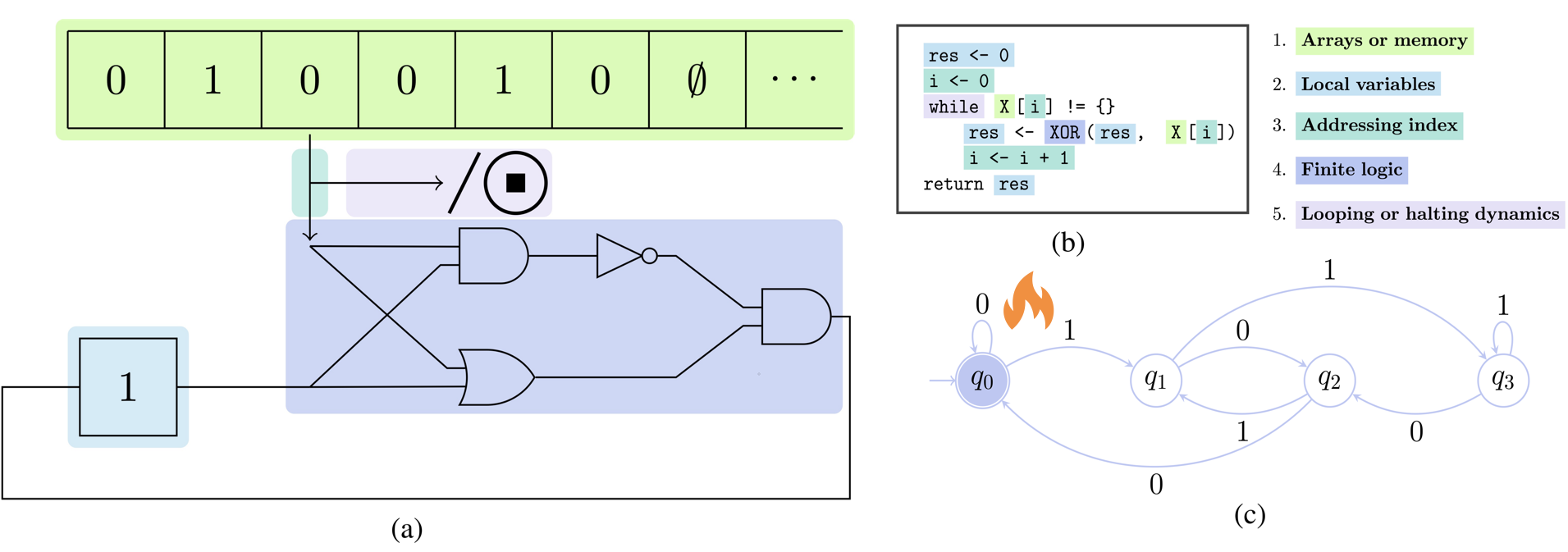} 
   \caption{Illustrations of computations with absolute irreversibility, unidirectional transitions, and stochastic computation times. An algorithm is executed by using a set of instructions, a finite control unit or local variables, a working memory to store the input or intermediate execution values, an address index, and mechanisms which specify when to loop and when to halt (the latter is of interest to us in thinking of stopping times). Both in computer science models and physical computers, the finite control unit generally corresponds to a circuit [as in (a), implementing the algorithm in (b)] or a DFA as in (c), here deciding whether input bit strings are divisible by four. In physical implementation of such devices which solve a myriad of computational tasks, energetic costs are inevitable.} 
     \label{fig:sketch}
\end{figure*}

The past decade also witnessed progress in thermodynamics of computation. Although many initial studies on energetic costs of computation have mostly concerned unit operations such as bit erasure~\cite{leff2002maxwell,kawai2007dissipation,proesmans2020finite,dago2021information}, which is too primitive to be pertinent to the formal models of computation in theoretical computer science (TCS), very recent work started to investigate energetic costs of implementing computational machines central to TCS, which perform tasks such as string matching algorithms (which are justifiably more complex than bit erasure)~\cite{Wolpert2020thermodynamics, ouldridge2022thermodynamics, kardes2022inclusive}. Figure~\ref{fig:sketch}(a) shows the general model of a computational machine (henceforth called a computer) which implements a basic algorithm presented in Fig.~\ref{fig:sketch}(b). 

An algorithm is a finite procedure for implementing a given task, which can be executed in various physical ways, e.g., while modifying the current on electrical wires or the structure of a DNA origami. 
Formally, an algorithm consists of the instructions to be performed (which is implemented by the dynamics of the computer), the local variables and the memory arrays (stored by the computer), as well as mechanisms to decide when to repeat steps and when to halt. A computer executes an algorithm on a given set of inputs, starting from a certain initial state, potentially following unidirectional transitions in its state space, and halting at an arbitrary stochastic time that depends on the computation. Hence a general thermodynamic model of computers which implement arbitrary algorithms should be able to account for the energetic costs of implementing computational processes (i) at arbitrary stopping times, with (ii) unidirectional (possibly deterministic) transitions, and (iii) ``absolute irreversibility" due to the computer being initialized to a designated start state.

However, most of the central results in stochastic thermodynamics do not directly apply to processes having the aforementioned three key ingredients of computational processes (stopping times, unidirectional transitions, and absolute irreversibility). In fact, a central assumption in much of stochastic thermodynamics is the condition of local detailed balance (LDB), which requires the system to have only bidirectional transitions, i.e., all transitions between any two states $i \to j$ with their reverse $j \to i$ have a finite, non-zero probability to occur in a finite time. On the other hand, LDB can be formally avoided by taking an inclusive Hamiltonian approach~\cite{jarzynskihamiltonian,Esposito_2010,Deffner13,ptaszynski2019entropy}, which has been applied recently to computational machines~\cite{kardes2022inclusive}. However, in general there may be hidden nonequilibrium (driven) degrees of freedom which need to be included in the  thermodynamic framework beyond a surrounding thermal bath. Moreover, even when assuming LDB, the ratio of transition probabilities between $i \to j$ and their reverse $j \to i$ can grow exponentially with the entropy exchanged with the environment, so that the reverse transition will not be effectively seen in relevant time-scales, leading to an effective unidirectionality.
Until now, little is currently known about the stochastic thermodynamics of systems with broken LDB reflecting unidirectional transitions~\cite{rahav2014integral,pal2017integral,DiTerlizzi2018KineticUR,busiello2020entropy,hiura2021kinetic,moslonka2022martingale} or athermal and nonequilibrium environments~\cite{kanazawa2015minimal,Esposito15,Manzano18,Dabelow19,Tociu19}. In addition, the recently established martingale theory of thermodynamics (see~\cite{Edgar2022} for a review) --–which formulates fluctuation theorems and second-law-like inequalities at generic stopping times--– has not yet been extended
to apply to systems with either unidirectional transitions or absolute irreversibility.

In this paper, we develop a nonequilibrium thermodynamics theory for computations with stochastic computational times that may have absolute irreversibility and unidirectional transitions~\footnote{ However, we emphasize that our theory is also applicable to the more standard scenario, such as when LDB is verified.}. Throughout this work, we focus on the intrinsic thermodynamic costs of computations, modelled by generic Markovian dynamics, with minimal or no details about their physical implementation. We derive fluctuation relations for key thermodynamic quantities applicable to all computational processes which can be modeled as discrete-time Markov chains (DTMC), with both unidirectional and bidirectional transitions, and restricted initial probability distributions. These relations hold at both fixed and stopping times, and so simultaneously extend the martingale theory for stochastic thermodynamics to the case of DTMCs with absolute irreversibility and unidirectional transitions. 

The thermodynamic meaning of our results is established by introducing a generic physical implementation of the computer as a periodically-driven process operated over a set of hidden degrees of freedom. This allow us to link dissipation from underlying (physical) to visible (computational) levels, and obtain quantifiers for the energetic costs of computations up to their halting time or between halting times of consecutive computations, and their statistics.  
Our results, while emphasized here for computational processes, can also describe a wide range of systems, including, e.g., biochemical processes with irreversible release of molecules. 

We illustrate our results in deterministic finite automata (DFA). Loosely speaking, a DFA is a system with a finite state space, initialized to a special start state $q_0$ at $t = 0$, and a logical computer by itself, which can solve basic computational tasks such as string matching. More importantly, it constitutes the “finite logic” part [see Fig.~\ref{fig:sketch}(b), (c)] of the engineered computers at use today, and formally corresponds to the finite logic component of Turing machines~(TM).

\subsection{Summary of our contributions and roadmap}

Consider a discrete-time computational task which is implemented by iteratively processing an input sequence of symbols $w$
through a maximum of $\tau$ iterations. We refer to $\tau$ as the  {\em limit time} of the computation. As an example, $\tau$ could be the length $\vert w \vert$ of a bit string 
$w$, and the computation could involve iteratively processing each
of those bits in sequence.

During such a computation, the state of the computer evolves in a stochastic manner, tracing a stochastic trajectory on a set of computational variables $\mathbf{x}_{[0,\tau]}=x_0,\dots,x_\tau$. The computation finishes either at the limit time or, if it
is earlier, at a stochastic computation time $\mathcal{T}$, which in
general will be a function of the precise input $w$.
Formally, computational times  $\mathcal{T}$ are specific examples of  {\em stopping times}. A stopping time is the first time that a stochastic trajectory meets a specific predefined criterion~\footnote{See  Ch. 4 in Ref.~\cite{Edgar2022} for  rigorous mathematical definitions and mathematical properties of stopping times,  and  examples of stopping times in physics (e.g. first-passage times).}.  In this work, we deal with stopping times  which associate to each specific trajectory $\mathbf{x}_{[0,\tau]}=x_0,\dots,x_\tau$ a stochastic time $\T\leq \tau$ that is always smaller or equal than the limit time.
As an example of stopping time in computation, $\T$ could be the first time a DFA reaches a prescribed computational state $A$. $\T$ could also be defined as the first time a DFA reaches a state $B\neq A$ once having left state $A$. Stopping times thus provide a flexible yet rigorous mathematical toolbox to tackle computations that last a stochastic amount of time. 

In Sec.~\ref{sec:framework}, we provide the elementary concepts of our framework, including a formal definition of DFAs; the description of computational processes (such as the running of a DFA) as Markov chains, and the physical implementation of those Markov chains. We also review the relevant thermodynamic quantities in this context.

In this paper we investigate the thermodynamic costs of computations,
paying particular attention to those with such stochastic durations. As described below, this will lead us to concentrate on a specific thermodynamic quantity, which we call the \textit{intrinsic mismatch cost.} The intrinsic mismatch cost associated with a stochastic trajectory $\mathbf{x}_{[0,\T]}$ that takes place in the interval $[0,\T]$ of stochastic duration $\T$ is defined as
\begin{equation}
    \Sigma  (\mathcal{T})=\sum_{t=0}^{\mathcal{T}-1}\left[ \ln\frac{\rho_t(x_t)}{r(x_t)} - \ln\frac{\rho_{t+1}(x_{t+1})}{r'(x_{t+1})} \right]. \label{e1}
\end{equation}
In Eq.~\eqref{e1}, $\rho_t(x)$ is the probability for the computer to be in state $x$ at time $t$ during the computation, whereas $r(x)$ is an arbitrary \emph{reference} probability distribution.
On the other hand, $\rho_{t+1}(x)$ and $r'(x)$ correspond to the distributions retrieved applying one iteration of the computer to $\rho_t(x)$ and $r(x)$ respectively. In particular, when the reference distribution minimizes the dissipation in the computer, 
it is called the \textit{prior distribution}, and
the associated quantity~\eqref{e1} is known as the {\em mismatch cost} of the computation up to time $\T$. This cost provides a lower bound on the entropy production incurred by any digital synchronous computer that implements this dynamics over values of the computational variables $x$, without any precise assumptions about the continuous-time process implementing each successive iteration of the computation \footnote{A version of this result even holds for quantum thermodynamic processes. For further discussion of the concept of mismatch cost in stochastic thermodynamics of computation   see~\cite{Kolchinsky17,Wolpert2019,kolchinsky2021dependence,riechers2021initial}.}.
Note that while the distributions $\rho_t$ and $\rho_{t+1}$ are indexed by the iteration number, changing as the computation proceeds, the distributions $r$  and $r'$ are the same for every iteration of the computation. 

Some of our most important results are fluctuation relations and inequalities for the statistics of $\Sigma(\T)$. These are three-fold universal, in that they are valid for all (i) reference distributions $r(x)$ [although we will make specific choices for it to give a concrete thermodynamic meaning to $\Sigma$]; (ii) stopping rules $\T$ that halt before $\tau+1$; (iii) discrete-time Markov chains (DTMCs), even those with restricted initial conditions, and / or with unidirectional transitions, and even DTMCs that are implemented by underlying non-Markovian continuous-time processes.

To illustrate these results we need to first introduce more definitions. Here and throughout this paper, we use $\langle A(\mathcal{T})\rangle \equiv  \mathbb{E}[A(\T)]$ for the expectation of any functional $A$ of $x_{[0,\T]}$ over many realizations of the computation, each ending at the possibly different time $\T$. In addition, we introduce the  \textit{stochastic distinguishability} at stopping times
\begin{equation}
    \delta_\tau(\mathcal{T})= \left.\ln  \frac{\rho_t(x_t)}{\bar{\rho}_{\tau - t}(x_t)}\right\vert_{t=\T}
\end{equation}
which quantifies the asymmetry of the computational process under time-reversal through its distribution and plays a key role for identifying possible reductions in thermodynamic costs along processes with stochastic duration~\cite{gambling2021}.
The distribution $\bar{\rho}_t(x)$ above is the probability for an {\em auxiliary computation} running backwards in time to be in state $x$ at time $t$ during the computation. Such auxiliary computation evolves with a transition matrix that is the Bayesian inverse of the original computation with respect to $r$, such that it leads to the perfect retrodiction in time of the distribution $r$~\footnote{An analogous notion of stochastic distinguishability was introduced in Ref.~\cite{gambling2021}, in the context of forward and time-reversed non-stationary dynamics.}.
In Sec.~\ref{sec:effective}, we provide a formal definition for the auxiliary computational process, that allow us to address the thermodynamics of computations ending at a fixed time at the fluctuating level, and discuss how to incorporate explicitly the role of absolute irreversibility and unidirectional transitions, which is crucial because conventionally formulated computational machines have precisely those features.

In Sec.~\ref{sec:stop} we present our main results for general computations starting and ending at stochastic halting times, which include fluctuation theorems and second-law inequalities that provide lower bounds to the average  dissipation in a computation.
 A central result of our work is the derivation of an  integral fluctuation relation at stopping times which  applies to arbitrary computations:
\begin{align}
   \langle e^{- \Sigma(\mathcal{T}) - \delta_\tau(\mathcal{T})} \rangle = 1 - \Gamma_\tau.
   \label{eq:IFTSTintro}
\end{align}
In this equation, $\Gamma_\tau$ quantifies absolute irreversibility at stopping times. It equals the functional $e^{-\delta_\tau(\T)}$ averaged over the restricted set of trajectories that can take place in the auxiliary dynamics but have zero probability in the original dynamics, see Eq.~\eqref{eq:Gamma}. This contribution introduces an unavoidable source of irreversibility that limits possible reductions in the dissipation of the computer.

Among other things, Eq.~\eqref{eq:IFTSTintro} provides a second-law inequality at stopping times for the intrinsic mismatch cost,
\begin{eqnarray} \label{eq:3}
  \langle \Sigma(\mathcal{T})\rangle   \geq - \langle \delta_\tau(\mathcal{T}) \rangle - \ln [1 - \Gamma_\tau].
\end{eqnarray}
As we show in this paper, the right-hand side of Eq.~\eqref{eq:3}, which sums up the net effects of time-asymmetry and absolute irreversibility, gives a universal lower bound not only for the intrinsic mismatch cost of the computation but also for the underlying average entropy production incurred by the computer. Moreover, for the particular case of a stationary reference distribution ($r=\pi$, with $\pi'=\pi$) $\langle\Sigma(\T)\rangle$, equals the average (discrete-time) nonadiabatic entropy production~\cite{esposito2010three,hatano2001steady}.

We remark that Eq.~\eqref{eq:IFTSTintro} follows from a stronger result  that we derive here. In particular $ e^{- \Sigma(t) - \delta_\tau(t)} $ is a {\em supermartingale} process, i.e., 
$ e^{- \Sigma(t) - \delta_\tau(t)} $ 
decreases with time when conditioned on an earlier part of the trajectory of states: $\langle e^{- \Sigma(t) - \delta_\tau(t)} \,|\, x_{[0,s]} \rangle  \leq  e^{- \Sigma(s) - \delta_\tau(s)}$,
where $t\geq s\geq 0$, see Eq.~\eqref{eq:superM}~\footnote{In general in this paper,
$\langle A(t)\vert x_{[0,s]} \rangle=\mathbb{E}(A(t) \,|\, x_{[0,s]})$ is an average of functional $A$ over all computational trajectories with the condition of $x_{[0,s]} = x_0,\dots, x_s$ to be fixed to a given sequence.}.

Another contribution in this paper arises when we apply the general martingale theory for thermodynamics~\cite{Edgar2022}, to extend our results to multiple, ordered stopping times. In particular, for the case of two stopping times $\T_1$ and $\T_2$ with $\T_2\geq \T_1$ we obtain another central result:
\begin{eqnarray}
\langle\Sigma(\T_2) + \delta_\tau(\T_2) \rangle \geq \langle \Sigma(\T_1) + \delta_\tau(\T_1)\rangle,    
\label{eq:stop-start-intro}
\end{eqnarray}
which provides a powerful second-law inequality applicable to both starting and ending stochastic times of computations. As an example, in the case of a DFA, $\mathcal{T}_1$ could be the first time that some particular state $A$ is reached, and $\mathcal{T}_2$ could be the first time state $B$ is reached after state $A$ has been reached. Alternatively, $\mathcal{T}_2$ could be the \textit{second} time state $A$ is reached after the system has first reached state $A$, then left it, and then returned. See Sec.~\ref{sec:examples} for numerical illustrations of these ideas in a specific minimal model of a DFA processing binary strings.

Moreover, from Eq.~\eqref{eq:stop-start-intro} a sandwich inequality for $\langle\Sigma(\T)\rangle$ can be derived~\footnote{Here $D (\rho_0\,||\,\bar{\rho}_\tau) = \sum_x \rho_0(x)\ln[\rho_0(x)/ \bar{\rho}_\tau(x)]\geq 0$, is the Kullback-Leibler (KL)  quantifying the ``distance'' between the initial distribution of the computer's state $\rho_0$ and the distribution $\bar{\rho}_\tau$ of the auxiliary computational process at the limit time $\tau$.},
\begin{equation} 
    D (\rho_0\,||\,\bar{\rho}_\tau)- \langle \delta_\tau(\mathcal{T})\rangle\leq\langle\Sigma(\mathcal{T})\rangle \leq  \langle \Sigma (\tau) \rangle -\langle\delta_\tau(\mathcal{T}) \rangle,
\end{equation} 
extending recent research in upper bounds and inverse thermodynamic uncertainty relations in stochastic thermodynamics given in~\cite{di2023variance,Hasegawa23,salazar2022upper}.

In addition to the aforementioned fundamental results, in Sec.~\ref{sec:acc} we also combine the  supermartingale property of $ e^{- \Sigma(t) - \delta_\tau(t)} $ with the fluctuation relation~\eqref{eq:IFTSTintro} to derive universal equalities and inequalities for the probability that a computation is completed within a certain amount of time. Such idea can be applied e.g. to compute the probability that a sequence of $\tau$ ordered computational states visited by a DFA during its evolution reaches an accept state.

Sec.~\ref{sec:stochasticstart} is then devoted to sketch how our theory can be applied to investigate the thermodynamics of multiple concatenations of runs of a DFA, where after each run ends the system is reset to an initial start state and the next run begins. We conclude with \cref{sec:discussion}, where we present our main conclusions and further discuss future research directions motivated by our findings. Mathematical details of the derivations, proofs, and extra discussions are left to corresponding appendices. 

It is worth remarking that all our contributions, while originally motivated from problems arising in the context of the kinds of computational machines central to computer science theory, are applicable to any periodic process implementing a time-homogeneous DTMC, one that results in trajectories $\mathbf{x}_{[0,\T]}$.

\section{Markovian computations}
\label{sec:framework}
In the following, we make the assumption that the implementation of a task on a given computer is realized through a physical process which induces Markovian (discrete) dynamics over a set of relevant computational states. The actual physical process being modeled will be a generic physical, chemical or biological system, whose dynamics can be described at a microscopic level over a set of hidden degrees of freedom~\cite{Kolchinsky17,Wolpert19}, here assumed to be not directly accessible. In particular, it is customary to model a computation as a continuous-time Markov chain (CTMC)~\cite{parrondo2015thermodynamics,sagawa2014thermodynamic,strasberg2015thermodynamics,Wolpert19,kolchinsky2020thermodynamic}. 

In ``synchronous'' physical computers --- such as all real-world digital computers --- this CTMC is driven externally following a periodic protocol induced, e.g., by the AC electric current powering a computer. Such underlying periodic driving might be ignored in modelling the computational process, by describing its evolution by coarse graining it in time, and this results in an effective model given by a time-homogeneous DTMC. Throughout this paper, we will work at such a coarse-grained level, and consider computational processes as generic DTMCs with time-independent transition probabilities. In doing so, we will map the underlying physical process to the DTMC dynamics of the (symbolic) computational states to formulate the actual physical dissipation in a thermodynamically consistent manner.

\subsection{Stochastic computational processes}
We consider computational processes described by a DTMC that can take values over a discrete set of $N\geq 1$ computational states $x_t\in \mathcal{X}$, with $t=0,1,2\dots $. For simplicity, we assume that the transition probabilities between the computational states are time independent (however, our results can be extended to time-dependent transition probabilities). 

We write $P(x_{t+1}\,|\,x_t)$ for the conditional probability of jumping to state $x_{t+1}$ given that the previous state was $x_{t}$ in a single time-step or iteration of the computational process. (Note that in a DTMC $x_t$ can be the same as $x_{t+1}$, allowing for time instances where the system dwells in a given state.) We write $\rho_t(x)$ for the probability of being in state $x$ at time $t$, given an ensemble of realizations of the Markovian process. The associated discrete-time master equation ${\rho}_{t+1}=\mathbf{W}{\rho}_t$, where ${\rho}_t$ is an $N\times 1$ column vector and $[\mathbf{W}]_{i,j} = P({x_{t+1} =i \,|\, x_t = j})$ is the transition probability matrix. The transition matrix $\mathbf{W}$ has at least one fixed point with distribution $\pi(x)$ such that $\mathbf{W} {\pi} = {\pi}$, and if aperiodic and irreducible, ${\pi}$ becomes the unique stationary distribution in the long time run, that is, $\lim_{t \rightarrow \infty} {\rho}_t = {\pi}$. However, what follows does not require ${\pi}$ to be unique.

Throughout the paper we will write $\tau$ for the {\em limit time}  of a computation, i.e., the maximum time that can be spent to execute a computation, and assume it to be fixed. The probability of a sequence $\mathbf{x}_{[0,\tau]}=x_0,x_1,\dots, x_\tau$ is
\begin{equation}
    P(\mathbf{x}_{[0,\tau]}) = \rho_0(x_0) \prod_{t=0}^{\tau-1} P(x_{t+1}\,|\, x_{t}).
    \label{eq:Ppath}
\end{equation}
Here we allow for arbitrary initial distributions $\rho_0(x_0)$ and transition probabilities $P({{x_{t+1} = j\,|\,x_{t}}} = i) = P(j \,|\,i)$. In particular, some of the transitions might be bidirectional ($i \leftrightarrow j$) and others unidirectional ($i \rightarrow j$). Bidirectional transitions are characterized by conditional probabilities verifying $P(i\,|\,j) > 0$ whenever $P(j\,|\,i)>0$, while for unidirectional ones we can have $P(i\,|\,j) = 0$ with $P(j\,|\,i)>0$. We notice that exactly because of the existence of unidirectional transitions, it is mandatory to relax the condition of local detailed balance, which is arguably among the most common assumptions adopted in the formulation of stochastic thermodynamics~\cite{Seifert2012}.

One of the main quantities of interest in stochastic thermodynamics is stochastic entropy production (EP) which equals the logarithm of the ratio between forward and time-reversed path probabilities of a thermodynamic process~\cite{Seifert12,Kawai07}. 
This quantity, however, generically depends on the details of the underlying physical process implementing the computation, hence is not directly accessible unless certain simplifying assumptions, such as the condition of local detailed balance.
Nevertheless, here we aim to obtain a thermodynamic description of the computational processes as deduced solely from the (discrete-time) dynamics of the visible variables defining the computation $x_t \in \mathcal{X}$. While our analysis holds for arbitrary DTMCs, we focus on digital synchronous computers, which undergo a time-homogeneous dynamics over discrete time, and which we connect to the underlying physical process generating it in a simple manner. This allows us to express and bound the entropy production of the computational task implemented by the DTMC, alongside the work and heat dissipated into the environment. For simplicity, we take the \textit{continuous}-time physical process that implements the time-stationary DTMC to be periodic and choose units so that the period of the physical process is $1$. 

As an example (and to help ground the reader's intuition), suppose that our time-homogenous DTMC is implemented by a time-inhomogeneous CTMC. It is well-known that in general, this requires that the CTMC evolves over an enlarged version of the DTMC's state space  $\mathcal{Y} \supseteq \mathcal{X}$, which includes ``hidden states'' in addition to the ``visible" states of the DTMC~\cite{Wolpert19,Owen19}. In particular, this is true when the DTMC is the update function of a computational machine.
Therefore our assumption that the continuous-time physical process is periodic implies that the time-inhomogeneous CTMC is periodic. As a result, the thermodynamics arising in any single iteration of the physical system (implementing the computational machine that starts at discrete time $t$ in some state $y(t) \in \mathcal{Y}$) is independent of $t$. In the following, for further simplicity (and to ensure a time-homogeneous DTMC) we also assume that non-computational degrees of freedom in $\mathcal{Y}$ are reinitialized within every single iteration to their (possibly nonequilibrium) initial states.

\subsection{Mismatch cost}
\label{sec:mc}

Enlarging our original description over $\mathcal{Y}$ to include all relevant physical variables of the computer is crucial to define the associated entropy production and other relevant thermodynamic costs of computation. Here we show that this can be done in a standard way. 
In particular, suppose that we are interested in some generic thermodynamic {\em average  cost function} that can be written as
 \eq{
 \label{eq:2mm}
\mathcal{C}(\tau) &= S(\varrho_\tau) - S(\varrho_0) + F \nonumber \\
    &:= S(G \varrho_0) - S(\varrho_0) + F
 }
 where $S(\varrho)=-\sum_i \varrho(i) \ln \varrho(i)$ is Shannon entropy, $\varrho_0$ is any initial distribution over the (extended set of) states of the system, $G$ is the linear map that transforms that distribution to an associated ending distribution $\varrho_\tau$, and $F$ is an arbitrary linear functional of the initial state. Note that $\mathcal{C}(\tau)$
 is an implicit function of $\varrho_0$.

As a canonical example, in CTMC-based stochastic thermodynamics obeying local detailed balance, the EP generated during a process is given by \cref{eq:2mm} by setting $F$ equal to the \textit{average entropy flow} to the environment:
 \eq{ \label{eq:EF}
F = \int_0^\tau dt \sum_{v} \sum_{i, j} \varrho_t(j) K^v_{ij}(t) \ln \left[\frac{K^v_{ji}(t)}{K^v_{ij}(t)} \right]
}
where $K^v_{ij}(t)$ is the rate matrix associated to thermal reservoir $v$,
and the rate matrix of the CTMC is $\sum_v K^v_{ij}(t)$ \footnote{Note that formula for $F$ is indeed linear in $\varrho_t$}. For different choices of $F$, $\mathcal{C}(\tau)$ gives different thermodynamic quantities besides EP, such as the drop in nonequilibrium free energy of the system during the process~\cite{kolchinsky2021dependence}, among many others.

For any cost $\mathcal{C}$ of the form in Eq.~\eqref{eq:2mm}, and {\em any} physical process represented by $G$, the \textit{prior} distribution is defined to be the initial distribution $\varrho_0$ that minimizes $\mathcal{C}(\tau)$. (It is called the prior because it is, formally speaking, a prior distribution for calculating the posterior probability of an initial state of a thermodynamic process given its final state~\cite{wolpert2016free}.) We write the prior as $\varrho_{\min}$. The  associated {\em average mismatch cost} is 
  \eq{
\mathcal{M}(\tau) := D(\varrho_0 \,||\, \varrho_{\min}) - D(G \varrho_0 \,||\, G \varrho_{\min})
\label{eq:2mmm}
 }
  where $D(\varrho_1\,||\,\varrho_2) = \sum_y \varrho_1(y)\ln[\varrho_1(y)/ \varrho_2(y)]$, denotes the Kullback-Leibler (KL) divergence between the distributions $\varrho_1$ and $\varrho_2$ for the case of a discrete random variable. Note that $\mathcal{M}(\tau)$ is implicitly a function of $\varrho_0, \varrho_{min}$ and the linear function $G$ --- but nothing else. It depends on \textit{no} other property of the process besides 
  $\varrho_{min}$ and $G$.
  
 As shown in Refs.~\cite{Kolchinsky17,Wolpert19,Wolpert2020thermodynamics,kolchinsky2021dependence} we have, for all $\varrho_0$
 \eq{
 \mathcal{C}(\tau) = \mathcal{M}(\tau) + \mathcal{R}(\tau),
 \label{eq:2M}
 }
where $\mathcal{R}(\tau)$ is an additive non-negative contribution independent of $\varrho_0$ called ``residual cost''. In the specific case in which $F$ is identified with the entropy flow, residual cost is often called ``residual EP''. See \cref{app:residual} for a discussion of residual cost,
and why we ignore it in this paper.

Expressions analogous to \cref{eq:2mm,eq:2mmm,eq:2M} hold for other state spaces, e.g., real-valued states, density matrices, etc. Moreover,there are no assumptions of detailed balance or the like in the derivation of \cref{eq:2mmm}; it holds purely for mathematical reasons. For the trajectory level version of mismatch cost in Eq.~\eqref{eq:2mmm} see \cref{sec:traj_level_mmc}.

By the data-processing inequality for KL divergence~\cite{CoverThomasBook},  $\mathcal{M}(\tau)$ is never negative. Moreover, it can be shown that the prior $\varrho_{\min}$ in Eq.~\eqref{eq:2mmm} has full support (see App. A in Ref.~\cite{Wolpert2020thermodynamics}), which ensures that the mismatch cost is finite. Note also that the mismatch cost formula~\eqref{eq:2mmm} is based on evaluating $\varrho$ at both the beginning and the end of the time interval $[0, \tau]$. This means this general formula applies to \textit{any} physical process that maps $\varrho_0$ to $\varrho_\tau$, for \textit{any} choice of $\mathcal{C}$, i.e., any choice of the linear functional $F$. All the (messy) physical details of the process and the precise choice of $F$ are buried in the prior and the residual cost. 

In the following, unless explicitly stated otherwise we will focus on the case in which the cost function $\mathcal{C}(\tau)$ in Eq.~\eqref{eq:2M} is the average entropy production of the continuous-time periodic process implementing the computation. If such a process is moreover Markovian, the entropy flow $F$ will be given by the canonical CTMC expression in Eq.~\eqref{eq:EF}. However, markovianity of the continuous-time periodic process is not a necessary assumption for our results (see also Appendix~\ref{app:coarse-grain}).

\subsection{Strictly positive lower bounds for \\ dissipation in periodic processes}

As described above, in real world (synchronous, digital) physical computers, the underlying physical process implementing each iteration of the computer is identical. This is true whether that physical process is a CTMC, a quantum operation, and so on. As noticed in Ref.~\cite{ouldridge2022thermodynamics}, this means that the prior $\varrho_{\min}$ for each iteration of the computer is the same~\footnote{Here we are interested in the marginal prior distribution over computational states  $x \in\mathcal{X}$ only, not over the set $\mathcal{Y}$ containing hidden states.}. Using also the fact that non-computational variables are reinitialized in every single iteration (period) of the computational process, we can write the overall mismatch cost for any computation that takes exactly $\tau$ iterations in terms of computational variables as:
\begin{eqnarray}
\sum_{t=0}^{\tau-1} \mathcal{M}(\rho_t) = \sum_{t=0}^{\tau-1} \left[D(\rho_t \,||\, \mu) - D(\rho_{t+1} \,||\, \mu^\prime) \right],
\label{eq:sum_mismatch_costs}
\end{eqnarray}
where $\rho_t(x) = \sum_{y \notin \mathcal{X}} \varrho_{t}(y)$ is the (marginal) distribution over computational states $x \in \mathcal{X}$ only, $\mu(x) = \sum_{y \notin \mathcal{X}} \varrho_{\min}(y)$ is the prior over computational states at the beginning of (every) iteration, and $\mu' = \mathbf{W} \mu$ the prior at the end of every iteration (i.e., it is $\mu$ evolved to the end of the iteration). More details about the derivation of Eq.~\eqref{eq:sum_mismatch_costs} for the DTMC are provided in Appendix \ref{app:coarse-grain}, where we explicitly elaborate its relation to the underlying EP in the continuous-time periodic process.

We note that if $\rho_0 = \mu$ in Eq.~\eqref{eq:sum_mismatch_costs}, then the first difference of KL divergences being summed equals $0$. However, unless $\mathbf{W}$ is degenerate (e.g., the identity matrix), $\mathbf{W} {\rho}_0 \ne {\rho}_0$, and therefore $\mathbf{W} \rho_0 \ne \mu$. This in turn means that the \textit{second} difference of KL divergences being summed in \eqref{eq:sum_mismatch_costs} does \textit{not} equal $0$ (so long as $\mathbf{W}$ is not logically invertible, i.e., not a permutation matrix). Therefore in this case, the overall sum will be strictly positive. This argument can be extended to prove that so long as $\mathbf{W}$ is not logically invertible (and $\rho_0$ is not a fixed point of the dynamics), the mismatch cost sum in Eq.~\eqref{eq:sum_mismatch_costs} is not zero (see Appendix \ref{app:mismatch}). 

Since the above reasoning is true for all actual $\mu$, we can lower bound the sum \eqref{eq:sum_mismatch_costs} by minimizing over all distributions $\lambda$ in the unit simplex $\Delta_X$, whether or not they are a valid prior in some physical scenario:
\eq{
\sum_{t=0}^{\tau-1} \mathcal{M}(\rho_t) \geq  \inf_{\lambda \in \Delta_X} \sum_{t=0}^{\tau-1} \left[D(\rho_t \,||\, \lambda) - D( \rho_{t+1} \,||\, \lambda^\prime)\right] > 0
 \label{eq:sum_mismatch_costs_min}
 }
with again ${\lambda}\, ' = \mathbf{W} {\lambda}$  (see also Ref.~\cite{ouldridge2022thermodynamics}). The precise prior $\mu$ in Eq.~\eqref{eq:sum_mismatch_costs} for the EP cost function will depend on the details of the precise physical process under consideration. On the other hand, the sum \eqref{eq:sum_mismatch_costs_min} \textit{is independent of those details}. We therefore obtain a strictly positive lower bound on EP, given \textit{in toto} by $\rho_0$ and $\mathbf{W}$. This strengthened second law arises solely from the fact that we have a periodic process with a non-logically invertible $\mathbf{W}$. Moreover, because minimization in \cref{eq:sum_mismatch_costs_min} is over all possible priors, it provides a lower bound on all costs that can be written as in \cref{eq:2mm}. We therefore refer to it as the \emph{minimal dissipation}.

As an example, suppose that our DTMC is the dynamics of a noise-free,
synchronous, digital computer, with update function $f : X \rightarrow X$. Plugging in Eq.~\eqref{eq:sum_mismatch_costs_min}, the minimal possible EP is
\eq{
   \label{eq:min_EP_comp}
&\min_{\lambda \in \Delta_X} \bigg( \sum_{t=0}^{\tau-1} \sum_{x \in \Omega_t}  \rho_0(f^{-t} (x)) \ln \left[ \frac{\rho_0(f^{-t}(x))}{\lambda(x)} \right] 
\nonumber\\
&\qquad\qquad - \rho_0(f^{-t-1}(x)) \ln \left[\frac{\rho_0(f^{-t-1}(x))}{\lambda(f^{-1}(x))} \right] \bigg)
}
The term $\Omega_t$ in \cref{eq:min_EP_comp} is the set of 
all states that have nonzero probability
if the update function is applied to the actual distribution $\rho_0$ a total of $t$ times. The term $\rho_0(f^{-t}) (x)$ in \cref{eq:min_EP_comp}
is the probability, under
$\rho$, of the entire set of those states in $X$ which, after $t$ 
iterations of (the periodic processes underlying) the update function $f$ of the digital computer, are in state $x$ (and similarly for $\rho_0(f^{-t-1} (x)$). Suppose that $f$ is not just a permutation of the states of the computational machine that lie in the support of $\rho_0$. Then \cref{eq:min_EP_comp} provides a strictly positive lower bound on the dissipation incurred by any physical device that implements that computation, $f$.

In the sense that it only depends on the conditional distribution $\textbf{W}$ and the initial distribution $\rho_0$, the bound for periodic processes in \cref{eq:sum_mismatch_costs_min} is similar to the generalized Landauer's bound. In particular, the thermodynamic uncertainty relations and speed limit theorems are also lower bounds on EP that depend on the initial distribution over states and the discrete time conditional distribution of the dynamics. However, unlike the lower bound above, those other bounds depend on other properties of the process besides the initial distribution and the conditional distribution giving the dynamics (for example current precisions or expected activities). In this sense, the minimal dissipation given in \cref{eq:sum_mismatch_costs_min} is more powerful than those other lower bounds on EP (a closed form of this result in terms of Jensen-Shannon divergence has been also reported very recently in Ref.~\cite{Tasnim23}).

In this paper, we calculate mismatch costs by summing the cost over single iterations of a computational machine operating periodically, as in Eq.~\eqref{eq:sum_mismatch_costs}. In general this does \textit{not} equal the standard mismatch cost for the entire computation, with an overall prior and a single drop in KL divergence between initial and final time $\tau$. We remark that, to the authors' knowledge, the necessary and sufficient conditions for this quantity to be larger than the one we use in this paper are not known. However there is a particularly interesting case in which these two expressions become the same, namely, when EP is minimized at the stationary state of the DTMC, i.e. the prior $\mu$ coincides with $\pi$. In such case we recover from Eq.~\eqref{eq:2M} the well-known decomposition of EP into adiabatic and non-adiabatic contributions~\cite{oono1998steady,hatano2001steady,esposito2010three,Esposito10PRE}, where mismatch cost reduces to non-adiabatic EP (also called excess EP~\cite{Edgar2022}) and the residual cost becomes  adiabatic EP (house-keeping heat~\cite{oono1998steady,hatano2001steady}). 

\subsection{Deterministic Finite Automata}
\label{sec:DFAdef}

An important class of computational machines that can be described within our framework are the deterministic finite automata (DFA). There are several different, very similar definitions of DFA, some of which overlap with common definitions of ``finite state machines''. To fix the discussion, here we adopt the following definition. A \textit{deterministic finite automaton} is a 5-tuple $\left(Q, \theta,  q_0, A, f\right)$
where:
\begin{enumerate}
\item $Q$ is a finite set of \textit{(logical) states};
\item $\theta$ is a finite (input) \textit{alphabet};
\item $q_0 \in Q$ is the \textit{start state};
\item $A \subseteq Q$ is the set of \textit{accept states}; and
\item $f : Q \times \theta \rightarrow Q$ is the \textit{update function},
mapping a current input symbol and the current logical state to a next
logical state.
\end{enumerate}

A finite string of successive input symbols, i.e., an input string $\omega \in \theta$, is sometimes called an \textit{(input) word}.  To operate a finite automaton on a particular input word, one
begins with the automaton in its start state, and feeds that state together with the first symbol in the input word into the update function, to produce a new logical state. Then one feeds in the next symbol in the input word (if any), to produce a next logical state.
Note that one can represent any given DFA's update function as a directed graph, where each edge $(q_1, q_2)$ taking logical state $q_1$ to state $q_2$ is labelled by the input symbols that would cause that transition (see \cref{fig:sketch} (c) and \cref{fig:oraux} for illustrations).

Our analysis of stochastic computational processes (as introduced above) in DFAs requires assigning probabilities to the input words (or to the symbols inside them) that are fed into the automaton, as well as to identify the computational states of the DTMC $\mathcal{X}$, which may coincide or not with the set $Q$ of logical states of the DFA (typically $\mathcal{X}$ may contain more variables as e.g. previously processes symbols). An important contribution of our work will be to show how one can do this analysis even though the dynamics of a DFA --- its update function --- is deterministic and often non-invertible  (i.e., unidirectional), and given that the initial distribution over states of the DFA (though not over the input words) is a delta function, centered on the start state (i.e., leading to absolute irreversibility).

 Physically, the (probabilistically generated) input word $\omega$ may be encoded in a tape whose symbols are read by the DFA  "head" one by one in each cycle of the computation, but are not modified by the automaton operation~\footnote{Here we will not consider the eventual erasure of the input tape after the computation ends, but we focus on the thermodynamic costs and dissipation due to the computing process itself, using the information contained in the tape. Such eventual erasure costs are independent of the mismatch costs considered here and might be calculated on a separate basis.}. In this way the input tape behaves as an (energy-less) information reservoir~\cite{barato2014stochastic}, whose Shannon entropy is keep constant during the computation. More formally, we can consider that tape as forming part of the physical states of the computation in the extended state space $\mathcal{Y}$ (a Cartesian-product factor of $Q$, $\omega$, and other physical variables depending on the implementation), but we will not generically include it within the computational states in $\mathcal{X}$~\footnote{ Notice that the DTMC transition probabilities $P(x_{t+1}|x_t)$ among the (coarse-grained) computational states in $\mathcal{X}$ are given by marginalizing a fine-grained conditional distribution $P(y_{t+1} | y_t)$ where $y = (x, s, ...)$ over $s$, namely the current symbol being read, and where the update function $f$ defining the DFA specifies $P(x_{t+1} | (x_s, s_t)$. As a consequence, the transition probabilities $P(x_{t+1}|x_t)$ are time-independent and fixed along the entire computational process.}. On the other hand, we will eventually incorporate some already processed symbols explicitly into $\mathcal{X}$, which are then assumed to be stored (and modified) in extra physical variables acting as a memory for the computer.

A typical question of interest in computer science is whether the DFA is in an accept state of the set $A$ after the last symbol from the input word is processed. If that is the case, one says that the automaton \textit{accepts} that input word. In this way any given automaton uniquely specifies a \textit{language} of all input words that that automaton accepts, which is called  a \textit{regular} language. Importantly, any particular DFA can process input words of arbitrary length~ \footnote{This means that one cannot model a given DFA as some specific (and therefore fixed width) circuit, in general. The DFA will have properties that are not captured by that circuit.
In this sense, individual DFAs are computationally more powerful than individual circuits.}, and in general may enter and exit its set of accepting states multiple times, before the end of the input word. While the definition of whether an input word is accepted only depends on whether the ending logical state is an accepting state, the statistics of whether, how often, and precisely when a given DFA enters an accept state (when fed words generated by some given distribution) can be of independent interest.

\section{Intrinsic thermodynamics of computations at fixed times}
\label{sec:effective}

The mismatch cost sum introduced in Eq.~\eqref{eq:sum_mismatch_costs} depends only on the computational degrees of freedom involved in the original DTMC dynamics and provides a lower bound on the average entropy production generated by the machine implementing the computation. It is hence a particularly useful candidate to assess the intrinsic (minimal) thermodynamic costs of computations. The prior $\mu(x)$ encodes the specific details of the physical implementation of the computational process. Concern for such details can be even avoided by considering the distribution $\nu(x)$ given by the infimum of Eq.~\eqref{eq:sum_mismatch_costs_min}, which still provides a useful (positive) bound on EP.

To begin, we construct a stochastic description based on thermodynamic quantities that can be computed by introducing an {\em auxiliary process}. This process is defined in terms of the 
``forward" discrete-time dynamics $P(j \,|\, i)$, the initial distribution of that dynamics, $\rho_0(x)$, and a \emph{reference distribution} $r(x)$ over computational states. The reference $r(x)$ is arbitrary, and in particular could be chosen  to obtain stochastic versions of the mismatch cost sum in Eq.~\eqref{eq:sum_mismatch_costs} [$r(x)=\mu(x)$] and the minimum dissipation in Eq.~\eqref{eq:sum_mismatch_costs_min} [$r(x)=\nu(x)$].

\subsection{Thermodynamic costs of periodic \\ computations at the fluctuating level}
\label{sec:ref}
We start by introducing the discrete-time auxiliary dynamics of the auxiliary process $\overline{\mathbf{W}}_{i,j}$, with transition probabilities defined from the ones in $\mathbf{W}$ by
\begin{align} \label{eq:auxiliary}
   \bar{P}(i\,|\,j) \equiv \frac{P(j \,|\, i) r(i)}{ r^\prime(j)},
\end{align}
where ${r}\,' = \mathbf{W} r $ is the reference distribution $r$ evolved for one iteration, i.e. $r'(j)=\sum_i P(j\,|\,i) r(i)$~\footnote{Note that by ``$P(j \,|\, i)$'' we do \textit{not} mean the Bayesian inverse of the forward transition matrix $P(i \,|\, j)$ --- rather we mean that forward transition matrix evaluated for a transposed choice of the initial and final states.}. This auxiliary process is a bona fide Markov chain with $\sum_i \bar{P}(i\,|\,j) = \sum_i [P(j\,|\,i)r(i)]/r^\prime(j) =1$ and $0 \leq \bar{P}(i \,|\,j) \leq 1$~\footnote{This can be easily checked from $\sum_j \bar{P}(i\,|\,j) r^\prime(j) = \sum_j P(j\,|\,i) r(i) = r(i)$, which immediately implies $[\bar{\mathbf{W}}] {r^\prime} =  {r}$.}. 
Moreover, ${r^\prime}$ transforms back into ${r}$ in a single iteration under $\overline{\mathbf{W}}$. That is, $\bar{W}$ corresponds to the Bayesian inverse of $W$ with respect to the reference distribution $r$, leading to perfect retrodiction for the distribution $r$~\cite{Jaynes,kolchinsky2020thermodynamic,Buscemi21,Surace2023stateretrieval}.

To fully specify the auxiliary dynamics we must specify its initial distribution;
here we will always set it to distribution of the original actual dynamics at its limit time,  i.e., $\bar\rho_0(x) = \rho_\tau(x)$. So the joint distribution of a trajectory $\mathbf{x}_{[0,\tau]}$ under the auxiliary dynamics is
\eq{
\bar{P}({\mathbf{x}_{[0,\tau]}}) = 
\bar\rho_0(x) \prod_{t=0}^{\tau - 1}\bar{P}(x_{t+1} | x_t)
\label{eq:16}
}
Note that this choice of the initial distribution of the auxiliary process is not restricted by the choice of $r$ in any way. Note as well that $\bar{P}(i\,|\,j)$ does not necessarily coincide with the transition probabilities induced by the time-reversed implementation of the underlying physical process, but it is solely defined from the distribution $r(x)$ and the original Markov chain transition probabilities.

Using Eqs.~\eqref{eq:auxiliary} and~\eqref{eq:16}, we can write the probability of a time-reversed discrete-time trajectory, $\Theta \mathbf{x}_{[0,\tau]} = x_\tau,x_{\tau-1},\dots, x_0$, under the auxiliary dynamics as
\begin{align}
     \bar{P}(\Theta \mathbf{x}_{[0,\tau]}) &= \bar\rho_0(x_\tau) \bar{P}({x_{\tau-1}\,|\, x_{\tau}}) \dots \bar{P}({x_0 \,|\, x_1})  \nonumber \\ 
    &= \rho_\tau(x_\tau) \prod_{t=0}^{\tau-1} P(x_{t+1}\,|\, x_{t})\frac{r(x_t)}{r'(x_{t+1})} .
    \label{eq:Ppathaux}
\end{align}
The ratio between the path probability to observe a given trajectory of states, and the path probability to observe its time reversal under the auxiliary dynamics is
\begin{align}
    \Sigma(\mathbf{x}_{[0, \tau]}) &\equiv~  \ln [{P}( \mathbf{x}_{[0,\tau]}) / \bar{P}(\Theta \mathbf{x}_{[0,\tau]})] \nonumber\\
    &= \sum_{t=0}^{\tau -1} \left[ \ln\frac{\rho_t(x_t)}{r(x_t)} - \ln\frac{\rho_{t+1}(x_{t+1})}{r'(x_{t+1})} \right],
    \label{eq:Sna}
\end{align}
providing us, for $r = \mu$, a stochastic version of the mismatch cost sum in Eq.~\eqref{eq:sum_mismatch_costs}, and for $r = \nu$, the minimal dissipation in Eq.~\eqref{eq:sum_mismatch_costs_min}. The functional $\Sigma(\mathbf{x}_{[0, \tau]})$ is an example of a ``$\Sigma-$entropic functional", as introduced in Ref.~\cite{Edgar2022}. 

The specific choice for the transition probability of the auxiliary dynamics introduced in Eq.~\eqref{eq:auxiliary} is crucial for avoiding divergences that would be induced by unidirectional links if we evaluate expressions like $\ln[P(i\,|\,j)/P(j\,|\,i)]$ --- expressions that appear in  most functionals associated with entropy production. This makes the functional $\Sigma$ given by Eq.~\eqref{eq:Sna} suitable to tackle fluctuations of Markovian processes with unidirectional transitions, which are precisely the (idealized) dynamics of many computational processes.

Here and in the following, as shorthand, we will often write trajectory-level quantities such as $\Sigma(\mathbf{x}_{[0, \tau]})$ simply as $\Sigma(\tau)$, with the precise trajectory left implicit. Following such shorthand notation, equation~\eqref{eq:Sna} can be decomposed as
\begin{equation}
    \Sigma(\tau) = \Delta S_\mathrm{sys}(\tau) - \Delta \phi(\tau).
    \label{eq:Sna2}
\end{equation}
where we write the change in stochastic Shannon entropy of the computer as
\eq{
\Delta S_\mathrm{sys}(\mathbf{x}_{[0, \tau]}) &:= -  \sum_{t=0}^{\tau -1} \left[ \ln\rho_t(x_t) - \ln\rho_{t+1}(x_{t+1}) \right]  \nonumber \\
    &=  - \ln \rho_\tau(x_\tau) + \ln \rho_0(x_0) ,
}
and write the change in the \textit{nonequilibrium potential} as
\eq{ \label{eq:phi}
\Delta \phi(\mathbf{x}_{[0, \tau]}) := \sum_{t =0}^{\tau-1} \left[-\ln r^\prime(x_{t+1}) + \ln r(x_t) \right]. 
}
Such non-equilibrium potentials have been fruitfully employed in steady-state  thermodynamics~\cite{hatano2001steady,Manzano15, Manzano18}, and account for the excess of entropy absorbed from the environment during the computation $\mathbf{x}_{[0,\tau]}$ whenever the state of the system $\rho_t$ differs from the distribution $r$ along its time evolution.

Suppose that the initial distribution $\rho_0(x)$ has full support. Then if we average Eq.~\eqref{eq:Sna} over ${P}( \mathbf{x}_{[0,\tau]})$ we get $\langle \bar{P}(\Theta \mathbf{x}_{[0,\tau]}) / {P}( \mathbf{x}_{[0,\tau]}) \rangle = 1$, which is an integral fluctuation relation~\cite{Seifert12}, $\left\langle e^{-\Sigma(\tau)} \right\rangle = 1 $. Moreover, $\langle \ln[{P}( \mathbf{x}_{[0,\tau]})/\bar{P}(\Theta \mathbf{x}_{[0,\tau]})]\rangle\geq 0 $ is a KL divergence, which can be rewritten in an appealing form as 
\begin{align} \label{ineq-fixed}
     \langle \Sigma(\tau)\rangle  = \sum_{t=0}^{\tau -1} \left[ D (\rho_t\,||\,r)  - D (\rho_{t+1}\,||\,r') \right] \geq 0.
\end{align}
We notice that for the choice $r = \mu$ we recover the expression for mismatch cost sum in Eq.~\eqref{eq:sum_mismatch_costs} while for $r = \nu$ we obtain Eq.~\eqref{eq:sum_mismatch_costs_min}, as expected. Crucially, for  the two choices $r = \mu$ and $r = \nu$, the quantity $\langle \Sigma(\tau) \rangle$ provides a lower bound on the total average entropy production incurred in the physical implementation of the computational process, and therefore we may refer to it as the \emph{intrinsic mismatch cost} associated to a given computation. We remark that here and above averages are over trajectories of fixed length $\tau$, that is, $\langle\Sigma(\tau)\rangle \equiv\mathbb{E}(\Sigma(\tau))$.

For more general choices of $r$, the quantity $\Sigma(\tau)$ can still be defined (as long as the distribution $r$ has full support over $\mathcal{X}$), however it cannot be guaranteed in general that $\langle \Sigma(\tau) \rangle$ would provide a lower bound on the underlying entropy production anymore.
In particular, by taking $r = \pi$, the stationary state of the DTMC, $\Sigma(\tau)$ becomes the discrete-time non-adiabatic entropy production for a relaxation process, whose average reads
\begin{eqnarray} \label{eq:non-adiabatic}
 \langle \Sigma(\tau)\rangle = D (\rho_0\,||\,\pi)  - D (\rho_\tau\,||\,\pi) \geq 0,   
\end{eqnarray}
thus we recover the expression for EP proposed by Spohn~\cite{Spohn78} (see also Ref.~\cite{Manzano18}). Remarkably in this case $\langle \Sigma(\tau) \rangle$ becomes non-extensive in time, contrary to the general case [c.f.~\eqref{ineq-fixed}]. As a consequence, the steady state $\pi$ of the DTMC (whenever aperiodic and irreducible) becomes the natural candidate for the prior $\nu$ providing the infimum in Eq.~\eqref{eq:sum_mismatch_costs_min} in the large time limit. Therefore we expect the non-adiabatic entropy production in Eq.~\eqref{eq:non-adiabatic} to provide the minimum dissipation of the computation in many cases of interest. However it is worth remarking that for ensuring the average non-adiabatic entropy production  to be a lower bound on the EP would require $\pi$ to share support with the initial distribution $\rho_0$ ---which would often not be the case in the computational context--- and $\pi$ being also invariant state in the time-reversed (underlying) physical dynamics of the computer~\cite{Manzano18}.

\subsection{The role of absolute irreversibility}

In many models of computation in TCS, the initial distribution $\rho_0(x)$ over the states of the computational machine is restricted to a subset of computational states in~$\mathcal{X}$. For instance, almost any automaton ---in particular, not just a DFA but also a TM--- starts 
in a single, predetermined state, $x_0$. Such a system
may have a delta-function initial distribution, $\rho_0(x) = \delta_{x, x_0}$. For such an initial distribution the quantity $ e^{-\Sigma(\tau)}  = \bar{P}(\Theta \mathbf{x}_{[0, \tau]})/P(\mathbf{x}_{[0, \tau]})$ may become ill-defined as there might be trajectories for which $P(\mathbf{x}_{[0,\tau]}) = 0$, but $\bar{P}(\Theta \mathbf{x}_{[0,\tau]}) > 0$, e.g., trajectories in the auxiliary dynamics that do only reach states different from $x_0$. This phenomenon has been often referred to as {\em absolute irreversibility}~\cite{Murashita2014, Funo2015}. 

Following the techniques in Refs.~\cite{Murashita2014, Funo2015,masuyama2018information} one can circumvent the divergence associated with absolute irreversibility by restricting the averages over sets of trajectories for which the intrinsic mismatch cost, $\Sigma(\tau)$, is well defined. Adopting the language of modern probability theory~\cite{Williams}, we call such sets {\em filtrations} (see also ~\cite{Edgar2022}). In particular, we denote $\mathcal{F}$ the filtration containing all possible trajectories $\mathbf{x}_{[0,\tau]}$ taking place in $[0,\tau]$. Similarly, we call  $\mathcal{F}_\mathrm{AI}$ the filtration containing all ``absolutely irreversible" trajectories, that is, trajectories for which $P(\mathbf{x}_{[0,\tau]}) = 0$, but $\bar{P}(\Theta \mathbf{x}_{[0,\tau]}) > 0$. On the other hand, we denote the complementary set of ``absolutely continuous" trajectories as $\mathcal{F}_\mathrm{AC}$, such that $\mathcal{F} = \mathcal{F}_\mathrm{AC} \cup \mathcal{F}_\mathrm{AI}$.
Using these definitions, an extended version of the integral fluctuation theorem (IFT) for the  intrinsic mismatch cost is shown,
\begin{equation} \label{eq:FTai}
\langle e^{-\Sigma(\tau)} \rangle = 1 - \gamma_\tau
\end{equation}
where $0 \leq \gamma_\tau \leq 1$  the total probability that the time-reversed picture of any absolutely irreversible trajectory (i.e. belonging to $\mathcal{F}_{\mathrm AI}$) occurs in the auxiliary dynamics 
\begin{equation} \label{eq:gamma}
\gamma_\tau = \sum_{\mathbf{x}_{[0, \tau]} \in \mathcal{F}_\mathrm{AI}} \bar{P}(\Theta \mathbf{x}_{[0, \tau]})\leq 1. \end{equation}
Applying Jensen's inequality $\langle e^{x}\rangle \geq e^{\langle x\rangle}$ to the IFT~\eqref{eq:FTai} we obtain a lower bound on the intrinsic mismatch cost, implying a minimum dissipation due to the restricted initial condition:
\begin{equation} \label{eq:second-law-ai}
   \langle  {\Sigma}(\tau) \rangle \geq  - \ln[1-\gamma_\tau] \geq 0,
\end{equation}
where the second inequality follows from $\gamma_\tau \geq 0$ and hence extends the applicability of Eq. \eqref{ineq-fixed} to systems showing absolute irreversibility. We remark that here absolute irreversibility arises because of the restricted initial distribution, but not because of the unidirectional transitions, since they have been flipped in the auxiliary dynamics according to Eq.~\eqref{eq:auxiliary}. Fluctuation theorems similar to Eq.~\eqref{eq:FTai} has been previously derived within the canonical framework of stochastic thermodynamics for entropy production~\cite{Murashita2014} and standard mismatch cost~\cite{kolchinsky2021dependence}, as well as in the inclusive Hamiltonian framework for entropy production~\cite{kardes2022inclusive}.

\section{Thermodynamics of computations at stochastic stopping times}
\label{sec:stop}

We now extend our analysis to investigate the thermodynamics of computations which first reach a 
computational state of interest at a time that varies depending on the random input provided to the computer. In doing so, we extend the martingale theory for stochastic thermodynamics~\cite{Edgar2022} to accommodate unidirectional transitions and arbitrary initial distributions leading to absolute irreversibility.

Consider a random sequence of $\tau$ bits sequentially fed into a computer (e.g. a  DFA) see also Fig.~\ref{fig:sketch}:
\begin{equation}
    \underbrace{0\;0\;0\;1\;0\;1\dots 0\;1\;1\;1}_{\tau {\rm\; bits}},
\end{equation}
with $\tau\geq 1$ being the \textit{word length}  processed by the machine. While processing a specific sequence, the computer jumps between its computational states, as described in Sec.~\ref{sec:DFAdef}.
We are interested in the thermodynamics of the (physical implementation of the) computer during the time from when it starts to a \textit{stopping time}, $\mathcal{T}$, that is until when a \textit{stopping condition} is met. For example we will often consider that the stopping condition is simply that the computer has for the first time reached an accept state. Note that this stopping time generally takes a different value when processing different words. Since the words are generated by sampling a distribution, this means that the stopping time is a random variable.

Generalizing from this case to give a fully formal definition, a \textit{stopping time} is the earliest instance when a particular condition concerning the entire trajectory generated by a stochastic process is met:
\begin{equation}
    \mathcal{T}(\mathbf{x}_{[0,t]}) :=  \inf \{  t\in [0,\tau] \;\vert\; \mathbf{x}_{[0,t]}\in \Omega \},\label{stop}
\end{equation}
where $\Omega\subseteq \mathcal{F}$ denotes the set of trajectories satisfying the stopping condition. For example, $\Omega$ might be the set of trajectories of a given DFA that have reached an accept state at least once.

Note that its definition in Eq.~\eqref{stop} involves a limit time~$\tau$. So the stopping time associated with each stochastic trajectory is a bounded random variable that obeys $0 \leq \mathcal{T}\leq \tau$. As shorthand, from now on we will typically just write ``$ \mathcal{T}$", leaving the precise trajectory $\mathbf{x}_{[0,\mathcal{T}]}$ implicit. It is also worth remarking that the computational machine does not necessarily stop functioning at $\T$, but this variable can just signal to us the time at which a specific computation is processed (e.g. accepting a word). We will therefore sometimes refer to $\T$ in this context as the \emph{computation time}, which is a particular instance of a (bounded) stopping time.

\subsection{Martingale theory with absolute irreversibility}

Inspired by \cite{gambling2021}, we now introduce the \textit{stochastic distinguishability} between the computational process and the auxiliary process. Stochastic distinguishability (with respect to time $\tau$) evaluated at time $t \leq \tau$ is defined as
\begin{equation} \label{eq:SD}
    \delta_\tau(t) := \ln  \frac{\rho_t(x_t)}{\bar{\rho}_{\tau - t}(x_t)} ,
\end{equation}
where $\bar{\rho}_{\tau - t}(x)$ is the probability distribution of the auxiliary process defined in \cref{eq:auxiliary}, evaluated at the conjugate time $\tau - t$ for the state $x_t$. 
(Recall that the auxiliary dynamics has initial distribution $\bar{\rho}_0(x) = \rho_\mathrm{\tau}(x)$, i.e., it is the distribution of the original dynamics at the limit time $\tau$.) 
Stochastic distinguishability is a measure of the asymmetry between the original and the auxiliary dynamics and plays a crucial role in martingale theory for stochastic thermodynamics of non-stationary processes~\cite{Edgar2022}. 

It will be useful to consider an associated process involving the stochastic distinguishability,
\begin{eqnarray}\label{eq:smprocess}
M_\tau(t)&:=& e^{-\Sigma(t) - \delta_\tau(t)} \nonumber \\
&=&\frac{\bar{\rho}_{\tau - t}(x_t)}{\rho_t(x_t)}\left[\prod_{s=0}^{t-1}\frac{r(x_s)}{\rho_s(x_s)} \frac{\rho_{s+1}(x_{s+1})}{r'(x_{s+1})}\right]. 
\end{eqnarray}
In the  second line of Eq.~\eqref{eq:smprocess} we used the first equality in Eq.~\eqref{eq:Sna} together with Eq.~\eqref{eq:SD} and the second line in Eq.~\eqref{eq:Sna}.
Notice that $M_\tau(\tau)=e^{-\Sigma(\tau)}$ because $\delta_\tau(\tau) = 0$. In general though $\delta_\tau(t) \neq 0$ for $t<\tau$, and so $M_\tau(t) \neq e^{-\Sigma(t)}$ for such $t$.

An important property of $M_\tau$ is that the expectation of $M_\tau(\tau)$ conditioned on a fixed trajectory ending at a time $0 \le t \le \tau$, satisfies
\begin{eqnarray} \label{eq:superM}
    \langle M_\tau(\tau) \,|\, \mathbf{x}_{[0,t]} \rangle  = M_\tau (t) ~[ 1 - \alpha_\tau (t)]
\end{eqnarray}
where we introduced the quantity defined by
\eq{ \label{eq:correctionai}
  \alpha_\tau(t)   &:= \sum_{\mathbf{x}_{[t+1, \tau]} \in \mathcal{F}_\mathrm{AI}} \frac{\bar{P}(\Theta \mathbf{x}_{[t, \tau]})}{ \bar{\rho}_{\tau - t}(x_t)} \leq 1, ~~~~ t<\tau 
}
and $\alpha_\tau(\tau) := 0$ (See Appendix~\ref{appS} for details.) Combining Eq.~\eqref{eq:superM} with the fact that $\alpha_\tau(t) \leq 1$ we establish that $M_\tau(t)=e^{-\Sigma(t) - \delta_\tau(t)}$ is a {\em supermartingale}:
\begin{equation} \label{eq:superM2}
     \langle M_\tau(\tau) \,|\, \mathbf{x}_{[0,t]} \rangle  \leq M_\tau(t),
\end{equation}
i.e., its conditional expectation given a fixed trajectory of length $t < \tau$ monotonically decreases over time.
Note that for $t=0$ one has $\Sigma(0) = 0$ and hence Eq.~\eqref{eq:superM} yields the IFT with absolute irreversibility [cf. Eq.~\eqref{eq:FTai}]: 
\eq{\label{eq:fixedFT}
&\langle e^{-\Sigma(\tau)} \rangle = \langle M_\tau(\tau) \rangle \nonumber 
= \sum_{x_0} \rho_0(x_0) \langle M_\tau(\tau) \,|\, x_0 \rangle \nonumber  \\
&= \sum_{x_0} \rho_0(x_0) M_\tau(0)~ [1 - \alpha_\tau (0)] =  1 - \gamma_\tau,
}
where we have used Eq.~\eqref{eq:gamma} in the last equality. 
 In addition, in the absence of absolute irreversibility,  $\mathcal{F}_\mathrm{AI}$ is the empty set and $\alpha_\tau(t) = 0$ for all $t\in[0,\tau]$. In such a case $M_\tau(t)$ in Eq.~\eqref{eq:superM} becomes a  martingale. Therefore in that limit we would be able to use the analysis in \cite{gambling2021,Edgar2022} on the thermodynamics of systems with stochastic stopping times. However, that analysis does not directly apply for generic initial states $\rho_0(x)$ without full support.

\subsection{Integral fluctuation relations with absolute irreversibility at stopping times}

 Fortunately, the fact that  $M_\tau (t)$ is a supermartingale  rather than a martingale when our system has absolute irreversibility does not prevent us from analyzing its thermodynamics at stopping times. To carry out such analysis, here we closely follow the derivation of Doob's optimal stopping theorem for martingales, generalizing it to apply to supermartingales that are written as in Eq.~\eqref{eq:superM}. 
 
 As elaborated in Appendix~\ref{appT}, this generalized form of the optimal stopping theorem provides an fluctuation theorem at stopping times, which is valid even in the presence of absolute irreversibility:
\begin{align}
 \langle e^{- \Sigma(\mathcal{T}) - \delta_\tau(\mathcal{T})} \rangle = \langle M_\tau(\mathcal{T}) \rangle = 1 - \Gamma_\tau,
   \label{eq:IFTST}
\end{align}
 where $\mathcal{T}\leq \tau$ is the (stochastic)
 stopping time,  
 $\Gamma_\tau \in [0, 1]$ is a contribution from absolute irreversibility, and therefore $ \langle e^{- \Sigma(\mathcal{T}) - \delta_\tau(\mathcal{T})} \rangle \leq 1$. Since $\langle . \rangle$ is an average over trajectories, and different trajectories have different stopping times, $\langle e^{- \Sigma(\mathcal{T}) - \delta_\tau(\mathcal{T})} \rangle$ involves averaging over (stochastic) values of $\mathcal{T}$. This introduces statistical coupling between the time $\mathcal{T}$ and the value $\Sigma(\mathcal{T})$.

The quantity $\Gamma_\tau$ appearing in~\eqref{eq:IFTST} is an average of the functional $e^{-\delta_{\tau}(\mathcal{T})}$ evaluated at stopping times $\mathcal{T}$ for trajectories leading to absolute irreversibility:
\begin{align}\label{eq:Gamma}
  \!\! \Gamma_\tau & :=  \sum_{\mathcal{T}=0}^\tau \sum_{ \mathbf{x}_{[0,\mathcal{T}]} \in \mathcal{F}_\mathrm{AI}^{(\mathcal{T})}} \bar{P}(\Theta \mathbf{x}_{[0,\mathcal{T}]}) \frac{\bar{\rho}_{\tau-\mathcal{T}}(x_\mathcal{T})}{{\rho}_\mathcal{T}(x_\mathcal{T})}.
\end{align}
To understand its meaning intuitively, first note that the second summation in $\Gamma_\tau$ is done over trajectories $\mathbf{x}_{[0,\mathcal{T}]}$ that belong to  $\mathcal{F}_\mathrm{AI}^{(\mathcal{T})}$, that is, trajectories verifying the stopping condition for the first time at $\mathcal{T}$, but that have zero probability to occur in the original process $P(\mathbf{x}_{[0,\mathcal{T}]}) = 0$, due to the restricted shape of the initial distribution $\rho_0(x)$. We notice also the presence of the distribution $\bar{\rho}_{\tau - \mathcal{T}}(x)$, which is due to $\delta_\tau(\mathcal{T})$.
That is, $\Gamma_\tau$ consists of the total probability of trajectories starting at the stopped point $x_\mathcal{T}$ according to distribution $\bar{\rho}_{\tau - t}(x)$, and not turning back to the set of states with $\rho_0(x) > 0$ under the auxiliary dynamics. Recall also that the reference distribution $r$ determining the precise meaning of $\Sigma(\mathcal{T})$ appears in \cref{eq:Gamma} only implicitly, due to the definitions of $\bar{P}$ and $\bar{\rho}_t$.

The inequality $\Gamma_\tau \leq 1$ is saturated when all trajectories are in the set $\mathcal{F}_\mathrm{AI} = \mathcal{F}$, for which the sum over all trajectories in Eq.~\eqref{eq:Gamma} is obtained, that is $\Gamma_\tau = \sum_{t=0}^{\tau} \sum_{\mathbf{x}_{[0,t]} \in \mathcal{F}^{(t)}} \bar{P}(\Theta \mathbf{x}_{[0,t]}) \bar{\rho}_{\tau - t}(x_t)/ \rho_t(x_t)= 1$. Moreover, we also have $\Gamma_\tau
\geq 0$, since it is a sum of probabilities. Whenever the initial distribution $\rho_0(x)$ is not restricted in the state space, we obtain $\Gamma_\tau = 0$, and recover the standard form of the fluctuation theorem at stopping times for non-stationary processes~\cite{gambling2021,Edgar2022}.

It is worth remarking here that our previous results for fixed times [Eqs.~\eqref{eq:FTai} and \eqref{eq:gamma}] can be directly obtained from Eqs.~\eqref{eq:IFTST} and \eqref{eq:stop-sl} by letting $\mathcal{T} = \tau$, i.e., when all trajectories are  stopped at the final time $\tau$, as we also discuss below in more detail. Our results thus provide an extension of Martingale theory to cover different versions of mismatch costs in physical scenarios with absolute irreversibility, where martingales can be transformed into super-martingales via the  correction term $\alpha_\tau(t)$ in Eq.~\eqref{eq:correctionai}, and stopping-time fluctuation relations can be derived from them. 

Moreover, using the fact that $M_\tau(t)$ is a supermartingale [c.f. Eq.~\eqref{eq:superM2}], we can also readily apply Doob's optional sampling theorem~\cite{Doob} for supermartingales to obtain (see Appendix~\ref{app:sampling}):
\begin{equation}
    \langle e^{-\Sigma(\T_2) - \delta_\tau(\T_2)} \rangle \leq \langle e^{-\Sigma(\T_1) - \delta_\tau(\T_1)}\rangle ,
    \label{eq:doob2st2}
\end{equation}
where $\T_1$ and $\T_2$ are two stopping times, ordered such that $P(\T_2 \geq \T_1)=1$, but otherwise arbitrary. Taking $\T_1 = \T$ and $\T_2 = \tau$, the above Eq.~\eqref{eq:doob2st2}, together with the FT for stopping times [Eq.~\eqref{eq:IFTST}] and fixed-times [Eq.~\eqref{eq:fixedFT}] implies:
\begin{align}
\Gamma_\tau &~= 1- \langle e^{-\Sigma(\T_2) - \delta_\tau(\T_2)} \rangle \nonumber \\ 
&~\leq  1 - \langle e^{-\Sigma(\tau)} \rangle = \gamma_\tau,
\end{align}
where we have used $\delta_\tau(\tau)=0$. The above inequality implies that the absolute irreversibility term at stopping times $\Gamma_\tau$ is always smaller than its fixed-time counterpart $\gamma_\tau$, that is, absolute irreversibility implies always greater dissipation at fixed-times than at stopping times.

\subsection{Second-law inequalities at stopping times: universal lower and upper bounds}

If we apply Jensen's inequality $\langle e^{x}\rangle \geq e^{\langle x\rangle}$ to the fluctuation theorem of Eq.~\eqref{eq:IFTST} we derive a second-law inequality at stopping times:
\begin{eqnarray} \label{eq:stop-sl}
  \langle \Sigma(\mathcal{T})\rangle   \geq - \langle \delta_\tau(\mathcal{T}) \rangle - \ln [1 - \Gamma_\tau].
  \label{eq:2LAI}
\end{eqnarray}
This sets a strict lower bound on the average dissipation incurred by a given computation up to an arbitrary stopping time $\T$, from its time-reversal-symmetry breaking (as quantified by $\langle \delta_\tau(\mathcal{T}) \rangle$) and the absolute irreversibility (as quantified by $\Gamma_\tau$). 

Moreover, $\Gamma_\tau \geq 0$ implies that $ -\ln [1 - \Gamma_\tau]\geq 0$. Therefore Eq.~\eqref{eq:stop-sl} also implies the simpler bound
\begin{eqnarray} \label{eq:stop-s2}
  \langle \Sigma(\mathcal{T})\rangle   \geq  - \langle \delta_\tau(\mathcal{T}) \rangle.\label{eq:2LAIw}
\end{eqnarray}
These inequalities suggest that $\langle \Sigma(\mathcal{T})\rangle$ might be negative whenever $\langle \delta_\tau(\mathcal{T})\rangle \geq - \ln [1 - \Gamma_\tau] \geq 0$, as we discuss in detail further below.

Any concave function [such as $\ln (x)$] of a supermartingale yields another supermartingale by Jensen's inequality. Therefore the supermartingale property of $M_\tau(t)$ also implies that $\ln[M_\tau(t)] = -\Sigma(t) - \delta_\tau(t)$ is supermartingale. So $\Sigma(t) + \delta_\tau(t)$ is a submartingale, i.e. it conditionally increases with time. If we now invoke Doob's optional sampling theorem for submartingales we get the inequality:
\begin{equation}
    \langle\Sigma(\T_2) + \delta_\tau(\T_2) \rangle \geq \langle \Sigma(\T_1) + \delta_\tau(\T_1)\rangle ,
    \label{eq:doob2st}
\end{equation} 
where again $\T_1$ and $\T_2$ are two ordered stopping times with $P(\T_2 \geq \T_1)=1$. This inequality has several implications, the most immediate one being a second law for intervals between two ordered stopping times $\T_1$ and $\T_2$:
\begin{equation}
 \langle \Delta \Sigma(\T_1,\T_2)\rangle \geq - [\langle \delta_\tau(\T_2) \rangle - \langle  \delta_\tau(\T_1) \rangle],
\label{eq:second-starting}
\end{equation} 
where $\langle \Delta \Sigma({\T_1,\T_2}) \rangle := \langle \Sigma(\T_2) \rangle - \langle \Sigma(\T_1) \rangle$. 
This inequality provides a result applicable to both stochastic stopping and starting times, bounding the entropy production incurred for computations that both start and end at stochastic times.

As an example, inequality~\eqref{eq:second-starting} provides a bound concerning the stochastic interval between the first time that a DFA enters an accept state, and the earliest \textit{subsequent} time that
it again enters an accept state, after having left the set of accept states 
in between. Then the time up to $\T_1$ can be interpreted as the time it took for the DFA to accept a first sub-string of the full input word, and the time between $\T_1$ and $\T_2$ can be interpreted as as the time it took for the DFA to accept a \textit{second} sub-string of the full input word, a sub-string which follows the first one. Again, the inequality in \cref{eq:second-starting} suggests that $\langle \Delta \Sigma(\T_1,\T_2)\rangle$ might eventually become negative for such a case, whenever there is an increasing time-reversal-asymmetry, i.e. for $\langle \delta_\tau(\T_2) \rangle > \langle  \delta_\tau(\T_1) \rangle$.

Moreover, for the choice $\T_1=\T$ and $\T_2=\tau$, the inequality~\eqref{eq:doob2st} gives us the following {\em upper} bound for the intrinsic mismatch cost at stopping times
\begin{eqnarray} \label{eq:improvement}
 \langle\Sigma(\mathcal{T})\rangle\leq \langle \Sigma (\tau) \rangle - \langle \delta_\tau(\mathcal{T}) \rangle.
\end{eqnarray}
The inequality~\eqref{eq:improvement} implies that whenever $\langle\delta_\tau(\T)\rangle \geq 0$, the intrinsic mismatch cost at stopping times will be upper bounded by its fixed-time counterpart, suggesting a drop in the thermodynamic costs of the computation at stopping times. On the other hand by taking $\T_1=0$ and $\T_2=\T$ in Eq.~\eqref{eq:doob2st}, we obtain an alternative second-law at stopping times, namely:
\begin{equation}
     \langle\Sigma(\mathcal{T}) \rangle  \geq  D (\rho_0\,||\,\bar{\rho}_\tau) - \langle \delta_\tau(\mathcal{T}) \rangle ,
    \label{eq:stop-s3}
\end{equation}
to be compared with Eqs.~\eqref{eq:stop-sl} and \eqref{eq:stop-s2}. Here we have used that $\Sigma(0)=0$ and 
\begin{equation}
    \langle \delta_\tau(0)\rangle = \sum_{x_0} \rho_0(x_0) \ln \frac{\rho_0(x_0)}{\bar\rho_\tau(x_0)}= D (\rho_0\,||\,\bar{\rho}_\tau).
    % \geq  0. 
    \label{eq:65}
\end{equation}

This inequality provides us an alternative lower bound on the intrinsic cost of the computation. We notice that, while we expect it to be less tight in general than Eq.~\eqref{eq:stop-sl}, it has the advantage of relying on the KL divergence between initial distribution $\rho_0$ and the final distribution in the auxiliary dynamics $\bar{\rho}_\tau$, which we expect to be more easily computable than $\Gamma_\tau$ in Eq.~\eqref{eq:Gamma}. 
Remarkably, combining Eqs.~\eqref{eq:improvement} and \eqref{eq:stop-s3} we find a sandwich inequality for the intrinsic mismatch cost at stochastic times,
\begin{equation} 
     D (\rho_0\,||\,\bar{\rho}_\tau) - \langle \delta_\tau(\mathcal{T})\rangle \leq  \langle\Sigma(\mathcal{T}) \rangle \leq \langle \Sigma (\tau) \rangle - \langle \delta_\tau(\mathcal{T}) \rangle ,
     \label{eq:sandwich}
\end{equation}
which provides both upper and lower bounds on $\langle \Sigma(\T) \rangle$.

The stopping time fluctuation relation in Eq.~\eqref{eq:IFTST} and the inequalities~\eqref{eq:stop-sl}-\eqref{eq:stop-s3} for the intrinsic thermodynamic costs in computational processes with stochastic stopping times provide our main results. In the following we further discuss their interpretation and some of their implications, while in Section~\ref{sec:examples} we investigate their applications to CS setups with some illustrative examples.

\subsection{Thermodynamic interpretation and implications}

The second-law inequality~\eqref{eq:stop-s2}, $\langle \Sigma(\mathcal{T})\rangle   \geq  - \langle \delta_\tau(\mathcal{T}) \rangle$ [as well the stronger versions~\eqref{eq:stop-sl} and~\eqref{eq:stop-s3}], suggests that both the intrinsic mismatch cost and the underlying entropy production incurred in a given computation may be negative on average when evaluated at stopping times. To understand how this is possible in light of the data-processing inequality we write $\langle\Sigma(\mathcal{T})\rangle$ explicitly as the functional~\eqref{eq:Sna} averaged over many trajectories that are stopped each at a stochastic time $\mathcal{T}$: 
\begin{eqnarray}
\langle \Sigma (\mathcal{T}) \rangle &=&\sum_{\mathcal{T}=0}^{\tau}
p(\mathcal{T}) \sum_{t=0}^{\mathcal{T}-1}\Big[\sum_{x_t}\rho_t(x_t\,|\,\mathcal{T}) \ln\frac{\rho_t(x_t)}{\mu(x_t)} \nonumber\\
&-&\sum_{x_{t+1}} \rho_{t+1}(x_{t+1}\,|\,\mathcal{T})\ln\frac{\rho_{t+1}(x_{t+1})}{\mu'(x_{t+1})} \Big].\label{eq:SigmaT}
\end{eqnarray} 
Here, $p(\mathcal{T})$ denotes the probability that the stopping time takes value $\mathcal{T}$. 
Similarly, $\rho_t(x\,|\,\mathcal{T})$  denotes the conditional probability that the process takes the value $x$ at time $t$ {\em given that} the stopping condition is met at time $\mathcal{T}$. Because $\rho_t(x\,|\,\mathcal{T})\leq \rho_t(x)$ in general,  the terms  $\sum_{x_t}\rho_t(x_t\,|\,\mathcal{T}) \ln[\rho_t(x_t)/\mu(x_t)]$ and  $\sum_{x_{t+1}}\rho_{t+1}(x_{t+1}\,|\,\mathcal{T}) \ln[\rho_{t+1}(x_{t+1})/\mu\,'(x_{t+1})] $ are {\em not} KL divergences in general, and thus not necessarily greater or equal than zero~(see also Ch.~8.3 in Ref.~\cite{Edgar2022}). This implies that $\langle \Sigma (\mathcal{T}) \rangle $ can in principle be negative. The second law at stopping times~\eqref{eq:stop-s2} permits $\langle \Sigma (\mathcal{T}) \rangle\leq 0 $ whenever $\langle \delta_\tau (\mathcal{T}) \rangle \geq 0$, yet it is not clear when this would be actually the case.

The explicit expression for the stochastic distinguishability at stopping times reads
\begin{equation} \label{eq:SDstex}
    \langle\delta_\tau(\mathcal{T}) \rangle=\sum_{\mathcal{T}=0}^{\tau}
\sum_{\displaystyle x_\mathcal{T}}p(\mathcal{T}) \rho_\mathcal{T}(x_\mathcal{T}\,|\,\mathcal{T}) \ln  \frac{\rho_\mathcal{T}(x_\mathcal{T})}{\bar{\rho}_{\tau - \mathcal{T}}(x_\mathcal{T})} .
\end{equation}
Equation~\eqref{eq:SDstex} also reveals that $\langle\delta_\tau(\mathcal{T}) \rangle$ {\em is not a KL divergence in general}, and thus can in principle take any sign, yet so far only examples where $\langle\delta_\tau(\mathcal{T}) \rangle\geq 0$ have been reported in the literature.  We remark that $\langle\delta_\tau(\mathcal{T}) \rangle$ is not a KL divergence unless $\mathcal{T}=\tau$, for which the process ``stops" at the deterministic limit time $\tau$, and one has that the joint stopping-time probability distribution
\begin{equation}\label{eq:detstop}
     p(\mathcal{T}) \rho_t(x_t\,|\,\mathcal{T}) = \begin{cases} 
0
    & \text{if } \mathcal{T} < \tau \\
&\\  
\rho_\tau(x_\tau)
    & \text{if } \mathcal{T} =\tau ,
\end{cases}
\end{equation}
i.e. it takes the value, at time $\tau$, of the solution of the Master equation. Plugging in Eq.~\eqref{eq:detstop} in Eq.~\eqref{eq:SDstex} one gets $\langle\delta_\tau(\mathcal{T}=\tau) \rangle= D (\rho_\tau\,||\,\bar{\rho}_0)=0 $ because $\bar{\rho}_0=\rho_\tau$.  Analogously for $\mathcal{T}=\tau$, intrinsic mismatch cost $\langle \Sigma (\mathcal{T}=\tau) \rangle $  takes the expression \eqref{ineq-fixed} thus retrieving non-negativity, $\langle \Sigma (\tau) \rangle\geq 0$. Note that other examples of negative entropy production at stopping times based on threshold criteria for work were first reported in Ref.~\cite{gambling2021} and for free energy more recently~\cite{albay2023winning}. Such {\em gambling demon}~\cite{gambling2021} effect is allowed whenever $\langle \delta(\mathcal{T})\rangle >0$, which is not guaranteed for arbitrary stopping conditions but possible for wise stopping strategies as shown experimentally in Refs.~\cite{gambling2021,albay2023winning}. 

We can obtain further insight on this effect by decomposing the intrinsic mismatch cost at fixed times $\tau$ in two terms, one associated to intervals $[0,\T]$ up to the stopping time $\T$ and $[\T, \tau]$ from the stopping time to the limit time $\tau$, that is:
\begin{eqnarray}
    \langle \Sigma(\tau) \rangle = \langle \Sigma(\T) \rangle  + \langle \Delta \Sigma(\tau, \T) \rangle \geq 0,
\end{eqnarray}
which follows from the fact that $\T$ is a single-valued function of the trajectory. Since $\langle \Sigma(\tau) \rangle \geq 0$, the above decomposition implies that, whenever $\langle \Sigma(\T) \rangle <0$, such a negative value must be compensated by an incremented mismatch cost $\langle \Delta \Sigma(\tau, \T) \rangle \geq \langle \Sigma(\tau) \rangle$ incurred in the interval $[\T, \tau]$, if no external action is taken on the system at time $\T$ to physically stop the dynamics. These considerations will be valid also in cases where the stopping condition is structurally imposed through the dynamical evolution of the computational process, e.g. using absorbing accept states to ``stop" the computation, as it is the case in some models of DFAs.

The role of absolute irreversibility as captured in the stronger inequality~\eqref{eq:stop-sl} with $- \ln[1- \Gamma_\tau] \geq 0$ makes more difficult the observation of negative average intrinsic mismatch cost, since it would require a higher time-reversal asymmetry in the dynamical evolution leading to large distinguishabilities $\langle \delta(\mathcal{T})\rangle > - \ln[1- \Gamma_\tau]$ [and similarly for inequality~\eqref{eq:stop-s3}]. Remarkably, however, the examples explored in Sec.~\ref{sec:examples} show how still dissipation can be reduced at stopping times thanks to a positive time-reversal asymmetry $\langle \delta(\mathcal{T})\rangle >0$, in agreement with Eq.~\eqref{eq:improvement} above. This reduction might be linked to the information needed to execute the stopping condition $\mathcal{T}$, similarly to what happens in feedback control scenarios~\cite{Ritort19,Ritort21}. However a general relation between these two quantities remains unknown.

The second-law inequality at stopping times~\eqref{eq:stop-s2} can be further rewritten using Eq.~\eqref{eq:Sna2} in a form reminiscent of Landauer's principle:
\begin{align}
- \langle \Delta \phi(\mathcal{T}) \rangle  \geq - \langle \Delta S_\mathrm{sys}(\mathcal{T}) \rangle -  \langle \delta_\tau(\mathcal{T}) \rangle,
\end{align}
where the l.h.s. accounts for the excess entropy flow dissipated into the environment as a consequence of a drop in Shannon entropy of the computational states, $-\langle \Delta S_\mathrm{sys}(\mathcal{T}) \rangle$. Again, whenever $\langle \delta_\tau(\mathcal{T}) \rangle > 0$, the above inequality suggests that the entropy flow to the environment may be eventually reduced. Here it is also worth noticing that even in the case in which trajectories are stopped when returned to the initial state (as in the DFA example  in~\cref{sec:examples}), the average system entropy change at stopping times, namely $\langle \Delta S_\mathrm{sys}(\mathcal{T}) \rangle = \langle S_\mathrm{sys}(\mathcal{T}) \rangle - S(\rho_0)$, with
\begin{align}
    \langle S_\mathrm{sys}(\mathcal{T}) \rangle =& - \sum_\mathcal{T} p(\mathcal{T}) \sum_{x_\mathcal{T}} \rho_\mathcal{T}(x_\mathcal{T}| \mathcal{T}) \ln \rho_\mathcal{T}(x_\mathcal{T}), \end{align} 
is non-zero even when  $x_\mathcal{T} = x_0$ for all $\mathcal{T}$ since in general the distribution $\rho_\mathcal{T}(x) \neq \rho_0(x)$, as corresponds to a relaxation process.

The second-law inequalities derived above not only can be applied to assess stochastic stopping times of a computation, but also to stochastic starting times, see Eqs.~\eqref{eq:doob2st} and \eqref{eq:second-starting}. This extension allow us to apply our theory to computations that may ``stop''
at multiple consecutive times $\T_1 < \T_2 < ... < \T_n$ (see Sec.~\ref{sec:examples} for a particular example in a DFA) or to the concatenations of simpler computations that start at a stochastic time, after the previous one is accomplished. We will further elaborate on the application of starting times to the computation of concatenated words with stochastic resetting in Sec.~\ref{sec:stochasticstart}.

\section{Application to deterministic finite automata}
\label{sec:examples}

In this section we analyze minimal yet insightful examples of  computations executed by deterministic finite automata~(DFA). A computational task for a DFA starts by it receiving a sequence of exogeneously generated symbols, an input string or an input word,~$\omega$. As the DFA iteratively processes the symbols of the input string, it makes associated transitions among its possible states. Here we first assume that the sequence of symbols to the DFA are produced in an independent identically distributed (i.i.d.) manner and so the time evolution over the DFA states while processing those strings can be modeled using a DTMC. Then we will move to the case of input symbols that are not produced in an i.i.d. manner, but from a Markovian source. In the following examples, we consider two minimal DFA models that processes binary strings. In the first example involving i.i.d. symbol sources, the DFA under consideration accepts strings which encode binary numbers divisible by four, e.g. $0$ (zero),  $100$ (four), $1100$ (twelve), etc. In the second example, involving  non-i.i.d. sources, we use a DFA that accepts strings which encode binary numbers divisible by three. In all cases, we assume that the input string behaves as an information reservoir~\cite{barato2014stochastic} whose symbols are not modified by the computation, hence not leading to further energy or entropy changes (see Sec.~\ref{sec:DFAdef}).

The state of the DFA when a stopping condition is reached (e.g., whenther the DFA enters a designated accept state) defines a computation that the DFA performs on that string. However this computation can be followed by further processing of input symbols up to a limit time $\tau$ (e.g. the DFA may exit the accept state in forthcoming iterations). In this sense our results for stopping times can be applied to various situations, for example: (i) computations generated by input words of fixed length $\tau$ where we ask about the value of thermodynamic quantities when visiting the accept state for the first (or the $n$-th) time; and (ii) computations that may actually end when visiting the accept state by some reason (e.g. the accept state is an absorbing state of the DFA or there exists an external mechanism that activates when the accept state is reached to stop the dynamics). In particular, we can always modify a given DFA by removing all edges of the associated directed graph that leave an accepting state. This turns the accept state into an absorbing state (or set of states, if there are more than one accepting states).

\subsection{Processing symbols from i.i.d. sources}
\label{sec:iid}
\begin{figure}[t]
\includegraphics[width=0.45\textwidth]{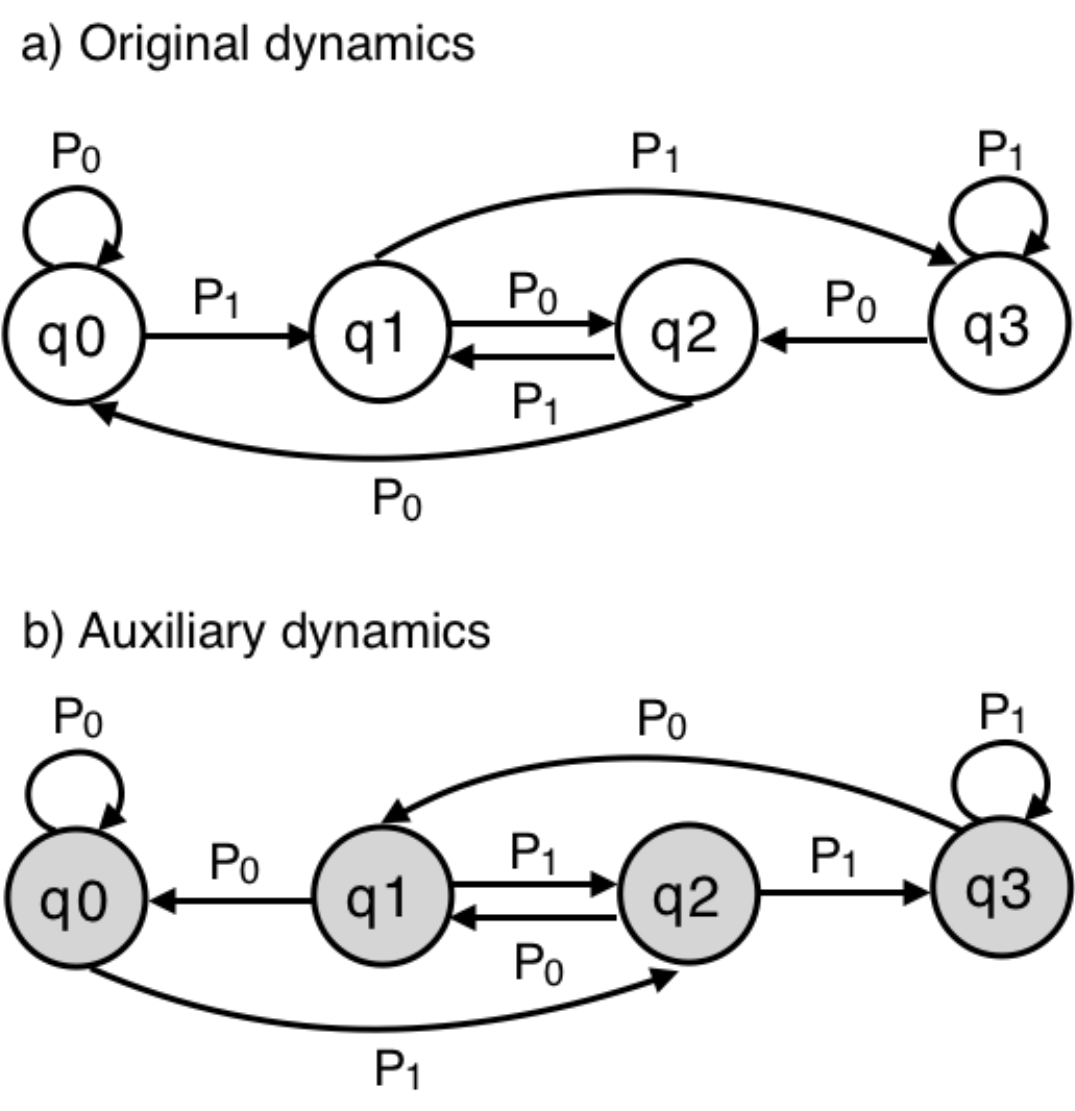}
\caption{(a) Discrete-time Markov chain (DTMC) associated with the DFA recognizing binary i.i.d. sequences that are multiple of four (see Fig.~\ref{fig:sketch}c). The transition matrix of such DTMC is given by Eq.~\eqref{eq:Wor} where $p_0$ and $p_1=1-p_0$ denote respectively the probability for a $0$ and a $1$ in the input string. (b) DTMC associated with the auxiliary dynamics associated with the stationary prior, with transition probability matrix obtained from Eq.~\eqref{eq:auxiliary} and given by Eq.~\eqref{eq:Waux}. \label{fig:oraux}}
\end{figure}

As mentioned above, consider the DFA from \cref{fig:sketch}, initialized to state $q_0$ with certainty, and that its computation starts by processing a stream of binary letters generated as an i.i.d. sequence of $0$s and $1$s, with $p_0\leq 1$ the probability to observe a $0$ and $p_1=1-p_0$ the probability to observe a $1$. Under this assumption, the time evolution of the DFA's states follows a DTMC over four computational states $q_0,q_1,q_2$ and $q_3$, with transition probabilities as indicated in Fig.~\ref{fig:oraux} (a). All together, the  Markov chain associated with the DFA's dynamics is characterized by its initial state 
\begin{equation}
    {\rho}_0=[1\quad 0\quad 0\quad 0]^{\dagger}
\end{equation} 
with $\dagger$ denoting here matrix transposition, and the transition matrix
\begin{equation}
    \textbf{W}= \begin{bmatrix}
p_0 & 0 & p_0 & 0\\
p_1 & 0 & p_1 & 0 \\ 
0 & p_0 & 0 & p_0\\ 
0 & p_1 & 0 & p_1
\end{bmatrix}.
\label{eq:Wor}
\end{equation}
It follows that for $t=1$ we have 
\begin{equation}
    {\rho}_1  = \mathbf{W} {\rho}_0=\begin{bmatrix}
p_0 \quad p_1\quad   0 \quad  0
\end{bmatrix}^{\dagger},
\end{equation}
whereas for larger times $t\geq 2$,
\begin{eqnarray}
    {\rho}_t  &=& \mathbf{W}^{t}{\rho}_0 
    =\begin{bmatrix}p_0^2 \quad p_0p_1 \quad p_0p_1 \quad p_1^2 
\end{bmatrix}^{\dagger} 
 \equiv{\pi},\label{eq:pi}
\end{eqnarray}
i.e., the dynamics already reaches the stationary state at the second iteration. 

For computing the auxiliary dynamics for this DFA's DTMC, we would need to identify the reference distribution $r(x)$ appearing in Eq.~\eqref{eq:auxiliary} as the prior $\mu(x)$ minimizing the mismatch cost sum in Eq.~\eqref{eq:2mm} or $\nu(x)$ leading to its minimum in Eq.~\eqref{eq:sum_mismatch_costs_min}. For simplicity, here we  assume $r(x) = \pi(x)$, the stationary state of the DFA dynamics. This is a reasonable assumption as long as the induced DTMC is aperiodic, irreducible, and $\pi$ has full support over the computational states. As discussed before, since $\Sigma$ becomes non-extensive in time in this case, there are reasons to expect minimal dissipation in the steady state (see also Refs.~\cite{Prigogine1945,Onsager1953,bertini2004minimum}). 

The auxiliary dynamics starts in ${\bar\rho}_0={\rho}_\tau$, which can take two possible values depending on the value of the final maximum time of the computation $\tau$: If $\tau=1$ we have ${\bar\rho}_0 =  {\rho}_1$, whereas for $\tau \geq 2$ we have ${\bar\rho}_0 = \pi$. 
Following Eq.~\eqref{eq:auxiliary} for the transition probability, with $r = \pi$, the
stationary distribution given in~\eqref{eq:pi}, we obtain the transition matrix associated with the auxiliary dynamics:
\begin{equation}
    \overline{\textbf{W}}= \begin{bmatrix}
p_0 & p_0 & 0 & 0\\
0 & 0 & p_0 & p_0 \\ 
p_1 & p_1 & 0 & 0\\ 
0 & 0 & p_1 & p_1
\end{bmatrix},
\label{eq:Waux}
\end{equation}
as illustrated in Fig.~\ref{fig:oraux} (b). It then follows that for computations ending at $\tau = 1$ we have $\bar{\rho}_1  = \overline{\mathbf{W}} \bar{\rho}_0 = \overline{\mathbf{W}} {\rho}_1 =\begin{bmatrix} p_0 \quad 0\quad   p_1 \quad  0
\end{bmatrix}^{\dagger}$. On the other hand, since by construction, the auxiliary dynamics Eq. ~\eqref{eq:auxiliary} will always preserve the steady state for $r = r'=\pi$, it follows that in the case $\tau \geq 2$, the auxiliary dynamics is stationary at all times $t$, that is $\bar{\rho}_t\,^{(\tau)}  =\overline{\mathbf{W}}^t \bar{\rho}_0 = \overline{\mathbf{W}}^t \pi = {\pi}$, with $\pi$ given by Eq.~\eqref{eq:pi}.

The intrinsic mismatch cost in Eq.~\eqref{eq:Sna}, evaluated over a trajectory $\mathbf{x}_{[0,\tau]}$, reduces in this case to the (discrete-time) stochastic non-adiabatic EP:
\begin{eqnarray} \label{eq:sigmaDFA} 
  \Sigma(\mathbf{x}_{[0, \tau]}) &=& \sum_{t=0}^{\tau -1} \left[ \ln\frac{\rho_t(x_t)}{\pi(x_t)} - \ln\frac{\rho_{t+1}(x_{t+1})}{\pi(x_{t+1})} \right] , \nonumber \\
 &=&~ \ln\frac{\rho_0(x_0)}{\pi(x_0)} - \ln\frac{\rho_{\tau}(x_{\tau})}{\pi(x_{\tau})} 
\end{eqnarray}
with $x_t \in \mathcal{X} = \{q_0,q_1,q_2,q_3 \}$ for all $t$, which only depends on the initial and final states.

\begin{figure}[t]
  \centering
\includegraphics[width=0.45\textwidth]{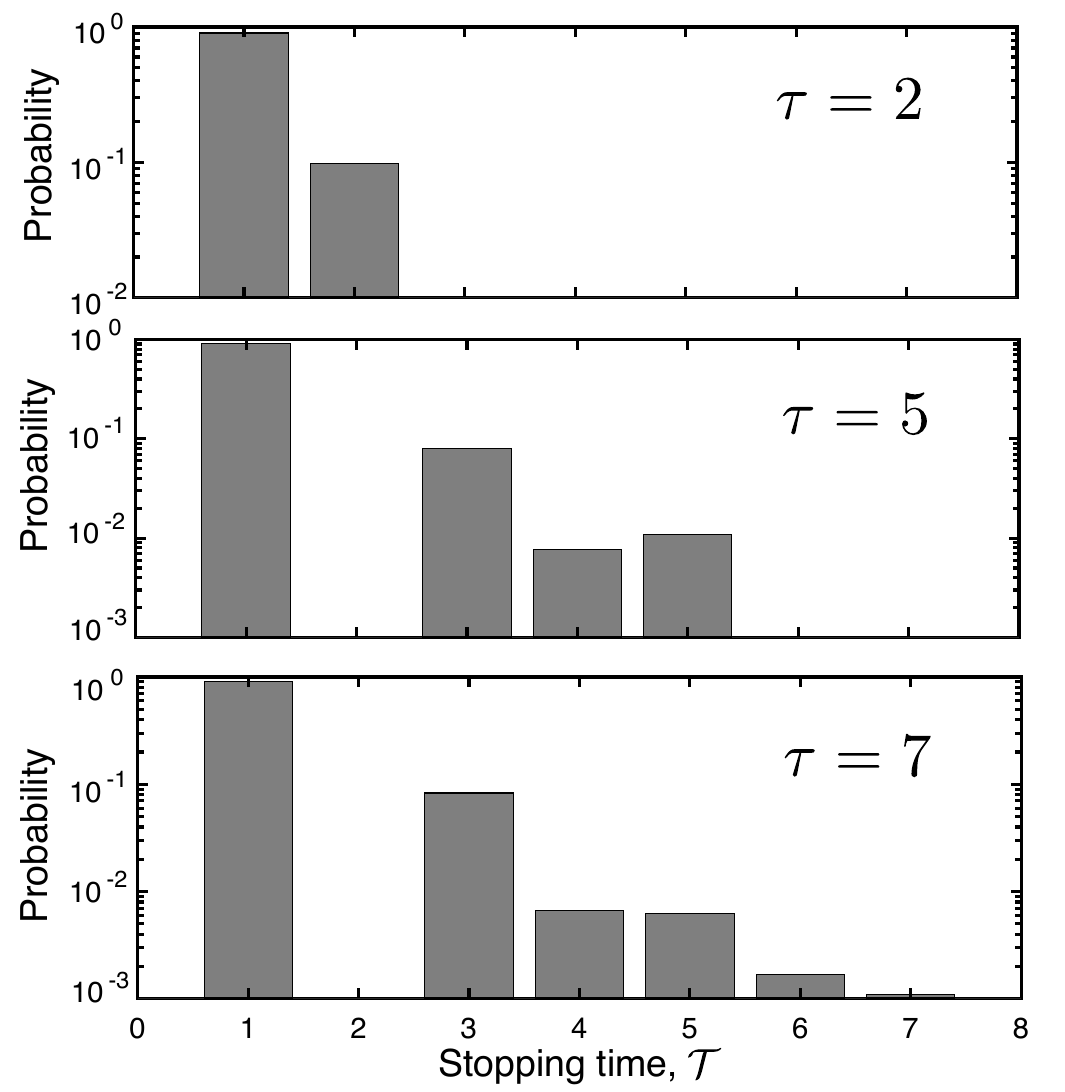}
\caption{Stopping-time statistics for the DFA recognizing binary numbers divisible by four, obtained from $10^4$ numerical simulations of the DTMC sketched in Fig.~\ref{fig:oraux}, with initial state $q_0$, and an absorbing accept state set also at $q_0$. Simulations are done by feeding the DFA with i.i.d. binary sequences with probability of observing the letter 0 given by $p_0=0.9$, obtained from Monte Carlo simulations of the discrete-time Markov chain sketched in Fig.~\ref{fig:oraux}c.}\label{fig:4aa}
\end{figure}

Having obtained the system probability distribution at all times for the original and auxiliary dynamics, we are now ready to compute thermodynamic quantities at stopping times. In particular we consider the family of stopping times
\begin{equation}
    \mathcal{T} = \min (\mathcal{T}_1,\tau)
    \label{eq:T1}
\end{equation}
with $\tau$ fixing a time horizon and $\mathcal{T}_1\geq 1$ the first time the DFA returns to the accept state $q_0$, hence accepting a word as a multiple of four (including ``$0$"). From numerical simulations, we obtained sample histograms  for the  stopping time $\mathcal{T}$ given by Eq.~\eqref{eq:T1} for three different choices of the limit time $\tau$, see Fig.~\ref{fig:4aa}. There we observe the first peak at $\mathcal{T}=1$ in the three plots, corresponding to the cases where the first incoming symbol is "$0$" and the word is then accepted. In order to allow longer accepted words we need $\tau > 2$, such that $\mathcal{T}_1 = 3$ (accepting four ``$100$") or $\mathcal{T}_1 = 4$ (accepting twelve ``$1100$"), etc. Notice however that with the stopping condition given in Eq.~\eqref{eq:T1} we do not capture the acceptance of some of the multiples of four like e.g. eighth ``$1000$'', since the stopping condition would be already verified at previous symbol of the string, ``$100$'', corresponding to four. Same happens for any other accepted number to which an arbitrary number of zeros are attached at the end. For assessing the acceptance of such numbers extra stopping conditions such as $\T_n$ i.e. the $n$-th time the DFA resturns to the accept state $q_0$, are needed (see example in Sec.~\ref{sec:non-idd}).

For all trajectories in which $\mathcal{T}=\mathcal{T}_1< \tau$, i.e., the word is accepted before the limit time $\tau$ is reached, we have $x_{\mathcal{T}_1}=q_0$, the accept state, and thus
\begin{equation}
     \Sigma(\mathcal{T}_1) 
    = -\ln  \rho_{\mathcal{T}_1}(q_0) = \begin{cases} 
-\ln p_0
    & \text{if } \mathcal{T}_1 = 1 \\
&\\  % blank row
-2\ln p_0
    & \text{if } \mathcal{T}_1 \geq  2 .
\end{cases}
\end{equation}
and the stochastic distinguishability in Eq.~\eqref{eq:SD} is:
\begin{equation} \label{eq:stodis2}
    \delta_\tau(\mathcal{T}_1) = \ln \frac{\rho_{\mathcal{T}_1}(q_0)}{\bar{\rho}_{\tau - \mathcal{T}_1}(q_0)} = \begin{cases} 
-\ln p_0
    & \text{if } \mathcal{T}_1 = 1 \\
&\\  % blank row
0
    & \text{if } \mathcal{T}_1 \geq  2 .
\end{cases}
\end{equation}
where we used the fact that (by construction) $\tau > \mathcal{T}_1 \geq 1$ and hence $\bar{\rho}_{\tau - \mathcal{T}_1} = \pi$.

If however $\mathcal{T}_1 \geq \tau$, the dynamics stops at the maximum time $\mathcal{T}=\tau$, independently of the state $x_\tau$, and we obtain:
\begin{equation}
     \Sigma(\tau) 
    =  - \ln\frac{\rho_\tau(x_\tau)\pi(q_0)}{\pi(x_\tau)} = \begin{cases} 
-\ln p_0
    & \text{if } \tau = 1 \\
&\\  % blank row
-2\ln p_0
    & \text{if } \tau\geq  2 .
\end{cases}
\end{equation}
Note that the case $\tau\geq 2$ is independent of $x_\tau$ because the system has already reached its stationary state, and thus $\Sigma(\tau) = -\ln \pi(q_0)=-2\ln p_0$ for all $x_\tau\neq q_0$. On the other hand, the stochastic distinguishability verifies $\delta_\tau(\tau) = 0$ since $\bar{\rho}_0 = \rho_\tau$ always.

Using the above calculations we obtain the average intrinsic mismatch cost at the stopping time~\eqref{eq:T1} for all $\tau \geq 2$ as:
\begin{equation} \label{eq:sigmaT}
\langle \Sigma(\mathcal{T}) \rangle = (p_0-2)\ln p_0 \geq 0,
\end{equation}
which follows from
\begin{align} \label{eq:sigmaT-proof}
\langle \Sigma(\mathcal{T}) \rangle  &= -P(\mathcal{T}_1=1)\ln p_0 - \sum_{t=2}^\tau P(\mathcal{T}_1=t) 2\ln p_0  \nonumber \\ &- P(\mathcal{T}_1> \tau) 2\ln p_0 = - p_0 \ln p_0 - (1 - p_0)2 \ln p_0 \nonumber \\ &= (p_0-2)\ln p_0,
\end{align}
where we have used $P(\mathcal{T}_1=1) = p_0$ and thus $P(\mathcal{T}_1 > 1) = 1 -p_0$. In addition, using  Eq.~\eqref{eq:stodis2}, we obtain the average stochastic distinguishability at the stopping time~\eqref{eq:T1} for all $\tau \geq 2$: 
\begin{eqnarray}
    \langle\delta_{\tau}(\mathcal{T})\rangle &=&  -p_0\ln p_0\geq 0, \label{eq:delta2}
\end{eqnarray}
where again we have used $P(\mathcal{T}_1=1) = p_0$. 
Notice that the above expressions remain also valid in the limit of large input word lengths, $\tau \rightarrow \infty$.

\begin{figure}[t]
  \centering
\hspace{-0.2cm}\includegraphics[width=0.5\textwidth]{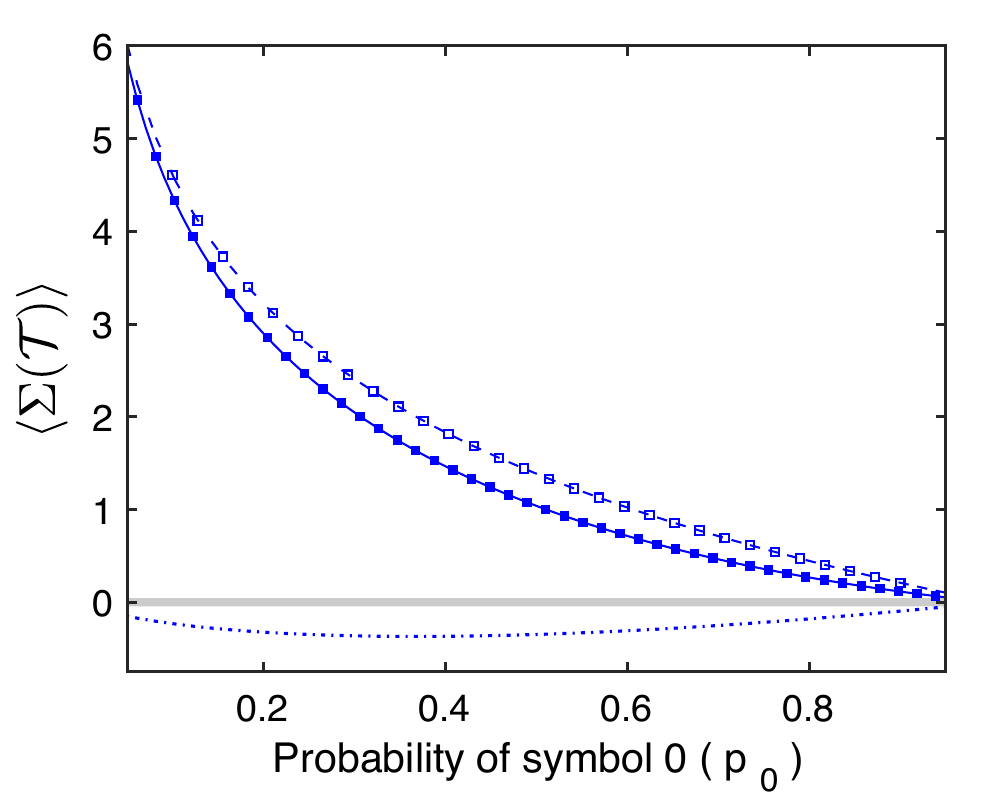}
\caption{Illustration of analytical results for the second law at stopping times applied to the discrete-time Markov chain model  of the DFA recognizing binary strings whose length is a multiple of four (see Fig.~\ref{fig:oraux} a). Here the computation stops at $\mathcal{T}=\min(\mathcal{T}_1,\tau)$ for the limit time $\tau = 2$, or earlier if the accept state is reached in one iteration. Symbols represent analytical results for the averages at the stopping time for the relevant thermodynamic quantities: intrinsic mismatch cost (non-adiabatic entropy production) $\langle \Sigma(\mathcal{T})\rangle$ given by Eq.~\eqref{eq:sigmaT} for prior equal to the stationary probability (blue filled squares); stochastic distinguishability  $\langle \delta_{2}(\mathcal{T})\rangle$, given by Eq.~\eqref{eq:delta2} (blue dotted line); fixed-time non-adiabatic entropy production $\langle \Sigma(\tau)\rangle$ evaluated over trajectories of the same length $\tau=2$ (open symbols); and the absolute irreversibility contribution $-\ln [1-\Gamma_2]$ given by Eq.~\eqref{eq:Gamma} (blue dashed line). The blue solid line  is given by the sum $-\langle \delta_{2}(\mathcal{T})\rangle-\ln [1-\Gamma_2]$ which in this example equals, $\langle\Sigma(\mathcal{T})\rangle$ thus saturating the second law~\eqref{eq:2LAI}. The horizontal black thick line is set to zero as a reference value.}\label{fig:4}
\end{figure}

To tackle the contribution from absolute irreversibility at stopping times, it is convenient to first identify which trajectories contribute to $\Gamma_\tau$ in Eq.~\eqref{eq:Gamma}. These are trajectories that are stopped at $\mathcal{T}\leq \tau$ (either with or without reaching $q_0$) and have zero probability to occur in the original dynamics. Note that the original dynamics is a Markov chain with initial state $\rho_0 (x) = \delta_{x,q_0}$.  The set of absolutely irreversible trajectories at stopping times consists of two sets: (i) trajectories that do not start in $q_0$  and reach $q_0$ with $\mathcal{T}\leq \tau $ in the original dynamics, (ii) trajectories of length $\tau$ that do not start at $q_0$ and do not reach $q_0$ in the original dynamics.

Let us now flesh out the list of such trajectories $\mathbf{x}_{[0,\mathcal{T}]}$ classified by the value of $\mathcal{T}$ for the special case $\tau = 2$:
\begin{itemize}
    \item $q_2 q_0$ reaches the accept state at $\mathcal{T}=1$ yet it has zero probability to occur in the original dynamics with $\rho_0(q_2)=0$.
    \item  $q_1 q_2 q_0$  and $q_3 q_2 q_0$ reach the accept state  at $\mathcal{T}=2$ yet they have zero probability to occur in the original dynamics since $\rho_0(q_1) = \rho_0(q_3) =0$.
    \item  $q_1 q_2 q_1$,    $q_1 q_3 q_2$, $q_1 q_3 q_3$,  $q_2 q_1 q_2$, $q_2 q_1 q_3$,   $q_3 q_2 q_1$,  $q_3 q_3 q_2$, and $q_3 q_3 q_3$
      are stopped at $\mathcal{T}=2$ without reaching the accept state. They have  zero probability in the original dynamics because their initial state is different from $q_0$.
\end{itemize}
All the sequences listed above are such that they would halt the computation at the stopping time $\mathcal{T}=\min(\T_1,2)$, they have non-zero probability in the auxiliary dynamics but zero probability in the original dynamics. 
In order to calculate the absolute irreversibility correction term $\Gamma_\tau$ in Eq.~\eqref{eq:Gamma} we thus need the probability of the above trajectories to occur in time-reversed order in the auxiliary dynamics. More precisely, one needs to compute $\bar{P}(\Theta \mathbf{x}_{[0,\mathcal{T}]})$ multiplied by $\bar{\rho}_{\tau - \mathcal{T}}(x_\mathcal{T})/\rho_\mathcal{T}(x_\mathcal{T}) = \pi(x_\mathcal{T})/\rho_\mathcal{T}(x_\mathcal{T})$, which in this case is equivalent to  modify their initial condition to $\pi(x_\mathcal{T})$, i.e. to compute the following path probabilities:
\begin{eqnarray} \label{eq:trajprobs}
     \bar{P}(q_0,q_2|q_0) \pi(q_0)    &=& p_0^2 p_1 \nonumber \\
     \bar{P}(q_0,q_2,q_1|q_0) \pi(q_0)    &=& p_0^2\; p_1 \;p_0\nonumber\\
     \bar{P}(q_0,q_2,q_3|q_0) \pi(q_0)    &=& p_0^2\; p_1 \; p_1\nonumber\\
     \bar{P}(q_1,q_2,q_1|q_1) \pi(q_1)    &=&  p_0 p_1 \; p_1\; p_0\nonumber\\
      \bar{P}(q_1,q_2,q_3|q_1) \pi(q_1)    &=& p_0 p_1  \; p_1 \; p_1\nonumber\\
       \bar{P}(q_2,q_1,q_2|q_2) \pi(q_2)    &=&  p_0 p_1 \; p_0\; p_1\nonumber\\
       \bar{P}(q_2,q_3,q_3|q_2) \pi(q_2)    &=&  p_0 p_1 \; p_1 \; p_1  \nonumber\\
       \bar{P}(q_2,q_3,q_1|q_2) \pi(q_2)    &=&  p_0 p_1 \; p_1\; p_0\nonumber\\
       \bar{P}(q_3,q_3,q_3|q_3) \pi(q_3)   &=&  p_1^2 \; p_1 \; p_1 \nonumber\\
        \bar{P}(q_3,q_3,q_1|q_3) \pi(q_3)    &=&  p_1^2 \; p_1\; p_0 \nonumber\\
         \bar{P}(q_3,q_1,q_2|q_3) \pi(q_3)    &=&  p_1^2 \; p_0 \; p_1.
\end{eqnarray}
Summing up all the contributions  in Eq.~\eqref{eq:trajprobs} leads us to the absolute irreversibility contribution [cf. Eq.~\eqref{eq:Gamma} for the general formula]:
\begin{eqnarray}
    \Gamma_2 &=& p_0 p_1 \left( p_0 + p_0^2 + 4 p_1 p_0+ 4p_1^2 \right) + p_1^4 \nonumber \\
    &=& 1-p_0^2.\label{eq:Gamma2}
\end{eqnarray}
Combining all the terms above, we observe that for the stopping time $\mathcal{T} = \min (\mathcal{T}_1,2)$:
\begin{equation}
    \langle \Sigma(\mathcal{T}) \rangle = -\langle\delta_{2}(\mathcal{T})\rangle - \ln [1-\Gamma_2].
    \label{eq:equality}
\end{equation}
In other words, the second law at stopping times given by Eq.~\eqref{eq:2LAI} is saturated over the stopping time given by Eq.~\eqref{eq:T1} for $\tau = 2$, as it is illustrated in Fig.~\ref{fig:4} for different values of the probability of incoming zeros, $p_0$. As can be appreciated in that figure, the positive sign of the term $\langle \delta_2(\mathcal{T})\rangle >0 $ implies that the intrinsic mismatch cost at stopping times $\langle\Sigma(\mathcal{T})\rangle = -2\ln p_0 + p_0\ln p_0$, see Eq.~\eqref{eq:delta2},  is smaller than its value at fixed times 
\begin{equation}
  \langle \Sigma (\tau ) \rangle = -2 \ln p_0 \geq \langle \Sigma (\mathcal{T} ) \rangle,  
\end{equation}
in spite of the presence of the absolute irreversibility contribution with $\Gamma_2$. For $\tau > 2$ we have $\Gamma_\tau \leq \Gamma_2$, which follows by combining the equality in Eq.~\eqref{eq:equality} with the generic bound in Eq.~\eqref{eq:stop-sl}. In any case, the inequality $\langle \Sigma (\tau ) \rangle\geq \langle \Sigma (\mathcal{T} ) \rangle $ holds for any limit time $\tau$ for this example.

When $p_0$ approaches $1$ (words with a high number of zeros) the dynamics cannot escape from the initial state $q_0$ and the steady state $\pi$ becomes equal to the initial distribution $\rho_0$. In this limit, the DTMC dynamics becomes fully stationary and hence the intrinsic mismatch cost becomes zero for every trajectory, the time-reversal asymmetry is lost, and the absolute irreversibility is no longer present, leading to a drop in the three quantities on the RHS of Fig. \ref{fig:4}. As we move away from that limit, the mismatch cost increases (both at stopping and fixed times), signaling the energetic costs incurred by the computational task, which grow as $p_0$ decreases. This can be justified by the fact that the dynamics on the DTMC spreads more easily over all computational states as $p_1$ increases (see Fig.~\ref{fig:4} a), leading to a greater distinction between initial and steady-state distributions. In this case we also observe non-zero stochastic distinguishability and an increasingly large absolute irreversibility term. In the limit $p_0 \rightarrow 0$ (words with a high number of ones) accepting a word becomes almost impossible, and hence the stopping occurs most probably at the maximum time $\mathcal{T} \simeq \tau$, leading again to zero stochastic distinguishability. We notice that in this limit $\pi$ tends to localize at state $q_3$ and hence it would lead to $\langle \Sigma(\tau) \rangle \rightarrow \infty$, which is not physically meaningful. The catch point is that in this limit the fixed point $\pi$ would not be equal to the prior $\mu$ or $\nu$ anymore.

\subsection{Uniform prior}
\label{sec:uniform}

\begin{figure}[t]
  \centering
\includegraphics[width=0.5\textwidth]{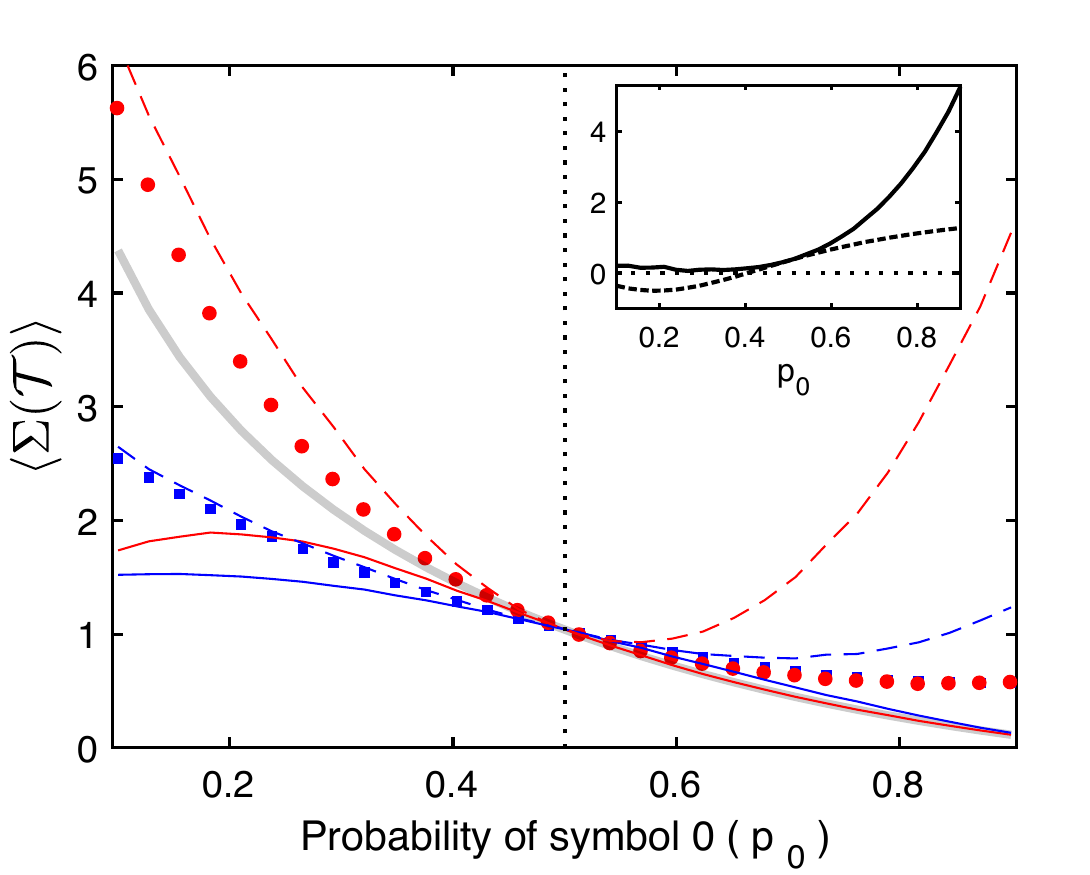}
\caption{Numerical results for the average intrinsic mismatch cost $\langle\Sigma(\mathcal{T})\rangle$ (symbols) at stopping times $\mathcal{T}=\min(\mathcal{T}_1,\tau)$ with uniform prior, evaluated for the DFA in Fig.~\ref{fig:oraux}a processing i.i.d. binary input data, as a function of the probability of input symbol $0$. We used different limit times:  $\tau = 5$ (blue filled squares), and  $\tau = 14$ (red filled circles). The solid lines correspond to the lower bound predicted by Eq.~\eqref{eq:stop-s3} and given by $D(\rho_0\,||\,\bar{\rho}_\tau)-\langle \delta_\tau(\mathcal{T}) \rangle$ for $\tau = 5$ (blue solid line), and $\tau = 14$ (red solid line), while the dashed lines are the corresponding upper bounds in Eq.~\eqref{eq:improvement}, $\langle \Sigma(\tau) \rangle - \langle \delta_\tau(\mathcal{T}) \rangle$ for same values of $\tau$.  
Averages are estimated from  $10^4$ numerical simulation for each parameter value. The thick gray line is the average cost at stopping times for stationary prior $\pi$, see Eq.~\eqref{eq:sigmaT} and Fig.~\ref{fig:4}, and the vertical dashed line is set to $p_0=1/2$ as a reference value. Inset: $\langle\Sigma(\tau)\rangle-\langle\Sigma(\T)\rangle $ (solid line) and $\langle\delta_\tau(\T)\rangle$ (dashed line) as a function of $p_0$ for $\tau=14$. The horizontal dotted line is set to zero as a reference value.
\label{fig:5}}
\end{figure}

We now implement the analysis in Sec.~\ref{sec:iid} for a different setting, where the strings are generated i.i.d. and we consider the same four-states DFA, now with a uniform prior distribution over its states
\begin{eqnarray}
    r  &=& 
    \begin{bmatrix}1/4 \quad 1/4 \quad 1/4 \quad 1/4 
\end{bmatrix}^{\dagger}
,\label{eq:ru}
\end{eqnarray}
The evolution of $r$ under $\mathbf{W}$ after one iteration yields
\begin{eqnarray}
    r'  &=& \mathbf{W} r =
    \begin{bmatrix}p_0/2 \quad p_1/2 \quad p_0/2 \quad p_1/2 
\end{bmatrix}^{\dagger} 
 .\label{eq:ru2}
\end{eqnarray}
Because $r$ changes after one iteration, we write $\Sigma$ as in Eq.~\eqref{eq:Sna2} for $\tau>0$
\begin{align}
  \Sigma(\mathbf{x}_{[0, \tau]}) 
 &=~ \ln\frac{\rho_0(x_0)}{\rho_{\tau}(x_{\tau})} + \sum_{t=0}^{\tau -1} \ln\frac{r'(x_{t+1})}{r(x_t)},
\label{eq:sigmaDFA2} 
\end{align}
where we the first term is the system entropy change $\Delta S_\mathrm{sys}(\mathbf{x}_{[0,\tau]})$ and the second one the nonequilibrium potential $\Delta \phi(\mathbf{x}_{[0,\tau]})$ in Eq.~\eqref{eq:phi}. Unlike for the stationary prior, now this term is extensive with time~[cf. Eq.~\eqref{eq:sigmaDFA}]. Note that in this case $\Sigma$ is no longer equal to the non-adibatic EP associated with the stochastic trajectory~$\mathbf{x}_{[0, \tau]}$.

The uniform distribution is not invariant under the map $\mathbf{W}$, hence the intrinsic mismatch cost $\Sigma$ associated with a stochastic trajectory $\mathbf{x}_{[0, \tau]}$ is extensive with time. This implies that, unlike for the case of stationary prior (see Sec.~\ref{sec:iid}), the averages of $\Sigma$ at fixed times $\tau$ as well as at stopping times with limit time $\tau$ [of the form of Eq.~\eqref{eq:T1}], will crucially depend on $\tau$. This is also the case for any other choice for the prior distribution which differs from the stationary distribution. 

In Fig.~\ref{fig:5} we show the intrinsic mismatch cost $\langle\Sigma(\mathcal{T})\rangle$ at the stopping time stopping time $\mathcal{T} = \min (\mathcal{T}_1,\tau)$ for the DFA with the uniform prior for two different values of $\tau$, and compare it with the case of stationary prior, Eq.~\eqref{eq:sigmaT}. We observe that the uniform prior leads to higher values for the intrinsic mismatch cost for high values of $p_0$, while for low $p_0$ values the tendency can be inverted. However when increasing $\tau$ sufficiently we always obtain a lower cost for the stationary prior, as expected from its non-extensivity. Indeed we observe a tendency for the mismatch cost at stopping times $\langle\Sigma(\mathcal{T})\rangle$ to saturate when increasing the limit time $\tau$, in contrast with the linear scaling of $\langle\Sigma(\tau)\rangle$ with $\tau$. In Appendix~\ref{app:scaling} we confirm this point by studying in more detail the scaling behaviour of these two quantities as a function of $\tau$.

We test the sandwich inequality in Eq.~\eqref{eq:sandwich} comprising the upper and lower bounds to $\langle\Sigma(\mathcal{T})\rangle$ in Eqs.~\eqref{eq:improvement} and \eqref{eq:stop-s3}, respectively. As can be appreciated in Fig.~\ref{fig:5}, both inequalities provide useful bounds that become tighter for small $\tau$, and are simultaneously saturated at the point $p_0=1/2$. This example also reveals that again there is a reduction of intrinsic costs at stopping times with respect to fixed times, that is $\langle\Sigma(\tau)\rangle \geq \langle\Sigma(\mathcal{T})\rangle$ holds over the entire parameter range of probability of symbol 0,  $p_0$, and the limit time $\tau$, as shown in the inset of Fig.\ref{fig:5}. This reduction is guaranteed by a positive value of the stochastic distinguishability $\langle \delta(\T) \rangle >0$ in the range $p_0 \geq 1/2$ [c.f. Eq.~\eqref{eq:improvement}] but, interestingly, it is also verified even for $\langle \delta(\T) \rangle <0$ as it happens for $p_0 \leq 1/2$.

\subsection{Beyond i.i.d. sources}
\label{sec:non-idd}
So far we have analyzed the statistics of a DFA processing inputs generated by a source of i.i.d. bits, which induces a Markovian dynamics for the time-evolution of the computational states. This is, however, one of the simplest possible computational processes, as e.g., regular languages recognized by DFAs are often composed of correlated words. To illustrate the applicability of our theory to computing thermodynamic costs of DFAs processing arbitrary strings from arbitrary languages, it is mandatory to consider DFAs processing non-i.i.d. sequences. 

In processing a generic non i.i.d. sequence, the dynamics over the computational states of a DFA is in general a non-Markovian process. However, one can extend the computational state space such that our formalism can be applied. For the analysis in this section it is important to remark the distinction between the “computational states” of the DTMC computational state space $\mathcal{X}$, and the states of the DFA. In particular, we refer to states of the DFA (as in the usual TCS definition) as logical states of the DFA, and remind that with “computational states” we refer to the sets of variables which describe the entire state-space for a computational process of interest, as introduced in Sec.~\cref{sec:framework}. 

Now consider that the (process generating the) input string itself is a DTMC characterized by time-independent transition probabilities $p(b_{i+1}|b_i)$ for the $(i+1)$’th bit to be equal to $b_{i+1}=\{0,1\}$ given that the $i’$th symbol (bit) of the string is $b_i=\{0,1\}$.  In this case, the logical state of, e.g., a three-state DFA $z_t=\{q_0,q_1,q_2\}$ processing this input string is not a DTMC, although by constructing the computational state space as the Cartesian product of $z_t=\{q_0,q_1,q_2\}$ and $b_t=\{0,1\}$, we encode the current computational state $x_t = \{z_t, b_t\}$ as the logical state of the DFA $z_t$ and the most recent input symbol fed to the DFA $b_t$. In this case, one is left with a DTMC with six possible computational states, for which our formalism can be readily applied to tackle the thermodynamic properties.

\begin{figure}[t]
\includegraphics[width=0.36\textwidth]{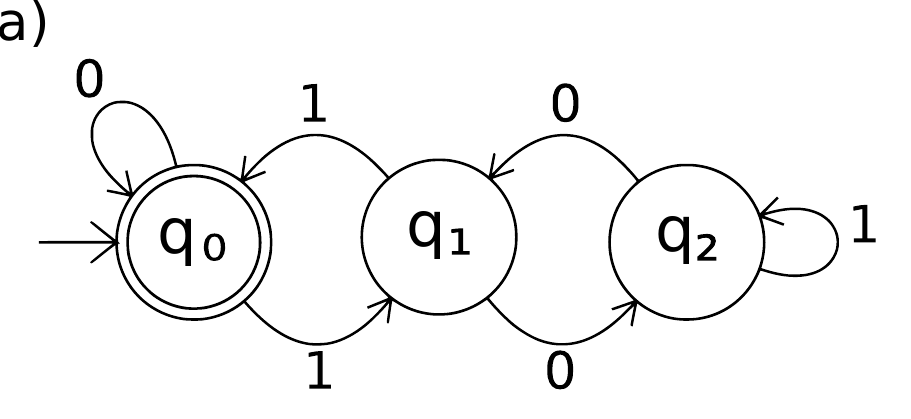}
\includegraphics[width=0.36\textwidth]{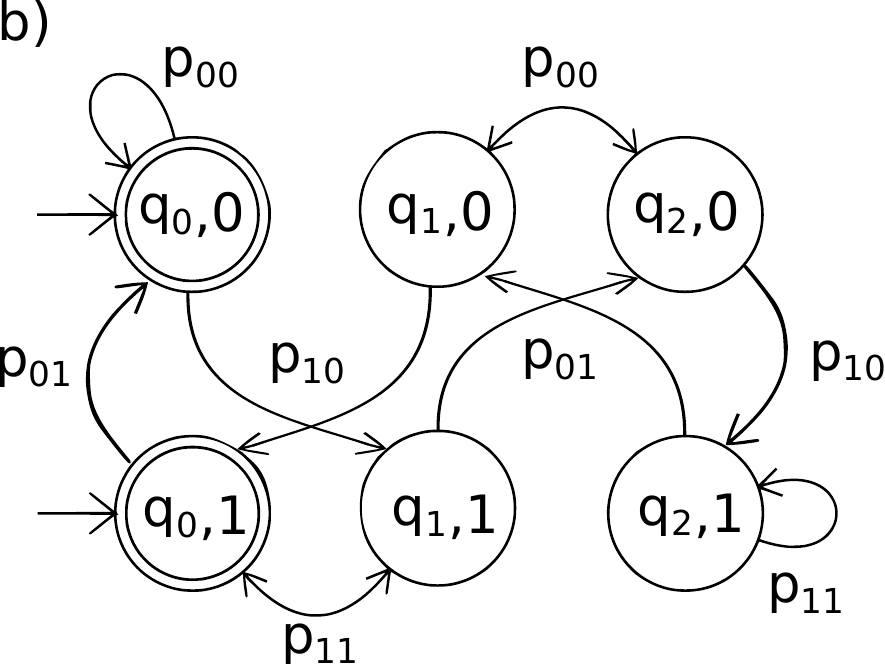}
\caption{(a) Minimal DFA that accepts binary multiples of three with three states $z_t=\{ q_0, q_1, q_2\}$ with same start and accept state $q_0$. (b) Associated DTMC, where the automaton processes input strings generated by a non-i.i.d. source of input symbols with probabilities depending only on the last processed symbol $b_t=\{ 0, 1\}$. See Eq.~\eqref{eq:W6} for the corresponding transition probability matrix. 
\label{fig:non-iid}}
\end{figure}

As an example, we consider the minimal DFA that accepts binary multiples of 3, shown in Fig.~\ref{fig:non-iid} (a), leading to the DTMC represented in Fig.~\ref{fig:non-iid} (b). The probabilities to obtain input bits $0$ or $1$ given the last intput symbol are fixed and denoted by $p(i|j):=p_{ij}$ for $i,j=\{ 0, 1\}$. They satisfy $p_{00} + p_{10} =1$, and $p_{01} + p_{11} =1$. Ordering the computational states distribution as $\rho_t = [\rho_t(q_0, 0), \rho_t(q_1, 0),\rho_t(q_2, 0), \rho_t(q_0, 1), \rho_t(q_1, 1), \rho_t(q_2, 1)]$, we obtain a $6 \times 6$ transition matrix given by:
\begin{equation}
    \textbf{W}= \begin{bmatrix}
p_{00} & 0 & 0 & p_{01} & 0 & 0\\
0 & 0 & p_{00} & 0 & 0 & p_{01} \\ 
0 & p_{00} & 0 & 0 & p_{01} & 0\\ 
0 & p_{10} & 0 & 0 & p_{11} & 0 \\
p_{10} & 0 & 0 & p_{11} & 0 & 0 \\
0 & 0 & p_{10} & 0 & 0 & p_{11} 
\end{bmatrix}. 
\label{eq:W6}
\end{equation}
We choose as the initial condition the probability distribution:
\begin{equation}
    {\rho}_0=[1/2 \quad 0\quad 0 \quad 1/2\quad 0 \quad 0]^{\dagger},
\end{equation}
with initial equal probabilities over the DTMC states corresponding to $q_0$ as the DFA start state and zero otherwise.
\begin{figure}[t]
  \centering
\hspace{-0.5cm}\includegraphics[width=0.47\textwidth]{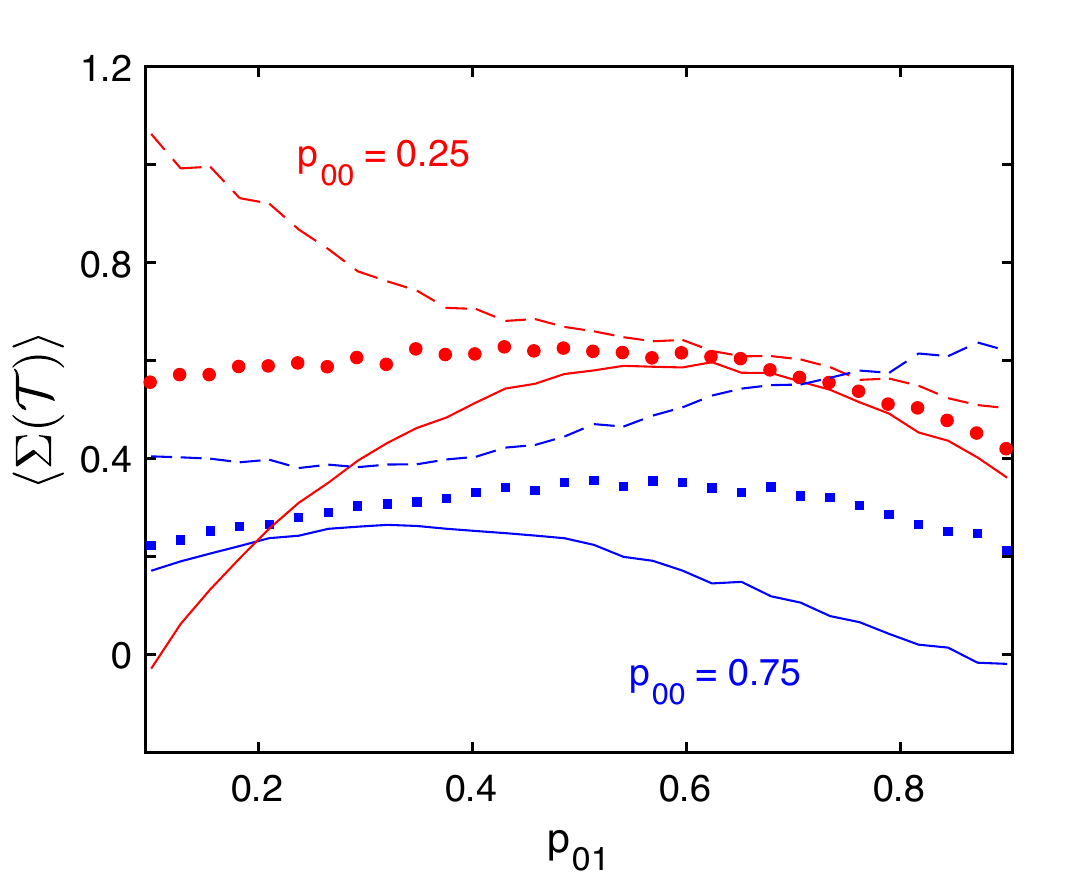}
\caption{Numerical results for the intrinsic mismatch cost $\langle\Sigma(\T)\rangle$ with uniform prior for the DFA in  Fig.~\ref{fig:non-iid}a) processing Markovian input strings, as a function of the input symbol transition probability. Here $\mathcal{T}=\min(\mathcal{T}_1,\tau)$, with $\mathcal{T}_1$ the first return time to any of the states $(q_0,0)$ or $(q_0,1)$, and $\tau=5$ a prescribed limit time. Symbols are numerical estimates for $\langle\Sigma(\T)\rangle$ obtained   from $10^4$ numerical simulations, for two different values of the transition probability $p_{00}$ of the input string containing two consecutive zeroes (see legend). Solid lines are estimates from the numerical simulations of the quantity    $ D (\rho_0\,||\,\bar{\rho}_\tau)-\langle \delta_\tau(\mathcal{T}) \rangle $, confirming the lower bound given by  Eq.~\eqref{eq:stop-s3}.
\label{fig:7}}
\end{figure}
We assume that the underlying entropy production is minimized for the uniform prior
\begin{eqnarray}
    r = [1/6 \quad 1/6 \quad 1/6 \quad 1/6 \quad 1/6 \quad 1/6]^\dagger,
\end{eqnarray}
which is transformed, after one iteration, into $r^\prime = \mathbf{W} r$:
\begin{align}
    r^\prime = [a/6 \quad a/6 \quad a/6 \quad b/6 \quad b/6 \quad b/6]^\dagger,
\end{align}
with $a:=p_{00} + p_{01}$ and $b:=p_{10} + p_{11}$. 

\begin{figure}[t]
\hspace{-0.2cm}\includegraphics[width=0.5\textwidth]{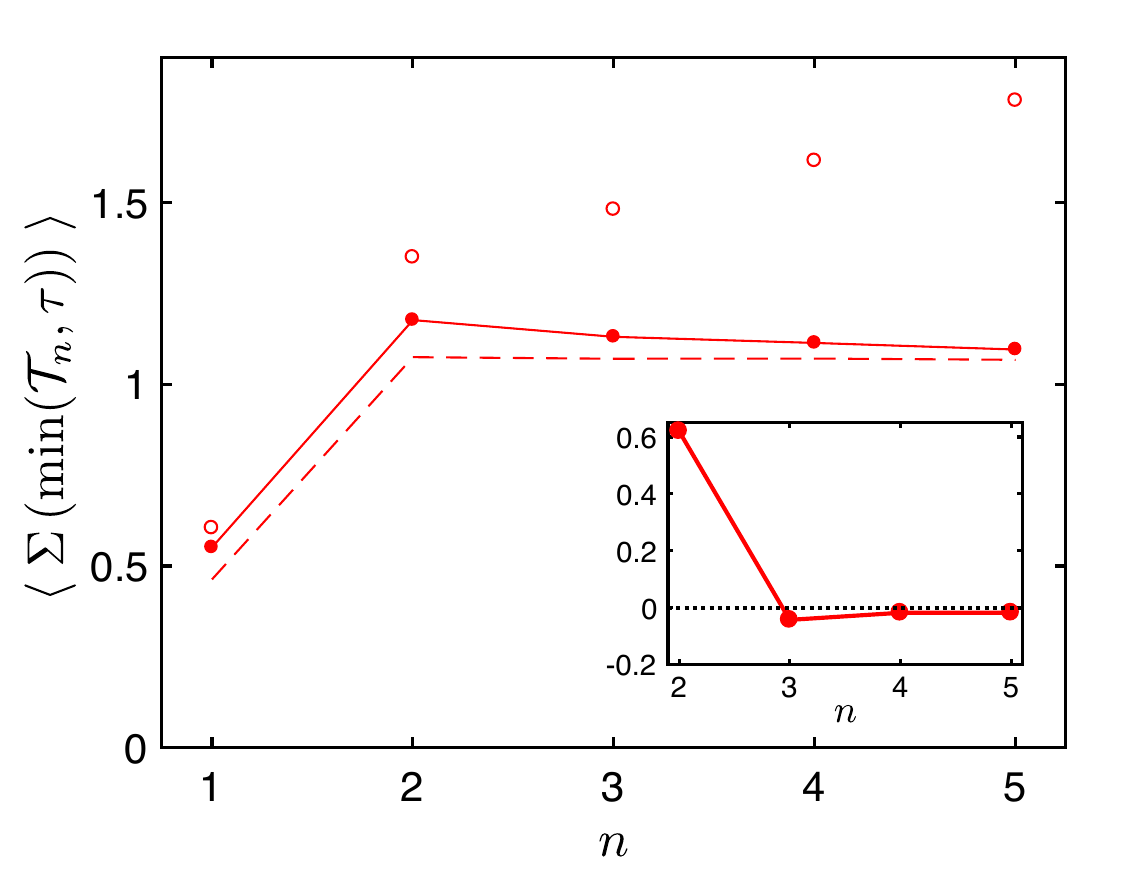}
\caption{Intrinsic mismatch cost up to the second, third, fourth, and fifth accepted word.  Numerical results for $\langle\Sigma(\min(\T_n,\tau))\rangle$ with uniform prior, with $\T_n$ the $n-$th return time to the accept states, and $\tau=40$ a prescribed limit time. Results are obtained for the DFA in Fig.~\ref{fig:non-iid}a) processing Markovian input strings, as a function of the $n$ input symbol transition probability, for parameter values:  $p_{00}=0.25, p_{01}=0.4$ (open symbols), and  $p_{00}=0.25, p_{01}=0.75$ (filled symbols). The lines are estimates from the numerical simulations of the quantity  $ D (\rho_0\,||\,\bar{\rho}_\tau)-\langle \delta_\tau(\mathcal{T}) \rangle $ for $p_{00}=0.25, p_{01}=0.4$ (red dashed line) and  for $p_{00}=0.25, p_{01}=0.75$ (red solid line).  Inset: 
Mismatch cost between returning times $\langle \Delta \Sigma (\T_n, \T_{n-1}) \rangle := \langle \Sigma(\min(\T_{n},\tau)) \rangle - \langle \Sigma(\min(\T_{n-1},\tau)) \rangle$ (red circles)  and $-[\langle \delta_\tau(\min(\T_{n},\tau)) \rangle - \langle \delta_\tau(\min(\T_{n-1},\tau)) \rangle]$ (red solid line)  as a function of $n$, see Eq.~\eqref{eq:second-starting}. The horizontal dotted line is set to zero as a reference value. 
}
\label{fig:8}
\end{figure}

Using the above definitions we can compute $\Sigma(\tau)$ in Eq.~\eqref{eq:Sna}, at arbitrary fixed times. Moreover, in order to evaluate thermodynamic quantities at stopping times, we embrace again the family of stopping times $\mathcal{T} = \min (\mathcal{T}_1,\tau)$ with $\tau$ the fixed time horizon and $\mathcal{T}_1\geq 1$ the first time the DFA returns to the accept state $q_0$ for either $b=\{0, 1\}$. 

We show numerical results in Fig.~\ref{fig:7}, where $\langle\Sigma(\mathcal{T})\rangle$, together with the corresponding upper and lower bounds given by Eqs.~\eqref{eq:sandwich} are plotted as a function of the probability $p_{01}=1 - p_{11}$ to obtain symbol $0$ after a symbol $1$, for different values of $p_{00}= 1 - p_{10}$. Again we obtain relevant bounds on the intrinsic mismatch cost at stopping times, which, interestingly, become tightest when $p_{00}=1-p_{01}$, i.e. when $p_{00} = p_{11}$ and $p_{10}=p_{01}$.
 This corresponds to the situation in which the input sequence is a Markovian process with homogeneous stationary probabilities, $p_0^{\rm st} = p_1^{\rm st}=1/2$. The fact that our bounds become tight for homogeneous input sequences was also observed for the i.i.d. example (see Fig.~\ref{fig:5}) and makes us conjecture that this phenomenon may be generic to correlated input sequence, maybe also non-Markovian.

As also commented for the previous examples, however, using a stopping time of the form $\mathcal{T} = \min (\mathcal{T}_1,\tau)$ allows us to describe computation times for the DFA to reach the accept state for the first time. That corresponds to the acceptance of only some of the multiples of three, e.g. ``0" (zero), ``11" (three), ``1001" (nine), but not other multiples like ``110" (six) or any other word that already contains an acceptable prefix. In order to explore thermodynamic costs associated to these words we now  consider more general stopping times $\T = \min(\T_n, \tau)$, where $\T_n$ is the $n$-th time the DFA returns to the accept state. Therefore $\T_2$ is related with the acceptance of words like ``110" (six) or ``10010" (eighteen), while $\T_3$ corresponds to accept words like ``1100" (twelve), among many others.

In Fig.~\ref{fig:8} we plot $\langle\Sigma(\mathcal{T})\rangle$ with  $\T = \min\{\T_n, \tau\}$ as a function of the return time to the accept state, $n = 1, 2, 3, 4 ,5$. We notice that different behaviors are obtained depending on the choice of input symbols probabilities, $p_{00}$ and $p_{01}$, leading to either increasing values of the intrinsic mismatch cost or a non-monotonic behavior. Interestingly considering different stopping times allows us to test inequality~\eqref{eq:second-starting} for two stopping times, which is shown in the inset of Fig.~\ref{fig:8} for $\T_1 = \min(\T_{n-1} , \tau)$ and $\T_2 = \min(\T_{n}, \tau)$ as a function of $n>1$.
In particular we observe that the mismatch cost between consecutive returning times to the accept state can be eventually negative for specific choices of parameters (probabilities $p_{00}$ and $p_{01}$), that is, $\langle \Delta \Sigma (\T_n, \T_{n-1}) \rangle < 0$ for $n= 3, 4, 5$,  owing to a reduction in the associated stochastic distinguishability and despite having $\langle \Sigma(\tau) \rangle \geq 0$ at fixed times.

\section{Universal equalities and inequalities for acceptance probabilities}
\label{sec:acc}

Our formalism can be further applied to address other issues in computer science theory, beyond automata literature, and besides second laws and fluctuation theorems at stopping times. Both in this Sec.~\ref{sec:acc} and in Sec.~\ref{sec:stochasticstart} we develop further theoretical predictions for key statistical properties of interest for computer science that may inspire numerical and experimental illustrations of future work. 
An example that we develop in this section is using our formalism to establish universal equalities and inequalities concerning the probabilities of acceptance or rejection of sets of distinct bit sequences when a given DFA is implemented.
In what follows we focus on a specific choice of such sets, namely, 1) the set of all strings or trajectories that end in an accept state before the limit time $\tau$ vs. 2) the set of all trajectories that do not end in an accept state before the limit time $\tau$. However, we emphasize that this formalism can be generalized to arbitrary pairs of sets of trajectories, by specifying suitable {\em filtrations} as done in martingale approaches.

Thus we will explore a class of simple examples of ``acceptance" statistics for binary words of length $\tau \geq 2$ that are processed by a computer. We will use the notation {\em accept} to signify that a computer reaches a prescribed accept state before the limit time $\tau$, and  {\em reject} otherwise. The probabilities $P_{\rm a}(\tau)$  denotes the probability for the computer to have reached the accept state within $[0,\tau]$, and  $P_{\rm r}(\tau)=1-P_{\rm a}(\tau)$ the probability for the complementary. Recall that for simple computer architectures (e.g., DFAs processing i.i.d. binary strings),  $P_{\rm a}(\tau)$ and $P_{\rm r}(\tau)$ can often be evaluated analytically or with Monte Carlo simulations. The approach we reveal below is complementary to Monte Carlo approaches in such simple computations, however we highlight its usefulness in revealing how such accept/reject statistics are related to thermodynamic quantities. Note that here those thermodynamic quantities can be used as a tool of calculation, determined completely by the computer update function and the distribution over input words. In particular, they need not correspond to any ``real'' thermodynamic quantities that one would measure in the laboratory. That is, our formalism provides a way to derive the relative probabilities of accepting or rejecting a string while sidestepping the conventional technical difficulties found in the traditional approaches to this issue \cite{Carlyle1971, Castro2016}. On the other hand, one can also interpret the results presented below as a way for obtaining information about the intrinsic thermodynamic costs of computations by looking at the acceptance probabilities (of languages solved by machines), which might be calculated by other means, such as Monte Carlo approaches.

So we consider again a stopping time $\mathcal{T}= \min(\mathcal{T}_1,\tau)$, which signifies the first time that the computer reaches the accept state, $\mathcal{T}_{{1}}$, or the limit $\tau$ in the case that the accept state is not visited before $\tau$.  So $\mathcal{T} < \tau$ if a word of length $\tau-1$ is accepted by the computer. Otherwise, $\mathcal{T}= \tau$, if the word is not accepted before $\tau$.
The probabilities that a word of length $\tau-1$ is accepted or not are then given by
\begin{eqnarray}
    P_{\rm a}(\tau) &=& P(\mathcal{T}< \tau),\label{eq:59}\\
    P_{\rm r}(\tau) &=& 1-P(\mathcal{T}< \tau) = P(\mathcal{T} = \tau),
\end{eqnarray} 
respectively. We now make use of our formalism to derive bounds for $P_{a}(\tau)$ and $P_{r}(\tau)$ in terms of thermodynamic quantities.

Using our fluctuation theorem at stopping times with absolute irreversibility, Eq.~\eqref{eq:IFTST}, $\langle M_\tau(\T) \rangle = 1- \Gamma_\tau$, we expand its l.h.s. into terms corresponding to accepted and rejected words as:
\begin{eqnarray}\label{eq:79}
   P_{\rm a}(\tau) \langle M_\tau(\T) | \T < \tau \rangle + P_{\rm r}(\tau) \langle M_\tau(\T) | \T = \tau \rangle,
\end{eqnarray}
with $\langle A(\T)\vert c(\T) \rangle =\mathbb{E}(A(\T) \,|\, c(\T))$ being the conditional average of functional $A$ over trajectories $x_{[0,\T]}$ given that the condition $c(\T)$ is fulfilled over the stopping time $\T$. Upon using $P_{\rm r}(\tau) =  1- P_{\rm a}(\tau)$, the decomposition~\eqref{eq:79}  gives us the following relation between the acceptance probability and the averages of the supermartingale $M_\tau(\T)$ at stopping times:
\begin{equation}
    P_{\rm a}(\tau) = \frac{1 - \Gamma_\tau - \langle M_\tau(\T) | \T = \tau \rangle}{ \langle M_\tau(\T) | \T < \tau \rangle-\langle M_\tau(\T) | \T = \tau \rangle}.
    \label{eq:pacbound3}
\end{equation}
Equality~\eqref{eq:pacbound3} generalizes analytical expressions obtained in previous works for absorption probabilities~\cite{Neri17,neri2019integral,Edgar2022} by including the absolute irreversibility contribution $\Gamma_\tau$. As can be appreciated in Eq.~\eqref{eq:pacbound3}, since $\Gamma_\tau \geq 0$, the role of absolute irreversibility is to decrease the acceptance probability $P_{\rm a}(\tau)$ of a word of lenght $\tau -1$ by the DFA. This can be intuitively understood from the fact that starting computation from a restricted set of initial states can only decrease the velocity at which the computational state space is explored, and hence the probability to reach a generic stopping condition before time $\tau$.

Since $P_{\rm a}(\tau)$ is a well-defined probability (i.e. $0 \leq P_{\rm a}(\tau) \leq 1$), we further obtain from Eq.~\eqref{eq:pacbound3} that one of the two following chain inequalities hold:
\begin{align}
    &\langle M_\tau(\T) | \T < \tau \rangle \geq 1 - \Gamma_\tau \geq \langle M_\tau(\T) | \T = \tau \rangle, \\
    &\langle M_\tau(\T) | \T = \tau \rangle \geq \langle M_\tau(\T) | \T < \tau \rangle \geq 1 - \Gamma_\tau \geq 0,
\end{align}
which provide us constrains on the values of $M_\tau(\T)$ for generic $\T$ of the form $\mathcal{T}= \min(\mathcal{T}_1,\tau)$. 

Analogously, we can exploit the second-law-inequality at stopping times~\eqref{eq:stop-s3}, namely $\langle\Sigma(\mathcal{T}) \rangle  \geq - \langle \delta_\tau(\mathcal{T}) \rangle + D (\rho_0\,||\,\bar{\rho}_\tau)$, to derive universal bounds for the finite-time acceptance probability. Indeed, average of the left-hand-side of this equation at the stopping time $\mathcal{T}= \min(\mathcal{T}_1,\tau)$ can also be decomposed into two terms, accounting, respectively, for accepted and rejected words of maximum length $\tau-1$:
\begin{eqnarray} \label{eq:72}   
  \langle\Sigma(\mathcal{T}) + \delta_\tau(\mathcal{T}) \rangle  &=& P_{\rm a}(\tau) C_{\rm a}(\tau) + P_{\rm r} C_{\rm r}(\tau), \,\,\,
\end{eqnarray}
where we have introduced the conditional averages
\eq{
\label{eq:67}
    C_{\rm a}(\tau)&=\langle\,\Sigma(\mathcal{T}) + \delta_\tau(\mathcal{T})\, | \mathcal{T}<\tau \rangle ,\\
\label{eq:68}
    C_{\rm r}(\tau)&=\langle\Sigma(\mathcal{T}) | \mathcal{T}= \tau  \rangle .
}
Note that in Eq.~\eqref{eq:68} we have used the fact that~$\delta_\tau(\tau)=0$. We refer to these two conditional averages as the average thermodynamic {\em costs} associated with the acceptance and rejection of words of length $\tau-1$, respectively. 

Combining Eqs.~\eqref{eq:stop-s3} and Eq.~\eqref{eq:72} we obtain the two following lower and upper bounds for the acceptance probability
\begin{eqnarray}
    P_{\rm a} (\tau)\geq \frac{D (\rho_0\,||\,\bar{\rho}_\tau) - C_{\rm r}(\tau)}{C_{\rm a}(\tau)-C_{\rm r}(\tau)},
\label{eq:pacbound}
\end{eqnarray}
valid whenever $C_{\rm a}(\tau) > C_{\rm r}(\tau)$, and similarly
\begin{eqnarray} \label{eq:pacbound2}
    P_{\rm a} (\tau)\leq \frac{ C_{\rm r}(\tau)- D (\rho_0\,||\,\bar{\rho}_\tau) }{C_{\rm r}(\tau)-C_{\rm a} (\tau)},
\end{eqnarray}
valid in the complementary case when $C_{\rm r}(\tau) > C_{\rm a}(\tau)$. These bounds express a constraint on the acceptance probability of a word with maximum lenght $\tau-1$ in terms of the average costs associated with the accepted and rejected words as defined in Eqs.~(\ref{eq:67}) and (\ref{eq:68}), and the KL divergence between the initial distribution of the computational state and the final distribution of the computational state under the auxiliary dynamics [see Eq.~\eqref{eq:65}]. 

Equation~\eqref{eq:pacbound} provides a meaningful bound whenever its r.h.s is non-negative and smaller than one, i.e. when $C_a(\tau)\geq D (\rho_0\,||\,\bar{\rho}_\tau) \geq C_r(\tau)$. On the other hand, the bound~\eqref{eq:pacbound2} is meaningful when $C_a(\tau)\leq D (\rho_0\,||\,\bar{\rho}_\tau) \leq C_r(\tau)$.  We expect the first condition to be satisfied if the probability of accepted words is large enough so that the associated cost $C_{\rm a}(\tau) $ is larger than the cost of rejected words $C_{\rm r}(\tau)$. So we expect the bound~\eqref{eq:pacbound} to be helpful for parameter values of the DFA and distribution over input words in which the acceptance rate is high. On the contrary when the probability of rejected words is large enough, we expect $C_{\rm r}(\tau) $ to be larger than $C_{\rm a}(\tau)$, and the bound~\eqref{eq:pacbound2} to be useful when the acceptance rate is low.

The above relations in Eq.~\eqref{eq:pacbound3} and Eqs.~\eqref{eq:pacbound} and \eqref{eq:pacbound2} concerning the acceptance probability of a word  can also be applied to any finite-horizon stopping time of the form $\mathcal{T}= \min(\mathcal{T}_c,\tau)$, where $\mathcal{T}_{\rm c}$ represents the time at which a given arbitrary condition $c$ is verified for the first time, e.g., the first time the accept state is reached twice, or the first time the accept state is reached after passing through any other arbitrary state (or sequence of them). Thus there is an ample flexibility in choosing the stopping condition $\T$, including the logical composition of any other set of conditions, e.g., $c = c_1 \cup c_2$ giving the first time either condition $c_1$ or condition $c_2$ are verified, or $c = c_1 \cap c_2$ for the fist time both $c_1$ and $c_2$ are simultaneously verified.

\section{Concatenating runs of a DFA with stochastic resetting}
\label{sec:stochasticstart}

In this section we further elaborate on how our results would be applied to sequences of computations separated by a {\em reset} of the dynamics which implements concatenated computational rounds. This is an interesting avenue where our results might be fruitfully combined in the future with the powerful analytical tools from the framework of {\em stochastic resetting}~\cite{evans2020stochastic,gupta2022stochastic,pal2017integral,tal2020experimental,roldan2016stochastic,bressloff2020modeling}. Let us consider a random sequence of symbols fed into a computer,
\begin{equation}
    0\;0\;0\;\sqcup\;0\;1\; 0\;1\;1\;1\; \sqcup\; 0\; 1\; 0 \; \dots,
    \label{eq:blanks}
\end{equation}
where $\sqcup$ is a blank symbol that flags the beginning of a new computation. For the example sequence~\eqref{eq:blanks}, a computation starts at the random starting time $\T_{\rm start}=5$ and ends at the stochastic ending time $\T_{\rm end}=10$ just before the next blank symbol arrives, thus generating the input word ``$010111$". During this computation, the computer begins computing at $\T_{\rm start}=4$ from its start state, and ends the computation either in an accept state or in another logical state. 

Now, stochastic starting times can be reformulated as stochastic stopping times [see also our results concerning multiple stopping times, Eq.~\eqref{eq:doob2st}].
In particular here the starting time $\T_{\rm start}$ is the first appearance of a blank symbol $\sqcup$. Whenever the probability of a blank symbol $p_{\sqcup}>0$ is greater than zero, then it is guaranteed that $P(\T_{\rm end} <\infty)=1$, i.e., there is a limit time $\tau$ that is a finite global upper limit to $\T_{\rm end}$. This is the setting which would correspond to, e.g., stochastic starting times that are drawn from distributions with bounded support, say from Bernoulli or Binomial distributions. Under such mild assumptions, it is then possible to establish thermodynamic constraints for computations starting at stochastic times.

Supposing $p_{\sqcup}>0$, we outline how the stopping-time fluctuation relations derived in our work can be applied to a computation of the example sequence~\eqref{eq:blanks}. First, we let the computer processes the sequence ``$000$", which implies visiting the accept state at least once.
At time $t=3$, the computer may or may not be in the start state depending on its update rules. Next at $t=4$, the state of the computer is reset to its start state from whichever  state $x_3$ it occupies at the previous time instance. Upon this, the computer processes the string $010111$ before the arrival of the next blank symbol, during which the logical state may or may not have reached the accept state. This leaves us with an ordered sequence of stopping times: $T_0=0$, $\T_1 = \min(\T_{\rm accept}^{(1)},\T_{\rm blank}^{(1)})$, $\T_2 = \T_{\rm blank}^{(1)}$, $\T_3 = \min(T_{\rm accept}^{(2)},T_{\rm blank}^{(2)})$, $\T_4 = T_{\rm blank}^{(2)}$, ..., $\T_{\infty} = \tau$, 
which obey
\begin{equation}
    \T_0\leq \T_1\leq \T_2\leq \T_3\leq \T_4\leq \dots\leq \T_\infty.
\end{equation}
Here above we have denoted by $\T_{\rm accept}^{(i)}$ the first return time to the accept state during  the computation of the $i-$th word. Similarly, $\T_{\rm blank}^{(i)}$ is the stochastic arrival time of the $i-$th blank symbol.  While the stochastic times $\T_{\rm accept}^{(i)}$ have the same structure as the stopping times  considered throughout our work,  the times $\T_{\rm blank}^{(i+1)}$ can be seen as stochastic starting times, which are also examples of stopping times for which our formalism applies.

\begin{figure}[t!]
\centering
   \includegraphics[width=\linewidth]{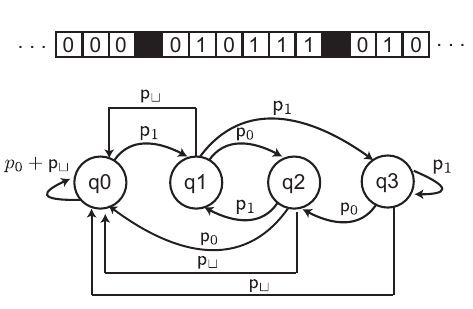} 
   \caption{An application of stochastic resetting to computer science. Top: illustration of a random tape of binary symbols interspersed by blank symbols that are processed during a computation. Bottom: discrete-time Markov chain associated with the deterministic finite automaton (DFA) recognizing binary words multiple of four (see Fig.~\ref{fig:sketch}c) with stochastic resetting to the start state. Along a stochastic computation, resetting takes place when a blank symbol is recognized by the DFA. For the model illustrated here, we have assumed that words are drawn from i.i.d. sequences with probabilities $p_0$, $p_1$ and $p_\sqcup$ of 0, 1 and blank symbols respectively, with $p_0+p_1+p_\sqcup=1$, however more complex scenarios could be envisaged in future work.  } 
     \label{fig:stochres}
\end{figure}

Figure~\ref{fig:stochres} provides an illustration of a DFA processing an i.i.d. sequence of bits interspersed by blank symbols. The DFA processing the symbols recognizes binary words multiples of four, as in the examples of Sec.~\ref{sec:examples}.
Assuming time-independent probabilities $p_0$, $p_1$ and $p_\sqcup$ for the occurrence of $0$,~$1$, and blank $\sqcup$ symbols respectively (with $p_0+p_1+p_\sqcup =1$), the DTMC associated with this computation can be represented by a discrete-time stochastic resetting process (see Fig.~\ref{fig:stochres}). In such processes, resetting takes place from each logical state to the start state $q_0$ at a stochastic starting time. This requires a suitable description of computation whose transition matrices include resetting events. For the DFA example considered here, such transition matrix takes the form 
\begin{equation}
    \textbf{W}= \begin{bmatrix}
p_0+p_\sqcup & p_\sqcup & p_0+p_\sqcup & p_\sqcup\\
p_1 & 0 & p_1 & 0 \\ 
0 & p_0 & 0 & p_0\\ 
0 & p_1 & 0 & p_1
\end{bmatrix},
\label{eq:Wor2}
\end{equation}
cf. Eq.~\eqref{eq:Wor} for the case where no resetting takes place, corresponding to $p_\sqcup=0$. The DTMC described by the transition matrix~\eqref{eq:Wor2} allows one to study multiple realistic computational scenarios where $\Sigma$ at stochastic starting and stopping times can be efficiently tackled. For example, one may consider that the processing of the input string by the DFA as a nonequilibrium stationary process with resetting and apply results from the  martingale theory for stationary processes (see Ch.~7 in Ref.~\cite{Edgar2022}). Alternatively, one can  apply the formalism in this work to establish bounds for the intrinsic mismatch costs of the computation between the first and the $n-$th arrival of a blank symbol, etc [see Eq.~\eqref{eq:second-starting}], similarly to Sec.~\ref{sec:examples}c.

\section{Discussion}
\label{sec:discussion}

In this work we  have shown how to extend stochastic thermodynamics to describe the minimal costs associated with a computer processing with a stochastic halting time, processing strings of arbitrary length. Our formalism applies to computations described by discrete-time Markov chains over a set of computational states that may have restricted initial conditions, unidirectional links, and start and/or stop at a stochastic time. We obtain quantifiers, which are collectively dubbed as the intrinsic mismatch cost of a computation, that lower-bound the entropy production incurred by the computer and that can be formulated at the fluctuating level. A key insight here is that these quantifiers, which provide a tool to probe the entropy production associated with computations at stopping times, can be entirely obtained from the DTMC evolution and the prior, without further details about their physical implementation. Notice that such an intrinsic cost is independent of the internal energy of the computational states, $x_t \in \mathcal{X}$, which can be indeed assumed to be equal for every computational state and constant over time. Still, non-zero entropy production through the irreversible dissipation of heat into the environment will be in general incurred for any physical computer which implements a given computation over such set of states.

Putting forward the modern martingale formalism of stochastic thermodynamics, we also unveiled a plethora of universal fluctuation relations and inequalities that are valid for the broad class of computations analyzed in this work. We obtained a main fluctuation theorem, Eq.~\eqref{eq:IFTST}, valid for settings which include arbitrary stopping times, unidirectional transitions, and absolute irreversibility. In doing that, we have extended the martingale theory for stochastic thermodynamics to account for this additional source of irreversibility in generic situations, which we expect to have broad applicability in nonequilibrium thermodynamics.

The rigor and flexibility of our theory for stopping times allowed us to formulate and interpret several second-law-like inequalities [Eqs.~\eqref{eq:stop-sl}--\eqref{eq:stop-s3}] at stochastic stopping times, as well as relations for the probabilities of acceptance/rejection of input data by a computer in terms of thermodynamic quantities [equality~\eqref{eq:pacbound3} and inequalities~\eqref{eq:pacbound}-\eqref{eq:pacbound2}]. In particular, the second law inequalities~\eqref{eq:stop-sl} and~\eqref{eq:stop-s3}, provides us useful lower bounds on the minimum dissipation incurred by a generic computation stopping at an arbitrary stopping time, while Eq.~\eqref{eq:improvement} establishes formally how stopping times can be used to reduce the thermodynamic costs of a computation by means of time-reversal-symmetry breaking. Moreover, we have also shown the relevance of accounting for absolutely irreversible sequences in providing accurate bounds for the intrinsic mismatch cost of the computation. In this sense, the bound we derived in Eq.~\eqref{eq:stop-sl} with the absolute irreversibility term $\Gamma_\tau$ is tighter, with respect to the alternative bound in Eq.~\eqref{eq:stop-s3}. However, computing $\Gamma_\tau$ might be challenging depending on the setting considered, specially for large limit times $\tau$. On the contrary, the alternative bound in Eq.~\eqref{eq:stop-s3}, would be much easier to compute (as it only depend on two probability distributions), while still providing a meaningful bound in all examples explored here.  

The framework developed in this paper can be readily applied for assessing thermodynamic costs of computations in a broad range of models of relevance in CS theory, including --but not being limited to-- deterministic finite automata. Our results apply to every computation implemented by a synchronous digital computer and remains valid independently of how the computational variables are defined. In particular they can include already processed input symbols (as in non-iid DFAs), stacks (as in pushdown automata models), or even entire words written on a random access tape (as in Turing machines). Hence our results provide a tool to classify abstract computational machines by their intrinsic (unavoidable) thermodynamic costs. Applying our framework to more complex models of computational machines such as pushdown automata or Turing machines halting at stochastic times is a natural step following the investigation initiated here.

Our results are amenable of experimental testing using state-of-the-art techniques in line with previous tests of Landauer's principle~\cite{berut2012experimental,jun2014high,proesmans2020finite} and other experimental platforms in stochastic thermodynamics~\cite{ciliberto2017} in setups ranging from colloidal particles~\cite{martinez2017colloidal} and nanoscale  devices~\cite{pekola2016maxwell,dago2022dynamics} to biopolymers~\cite{Ritort21}. Regarding the determination of the prior, in some cases the experimentalist will have designed the system in a sufficiently detailed manner such that it is possible to calculate the prior, or at least approximate it with reliable numerical estimates. In other cases, the experimentalist can estimate the prior by repeatedly running the system and observing the resultant behavior. In any case, our results have nonzero lower bounds that apply no matter what the prior is.

It would also be very interesting in the future to extend the framework developed here by combining analytical tools from  stochastic resetting (e.g. renewal theory and first-passage-time ideas~\cite{redner2001guide}) with computer science methods. This will allow to obtain tight bounds for the statistics of starting time and entropy production bounds in specific models of DFAs and TMs processing regular languages, as follows from the ideas sketched in Sec.~\ref{sec:stochasticstart}.
Also, we note that even if current digital devices are \textit{very close} to periodic, they are not exactly so. In other words, they are some first-order perturbation away from being periodic, which suggests other avenues for future work. In general, the prior is a function of the physical process implementing the computation. For example, it is a function of the time-dependent rate matrix in the case of a CTMC. Given this, we {might} be able to use the envelope theorem (often used in game theory) to calculate how much the prior can change under first order perturbations away from an exactly periodic process. That in turn might allow us to modify \cref{eq:sum_mismatch_costs_min} to involve some infinitesimal first-order perturbation parameter $\epsilon$ characterizing how much the process differs from being exactly periodic.

However, the results developed in this paper also provide new insights in the field of nonequilibrium thermodynamics. An important consequence of our work is the finding that the auxiliary dynamics introduced in Eq.~\eqref{eq:auxiliary} is suitable to treat processes at stopping times that may have unidirectional transitions and absolute irreversibility, hence making our framework applicable to generic situations where local detail balance is broken. Similar auxiliary dynamics has been invoked in the literature such as so-called "dual", ``dual-reversed" or "adjoint" dynamics, in the context of fluctuation theorems, see e.g.~\cite{chernyak2006path,esposito2010three,Verley12,Manzano15,baiesi2015inflow,Manzano18}. In particular, as shown above, if the process admits a well-behaved stationary solution, we can obtain from $\Sigma$ the so-called non-adiabatic (or excess) entropy production~\cite{esposito2010three,Esposito10PRE}. Within such scenario, our work is another brick in the wall or recent progress highlighting the role of non-adiabatic entropy and excess heat in characterizing the efficiency~\cite{datta2022second} and calorimetry~\cite{10.1063/5.0142694,dolai2023calorimetry} of active nonequilibrium systems.

We also expect our results to have potential applications outside statistical mechanics and computer science, e.g. in biological physics, for instance within the field of biomolecular computation~\cite{benenson2012biomolecular}, enzyme kinetics~\cite{cornish2013fundamentals}  and information processing in biology~\cite{dubuis2013positional}. As a minimal model, consider a minimal Michaelis-Menten scheme for enzyme kinetics in which an enzyme $E$ transforms a substrate molecule $S$ into a product molecule $P$.  A typical assumption is that the conversion of the substrate into a product takes place through an irreversible chemical reaction 
$ E+S\xrightleftharpoons[k_1^-]{k_1^+} ES \xrightharpoonup{k_2^+} E+P$,
where $k_i$'s here are suitable transition rates. Within this model, the  enzyme's state during enzymatic cycles follows a continuous-time Markov jump process with one irreversible transition.  In previous works, the presence of the irreversible transition has been circumvented by considering virtual processes with a very slow transition rate. However our formalism can be readily applied to describe the stochastic thermodynamics of such enzymatic reaction, inasmuch that it does not require the presence of bidirectional transitions. Similarly we expect our formalism to be suitable to describe fluctuations of biological populations in processes that include totally irreversible transitions such as cell-fate decisions~\cite{li2013quantifying,wang2015landscape} cell death and apoptosis~\cite{ban2023proliferative}, among others.  For such systems, our approach puts forward   recent approaches to estimate dissipation developed within in the field of active matter~\cite{battle2016broken,fodor2016far,roldan2021quantifying} which did not contemplate the presence of unidirectional transitions which are commonplace in biophysical modelling~\cite{morin2015mechano,dutta2020stochastic}.

\begin{acknowledgements}
We thank Artemy Kolchinsky for  useful comments on the first version of the  manuscript. GM acknowledges 'Ram\'on y Cajal' program (RYC2021-031121-I), the CoQuSy project (PID2022-140506NB-C21) and support from the María de Maeztu project (CEX2021-001164-M) funded by MCIN/AEI/10.13039/501100011033 and European Union NextGenerationEU/PRTR. ER acknowledges fruitful discussions with L\'ea Bresque and financial support from PNRR MUR project PE0000023-NQSTI. DHW acknowldges US NSF Grant CHE-1648973. DHW also thanks the Santa Fe Institute for support.
\end{acknowledgements}

\appendix

\section*{APPENDIX}

\section{Residual cost}
\label{app:residual}

Suppose we are given a physical process which implements a single-valued function $f$ over a space $X$. The \textit{islands} of $f$ are defined as the elements of the partition of $X$ given by the pre-images of $f$. Formally, if we write the image of $f$ as $f(X)$, the islands of $f$ are the sets $\left\{f^{-1}(x): x \in f(X)\right\}$. We write the set of islands of a function $f$ as $L(f)$. 

In this appendix we will  expand the residual cost $\mathcal{R}(\varrho_0)$ arising in \cref{eq:2M} in terms of islands, discuss some properties of the residual  cost, and then justify why this cost can be ignored in our investigation in this paper. For simplicity we will restrict attention to the case where the linear term $F$ in \cref{eq:2mm} is expected entropy flow generated in a process, so that $\mathcal{C}(\tau)$ is the EP generated in the process up to time $\tau$. However, all of the discussion extends to other choices of $F$, with the obvious modifications.

\cref{eq:2M} provides entropy production as a sum of two terms, the mismatch cost and the residual cost, where $\mathcal{R}(\tau)$ corresponds to the residual cost. 
For any conditional distribution $G$, and any island $c \in L(G)$, we define the prior within that island as \cite{Wolpert2019}: 
\begin{equation}
\varrho_{\min}^c \in \underset{\varrho: \operatorname{supp}(\varrho) \in \Delta_c}{\arg \min } \mathcal{C}(\tau)
\end{equation}
where the subscript means that $\varrho$ is a distribution whose support is restricted to the unit simplex over the island $c$, $\Delta_c$, viewed as a subset of the state space $\mathcal{Y}$.
The associated minimum EP in that island is written as
\begin{equation}
\mathcal{C}_{\min }^c(\tau):=\min _{\varrho: \operatorname{supp}(\varrho) \in \Delta_c} \mathcal{C}(\tau).
\end{equation}
Now introduce an arbitrary distribution over islands, $q(c)$, and define
\begin{equation}
\varrho_{\min}:=\sum_{c \in L(G)} q(c)~ \varrho_{\min}^c.
\end{equation}
Then the residual EP of the physical process that implements $G$ is 
\eq{
\mathcal{R}(\tau) = \sum_{c \in L(G)} p_\tau(c)~\mathcal{C}_{\min}^c(\tau)
}
as shown in~\cite{Wolpert2019}. Therefore the EP for the process starting with distribution $\varrho_0$ is
\begin{equation}
\mathcal{C}(\tau)=D(\varrho_0 \| \varrho_{\min})-D(G \varrho_0 \| G \varrho_{\min}) + \sum_{c \in L(G)} p_\tau(c) \mathcal{C}_{\min}^c.
\end{equation}

Since expected EP $\mathcal{C}(\tau)$ is non-negative, by definition the residual cost of any island 
$c$, $\mathcal{C}_{\min}^c$, is non-negative. So the total residual cost given by an expectation over
all islands, which is the residual cost for the entire thermodynamic process characterized by $G$, is also non-negative. Like priors, residual costs of islands in general will differ from one cost function $\mathcal{C}$ to the next. 

In general, as the iteration $t$ of a periodic process changes, the distribution $p_t(c)$ over the islands $c$ will change. Therefore so will the associated total expected residual cost. However, since that total residual cost is always non-negative, all the lower bounds on EP in the main text that consider only mismatch cost apply. This is true even if the residual costs of the islands are strictly positive. In particular, \cref{eq:min_EP_comp} will still be a lower bound on the EP generated in the process.

On the other hand, if we write down the formula for minimal total residual cost which is analogous to \cref{eq:min_EP_comp}, minimizing over the residual costs of each island, we just get zero,
by taking those costs to all equal zero. So unless we fix the physical details underlying the process, and therefore fix the residual costs of the islands to be strictly positive, our analysis of lower bounds isn't changed by the existence of residual costs. This is why such quantities are ignored in the main text.

As a final comment, note that in general both the prior and residual costs will vary with $\tau$. However, the same mismatch cost formula bounds dissipation for any such choice of $\tau$, once one plugs in the appropriate prior. This need not be true for residual cost, in the sense that the islands might change for different choices of $\tau$.

\section{Trajectory level mismatch cost}
\label{sec:traj_level_mmc}

Define $[G_t \varrho](y)$ as the distribution $\varrho$ evolved through the (linear) dynamics $G$ up to time $t$, and then evaluated at state $y \in \mathcal{Y}$. Using this, we can define a trajectory-level version of mismatch cost (and associated instance-level version), as
\eq{
& m_{\varrho_0}(Y_{[0,\tau]}) := 
\ln \left[\frac{\varrho_0(y_0)}{\varrho_{\min}(y_0)}\right] - 
\ln \left[\frac{G\varrho_0(y_0)}{G\varrho_{\min}(y_0)}\right] \\
   &= \sum_{t=0}^{\tau-1}
        \ln \left(\frac{[G_t \varrho_0](y_t)}{[G_t \varrho_{\min}](y_t)}\right) - \ln \left(\frac{[G_{t+1} \varrho_0](y_{t+1})}{[G_{t+1}\varrho_{\min}](y_{t+1})}\right). \nonumber
}
By construction the expected value of $m_{\rho_0}(Y_{[0,\tau]})$ over all trajectories $Y_{[0,\tau]}$ equals the (ensemble level) mismatch cost, $\mathcal{M}(\rho_0)$ given in \cref{eq:2mmm}. In
the context of inclusive thermodynamics, such definitions allowed the derivation of fluctuation theorems for mismatch cost~\cite{kolchinsky2021dependence}.

\section{Entropy production bounds from discrete-time evolution over computational states}
\label{app:coarse-grain}

We focus on the case in which the cost function $\mathcal{C}(\tau)$ introduced in Eq.~\eqref{eq:2mm} corresponds to the entropy production, that is, we identify $F$ as the entropy flow due to heat exchange of the system with (one or several) heat baths, and under the eventual presence of non-conservative forces~\cite{Seifert12}. In the case in which the continuous-time periodic process is also Markovian, that is, it can be described with a set of rate matrices $K_{ij}^v(t)$ associated to the different reservoirs, the entropy flow $F$ will take the form presented in Eq.~\eqref{eq:EF}. Otherwise, the explicit expression of the entropy flow $F$ might be more involved~\cite{Strasberg2017,Whitney2018,Cockrell2022}. 

In any case, the average entropy production in the continuous-time physical process can be written in terms of a Kullback-Leibler divergence between path probabilities of trajectories $\mathbf{y}_{[0,\tau]} = \{y_t ~|~ 0\leq t \leq \tau \}$ in the extended state space, $y_t \in \mathcal{Y}$, and their time-reversed counterparts $\Theta {\mathbf{y}}_{[0,\tau]} = \{ y_{\tau -t} ~|~ 0 \leq t \leq \tau\}$ as~\cite{Kawai07,Gomez08,Roldan10}:
\begin{eqnarray}
    \mathcal{C}(\tau) = \langle S_\mathrm{tot}(\tau) \rangle = D[\mathds{P}(\mathbf{y}_{[0,\tau]}) || \tilde{\mathds{P}}(\Theta{\mathbf{y}}_{[0,\tau]})],
\end{eqnarray}
where $\mathds{P}$ and $\tilde{\mathds{P}}$ stand for the probability measures in the forward and time-reversed driven processes, respectively.

By introducing the trajectories $\mathbf{x}_{[0, \tau]} = x_0, x_1, ..., x_\tau$ on discrete time, visible space dynamics with corresponding probability $ P(\mathbf{x}_{[0,\tau]})$ as introduced in Eq.~\eqref{eq:Ppath}, we can bound the above Kullback-Leibler divergence as
\begin{align} \label{eq:Stot2}
 \langle S_\mathrm{tot}(\tau) \rangle &= D[\mathds{P}(\mathbf{y}_{[0,\tau]}) || \tilde{\mathds{P}} (\Theta \mathbf{y}_{[0,\tau]})] \nonumber \\
     &=  D[ \mathds{P}(\mathbf{y}_{[0,\tau]} | \mathbf{x}_{[0,\tau]}) || \tilde{\mathds{P}} (\Theta \mathbf{y}_{[0,\tau]}| \Theta \mathbf{x}_{[0,\tau]}) ] \nonumber \\
    &~~~+  D[P(\mathbf{x}_{[0,\tau]}) || \tilde{P}(\Theta \mathbf{x}_{[0,\tau]})]
    \nonumber \\
    & \geq  D[P(\mathbf{x}_{[0,\tau]}) || \tilde{P}(\Theta \mathbf{x}_{[0,\tau]})] =: \langle \breve{S}_\mathrm{tot}(\tau) \rangle,
\end{align}
where we used chain rule for Kullback-Leibler divergence to write the second equality and denoted the coarse-grained path probabilities for forward and time-reversed driving protocols by ${P}(\mathbf{x}_{[0,\tau]}) = \int d{\mathbf{y}_{[0,\tau]}} {\mathds{P}}(\mathbf{y}_{[0,\tau]}) \prod_{n=0}^{\tau} \delta(y_{n} - x_n)$ and $\tilde{P}(\Theta \mathbf{x}_{[0,\tau]}) = \int d{\mathbf{y}_{[0,\tau]}} \tilde{\mathds{P}}(\Theta \mathbf{y}_{[0,\tau]}) \prod_{n=0}^{\tau} \delta(y_{n} - x_n)$ respectively.  

The quantity $\langle \breve{S}_\mathrm{tot}(\tau) \rangle$ defined in the r.h.s. of Eq.~\eqref{eq:Stot2} is a coarse-grained version of the average entropy production defined only over the DTMC dynamics and reduced state space $\mathcal{X}$ verifying $\langle {S}_\mathrm{tot}(\tau) \rangle \geq \langle \breve{S}_\mathrm{tot}(\tau) \rangle$. We can now decompose the coarse-grained entropy production $\langle \breve{S}_\mathrm{tot}(\tau) \rangle$ into the sum of mismatch and residual costs as in Eq.~\eqref{eq:2M} at the DTMC level. In particular, by assuming that all non-computational degrees of freedom $y_t \notin \mathcal{X}$ are reinitialized to their starting values in every cycle of the periodic computational process, the prior probability minimizing $\breve{S}_\mathrm{tot}$ in a single cycle will be the same for all cycles and can be easily obtained from marginalization of the fine-grained prior $\varrho_{\min}(y)$ as $\mu(x) = \sum_{y \notin \mathcal{X}} \varrho_{\min}(y)$.

As a consequence the coarse-grained DTMC mismatch cost over a single cycle of the process reads:
\begin{eqnarray} \label{eq:M}
\mathcal{M}(\rho_t) := D(\rho_t || \mu) - D(\rho_{t+1} || \mu^\prime) \geq 0,
\end{eqnarray}
with $\mu^\prime$ the prior distribution evolved during a single iteration of the DTMC, $\mu^\prime = \mathbf{W} \mu$. Analogously $\mathcal{R}(\rho_t) := \langle \breve{S}_\mathrm{tot}(\rho_t) \rangle - \mathcal{M}(\rho_t) =  \min_{\rho_t} \langle \breve{S}_\mathrm{tot} (\rho_t)\rangle \geq 0$ represents the residual costs during a single iteration, which corresponds to the entropy production when the computational system starts the evolution in distribution $\mu$. Hence for every single iteration $\langle \breve{S}_\mathrm{tot}(\rho_t) \rangle = \mathcal{M}(\rho_t) + \mathcal{R}(\rho_t)$, from which it follows that the above mismatch cost provides a lower bound to the average entropy production during a single cycle. 
Finally, by summing the mismatch costs per iteration \eqref{eq:M} over all iterations up to the final time $\tau$, we recover Eq.~\eqref{eq:sum_mismatch_costs} for the sum of mismatch costs during the entire computation (that we dub intrinsic mismatch cost), which verifies the chain inequality:
\begin{eqnarray}
 \sum_{t=0}^{\tau-1} \mathcal{M}(\rho_t) \leq \langle \breve{S}_\mathrm{tot}(\tau) \rangle \leq \langle S_\mathrm{tot}(\tau) \rangle,   
\end{eqnarray}
hence providing a lower bound on the average entropy production during the entire computation.

\section{Strict positivity of the mismatch cost sum given by Eq.~\eqref{eq:sum_mismatch_costs}.}
\label{app:mismatch}
Here we provide a proof of the strict positivity of the mismatch cost sum in Eq.~\eqref{eq:sum_mismatch_costs}, that is $\sum_{t=0}^{\tau-1} \mathcal{M}(\rho_t) > 0$.

We know that, in general, $D( \rho_0 \,||\, \mu) - D(G \rho_0 \,||\, G \mu) = 0$ if and only if $\rho_0 = \mu$. However, if in fact $G$ is not logically invertible, and yet $\rho_0 = \mu$, then $G \rho_0 \ne \mu$. This means that either $D( \rho_0 \,||\, \mu) - D(G \rho_0 \,||\, G \mu) = 0$ or $D(G \rho_0 \,||\, \mu) - D(G^2 \rho_0 \,||\, G \mu) = 0$ --- but not both. Therefore, so long as $G$ is not logically invertible, the sum in \eqref{eq:sum_mismatch_costs} is not zero.

\section{Proof of the supermartingale property~\eqref{eq:superM2}}
\label{appS}

Here we provide a detailed proof of the supermartingale property of the process $M_\tau(t)$ defined by Eq.~\eqref{eq:smprocess}:
\begin{align}
\label{eq:D1}
    &\langle M_\tau(\tau) \,|\, X_{[0,t]} \rangle \equiv  \sum_{X_{(t,\tau]} \in \mathcal{F}_\mathrm{AC}} P(X_{[0,\tau]} \,|\, X_{[0,t]}) M_\tau(\tau) \\ 
\label{eq:D2}
    &= \sum_{X_{[t+1,\tau]} \in \mathcal{F}_\mathrm{AC}} \frac{\bar{P}(\Theta X_{[0,\tau]})}{P(X_{[0,t]})} \\
    &=~ e^{-\Sigma(t)} \sum_{X_{[t+1,\tau]} \in \mathcal{F}_\mathrm{AC}} \frac{\bar{P} (\Theta X_{[0,\tau]})}{\bar{P}(\Theta X_{[0,t]})} \nonumber \\ &= e^{-\Sigma(t) - \delta_\tau(t)}  \sum_{X_{[t+1,\tau]} \in \mathcal{F}_\mathrm{AC}} \frac{\bar{P}(\Theta X_{[t, \tau]})}{\bar{\rho}_{\tau - t}(x_t)}  \nonumber \\ 
&= M_\tau(t) ~ \left[1 - \sum_{X_{[t+1, \tau]} \in \mathcal{F}_\mathrm{AI}} \frac{\bar{P}( \Theta X_{[t, \tau]})}{\bar{\rho}_{\tau - t}(x_t)} \right] \nonumber  \\ 
&= M_\tau (t) ~\left[ 1 - \alpha_\tau(t)\right] ~\leq M_\tau (t) \label{eq:superapp}.
\end{align}
To obtain Eq.~\eqref{eq:superapp} we first used the definition of conditional probability in \cref{eq:D1} to derive \cref{eq:D2}. We then multiplied and divided by $\bar{P}(\Theta X_{[0,t]})$, in order to obtain 
the $e^{-\Sigma(t)}$ multiplicative factor. After that we expanded the remaining probabilities $\bar{P}(\Theta X_{[0,\tau]})$ in the numerator and $\bar{P}(\Theta X_{[0,t]})$ in the denominator, and multiplied the resulting expression with $\rho_t(x_t)/\bar{\rho}_{\tau - t}(x_t)$ to obtain the expression
involving  $M_\tau(t)$ (recall the definition of $M_\tau(t)$ in Eq.~\eqref{eq:SD}). Finally, we transformed the sum to involve
$\mathcal{F}_\mathrm{AI}$, the complementary filtration of $\mathcal{F}_\mathrm{AC}$, and invoked the fact that $\sum_{X_{[t,\tau]} \in \mathcal{F}} \rho_\tau(x_\tau) \bar{P}({x_{\tau-1} \,|\, x_\tau}) ...~ \bar{P}({x_t \,|\, x_{t+1}}) = \bar{\rho}_\mathrm{\tau - t}(x_t)$ when the sum is taken over the complete set of trajectories $\mathcal{F}$.

\section{Stopping-times fluctuation theorem with absolute irreversibility in Eq.~\eqref{eq:IFTST}} \label{appT}
In this appendix we give a detailed proof of the main stopping-times fluctuation theorem with absolute irreversibility presented in Eq.~\eqref{eq:IFTST}. For the proof, it is convenient to split the filtration $\mathcal{F}$ into subsets of filtrations ${\mathcal{F}^{(t)}}$ containing all trajectories that are stopped at time $t$, i.e. for which $M_\tau(t) = M_\tau(\mathcal{T})$. We take $t= 0, 1 ,..., \tau$, where $\tau$ is the maximum allowed time. Notice that we enforce the dynamics to stop at $\tau$ if not previously done (however one can later take $\tau \rightarrow \infty$ whenever $M_\tau(\tau)$ remains bounded). Therefore we have $\mathcal{F} = \mathcal{F}^{(1)} \cup ~...~ \cup \mathcal{F}^{(\tau)}$. Now, in analogy to the previous section, we define for each (discrete) instant of time $t$, the sets $\mathcal{F}_\mathrm{AC}^{(t)}$ and $\mathcal{F}_\mathrm{AI}^{(t)}$ of trajectories that are both stopped at $t$ and which are either allowed $[P(X_{[0,t]}) >0$ or $P(X_{[0,t]}) = P(\Theta X_{[0,t]}) = 0]$ or only forbidden in the original dynamics $[P(X_{[0,t]}) =0$ with $\bar{P}(\Theta X_{[0,t]})>0]$, respectively. This implies that $\mathcal{F}^{(t)} = \mathcal{F}_\mathrm{AC}^{(t)} \cup \mathcal{F}_\mathrm{AI}^{(t)}$ at any $t = 0, 1 ... \tau$. Then we have:
\begin{align} 
    &\langle M_\tau(\mathcal{T}) \rangle \equiv \sum_{t=0}^\tau ~ \sum_{X_{[0,t]} \in \mathcal{F}_\mathrm{AC}^{(t)}} P(X_{[0,t]})~ M_\tau(t) \nonumber \\ &= \sum_{t=0}^\tau ~\sum_{X_{[0,t]} \in \mathcal{F}_\mathrm{AC}^{(t)}} P(X_{[0,t]}) ~\Big[\langle M_\tau(\tau) \,|\, X_{[0,t]} \rangle +   M_\tau(t) \alpha_\tau(t) \Big] \nonumber \\
     &= \langle M_\tau(\tau)  \rangle + \sum_{t=0}^\tau ~\sum_{X_{[0,t]} \in \mathcal{F}_\mathrm{AC}^{(t)}}   P(X_{[0,t]}) M_\tau(t) ~\alpha_\tau(t) \nonumber \\ 
     &= 1 - \gamma_\tau + \sum_{t=0}^\tau ~\sum_{X_{[0,t]} \in \mathcal{F}_\mathrm{AC}^{(t)}}  P(X_{[0,t]}) M_\tau(t) ~\alpha_\tau(t) \label{eq:subtle}  \\
     &=1 - \sum_{t=0}^\tau \sum_{X_{[0,t]} \in \mathcal{F}_\mathrm{AI}^{(t)}} \bar{P}(\Theta X_{[0,t]}) = 1 - \Gamma_\tau. \label{eq:FTSep}
\end{align}
where we used the fluctuation theorem at fixed times with absolute irreversibility, c.f. Eq.~\eqref{eq:FTai}, and in the last line we defined the correction term for stopping times:
\begin{equation}
   \Gamma_\tau \equiv  \sum_{t=0}^\tau \sum_{X_{[0,t]} \in \mathcal{F}_\mathrm{AI}^{(t)}} \bar{P}(\Theta X_{[0,t]}) \frac{\bar{\rho}_{\tau-t}(X_t)}{{\rho}_t(X_t)} \geq 0
\end{equation}
summing up the probabilities for all the (reversed) stopped AI trajectories.

To reach Eq.~\eqref{eq:FTSep} from Eq.~\eqref{eq:subtle} we used :
\begin{align}
    &\sum_{t=0}^\tau ~\sum_{X_{[0,t]} \in \mathcal{F}_\mathrm{AC}^{(t)}}  P(X_{[0,t]})  M_\tau(t) ~\alpha_\tau(t) \nonumber \\
    &=\sum_{t=0}^{\tau} \sum_{X_{[0,t]} \in \mathcal{F}_\mathrm{AC}^{(t)}} \frac{\bar{P}(\Theta X_{[0,t]})}{{{\rho}_t(X_t)}} \sum_{X_{(t, \tau]} \in \mathcal{F}_\mathrm{AI}} \bar{P}(\Theta X_{[t,\tau]}) \nonumber \\
    &=\sum_{t=0}^{\tau} \sum_{X_{[0,t]} \in \mathcal{F}_\mathrm{AC}^{(t)}} \frac{\bar{P}(\Theta X_{[0,t]})}{{{\rho}_t(X_t)}} 
    [ \bar{\rho}_{\tau-t}(X_t) - \sum_{X_{(t,\tau]} \in \mathcal{F}_\mathrm{AC}} \bar{P}(\Theta X_{(t, \tau]})] \nonumber \\ 
    & = \sum_{t=0}^{\tau} \sum_{X_{[0,t]} \in \mathcal{F}_\mathrm{AC}^{(t)}} \bar{P}(\Theta X_{[0,t]}) \frac{ \bar{\rho}_{\tau-t}(X_t)}{{\rho}_t(X_t)} -  \sum_{X_{[0,\tau]} \in \mathcal{F}_\mathrm{AC}} \bar{P}(\Theta X_{[0,\tau]}) \nonumber \\ 
    &= \sum_{t=0}^{\tau} \sum_{X_{[0,t]} \in \mathcal{F}_\mathrm{AC}^{(t)}} \bar{P}(\Theta X_{[0,t]})\frac{ \bar{\rho}_{\tau-t}(X_t)}{{\rho}_t(X_t)} - 1 + \gamma_\tau \nonumber \\
    &= - \sum_{t=0}^{\tau} \sum_{X_{[0,t]} \in \mathcal{F}_\mathrm{AI}^{(t)}} \bar{P}(\Theta X_{[0,t]})\frac{ \bar{\rho}_{\tau-t}(X_t)}{{\rho}_t(X_t)} + \gamma_\tau \nonumber \\ 
    &= -\Gamma_\tau + \gamma_\tau,
\end{align}
where in the last line we used that $\mathcal{F}_\mathrm{AC}^{(t)} \cup \mathcal{F}_\mathrm{AI}^{(t)} = \mathcal{F}^{(t)}$ and $\sum_{t=0}^{\tau} \sum_{X_{[0,t]} \in \mathcal{F}^{(t)}} \bar{P}(\Theta X_{[0,t]}) \bar{\rho}_{\tau-t}(X_t)/{\rho}_t(X_t) = 1$, since the trajectories are stopped with certainty in the interval $[0,\tau]$ and $\bar{P}(\Theta X_{[0,t]}) \bar{\rho}_{\tau-t}(X_t)/{\rho}_t(X_t)$ is a normalized path probability.

\section{Ordered stopping-times fluctuation relation in Eq.~\eqref{eq:doob2st2}} \label{app:sampling}

In this appendix we derive the stopping-times fluctuation relation for two ordered stopping times in Eq.~\eqref{eq:doob2st2} via Doob's optional sampling theorem. Let us start by considering a finite but otherwise generic stopping time $\T$ of the form in Eq.~\eqref{stop}, such that $0 \leq \T \leq \tau$. Doob's optional sampling theorem for supermartingales reads~\cite{Doob} (see also Ref.~\cite{Edgar2022}):
\begin{equation} \label{eq:OST}
    \langle A(\T) | \mathbf{x}_{[0,t]} \rangle \leq  A(\min\{\T, t\}),
\end{equation}
where $A$ is a supermartingale over the trajectories $\mathbf{x}_{[0,\tau]}$.

In the following we will apply Eq.~\eqref{eq:OST} to the supermantingale process $M_\tau(t) = e^{-\Sigma(t) - \delta_\tau(t)}$ introduced in Eq.~\eqref{eq:smprocess}, for two ordered stopping times $\T_1$ and $\T_2$ such that $P(\T_2 \geq \T_1) = 1$. In this case we have: 
\begin{eqnarray} \label{eq:OSTstop}
\langle M_\tau(\T_2) | \mathbf{x}_{[0,\T_1]} \rangle \leq  M_\tau(\T_1),    
\end{eqnarray}
where we called $\T_2 = \T$ and $\T_1 =  \min\{\T, t\}$. Notice that above $\mathbf{x}_{[0,t]}$ can be replaced by $\mathbf{x}_{[0,\T_1]}$ since $\T_1 \leq t$ by construction and Eq.~\eqref{eq:OST} is valid for any generic $t \leq \tau$. The average over stopping times of $M_\tau(\T_1)$ then verifies:
\begin{align}
    \langle M_\tau(\T_1) \rangle &= \sum_{\T_1 = 0}^\tau \sum_{\mathbf{x}_{[0,\T_1]}} P(\mathbf{x}_{[0,\T_1]}, \T_1) M_\tau(\T_1) \nonumber \\ 
    &= \sum_{\T_1 = 0}^{\T_2} \sum_{\mathbf{x}_{[0,\T_1]}} P(\mathbf{x}_{[0,\T_1]}, \T_1) M_\tau(\T_1)  \nonumber \\ 
    &~~~+ \sum_{\T_1 = \T_2}^{\tau} \sum_{\mathbf{x}_{[0,\T_1]}} P(\mathbf{x}_{[0,\T_1]}, \T_1) M_\tau(\T_1) \nonumber \\
    &\geq  \sum_{\T_1 = 0}^{\T_2} \sum_{\mathbf{x}_{[0,\T_1]}} P(\mathbf{x}_{[0,\T_1]}, \T_1) \langle M_\tau(\T_2) | \mathbf{x}_{[0,\T_1]} \rangle \nonumber \\ &= \langle M_\tau(\T_2) \rangle, \label{eq:proofoptimal}
\end{align}
where we used that $\T_2 \geq \T_1$ and Eq.~\eqref{eq:OSTstop} to reach the inequality.  The last line follows from the fact that we are summing the conditional average over all possible trajectories for stopping times $\T_1$ between $0$ and $\T_2$. Finally by replacing in Eq.~\eqref{eq:proofoptimal} the explicit expression for $M_\tau(t)$ in Eq.~\eqref{eq:smprocess}, we directly recover Eq.~\eqref{eq:doob2st2}. The inequality in Eq.~\eqref{eq:proofoptimal} is saturated either for $\T_2 = \T_1$ or when the supermartingale $M_\tau(t)$ becomes a martingale, that is, in the absence of absolute irreversibility.

\section{Scaling of thermodynamic averages with limit time in the DFA example} 
\label{app:scaling}
As pointed in Sec.~\ref{sec:examples}b for the minimal DFA in Fig.~\ref{fig:oraux}a), we observe a tendency for the intrinsic mismatch cost at stopping times $\langle\Sigma(\mathcal{T})\rangle$ to saturate when increasing the limit time $\tau$. This is in stark contrast with the scaling behaviour of the fixed-time average $\langle\Sigma(\tau)\rangle$ with $\tau$. 

\begin{figure}[tbh]
  \centering
\includegraphics[width=0.45\textwidth]{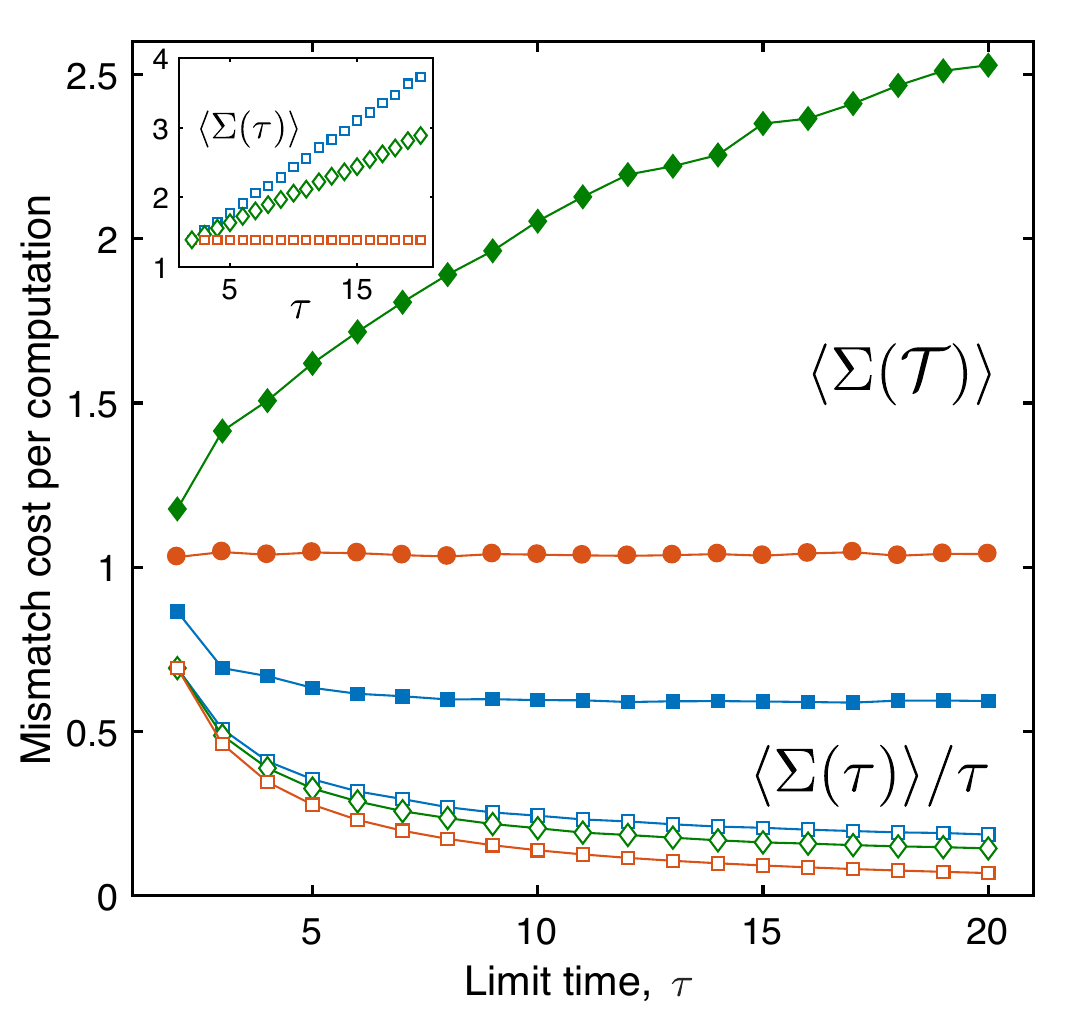}
\caption{Numerical results for the intrinsic mismatch cost with uniform prior as a function of the limit time $\tau$. We show three different values of the probability of symbol $0$: $p_0=0.75$ (blue squares), $p_0=0.5$ (red circles), and $p_0=0.3$ (green diamonds). Filled symbols correspond to the stopping-time average $\langle\Sigma(\mathcal{T})\rangle$. We include for comparison the corresponding fixed-time ensemble average rate $\langle \Sigma(\tau)\rangle/\tau$ without absorbing conditions (open symbols). The open symbols in the inset show the value of $\langle \Sigma(\tau)\rangle$. Parameters of the simulations are as in Fig.~\ref{fig:5}.
}
\label{fig:6}
\end{figure}

In this appendix we provide extra numerical evidence for the scaling behavior of mismatch cost at stopping and fixed times, which is shown in Fig.~\ref{fig:6}. There we observe that indeed the average intrinsic mismatch cost at fixed times scales linearly with $\tau$, that is $\langle\Sigma(\tau)\rangle\sim \tau$, as illustrated in the inset of Fig.~\ref{fig:6}. Moreover we obtain that $\langle\Sigma(\tau)\rangle/\tau$ decreases monotonically with $\tau$ up to a saturating positive value, yet its scaling behaviour is rather insensitive to the statistics of the input strings (see open symbols in Fig.~\ref{fig:6}). This point makes us question whether the fixed-time average $\langle\Sigma(\tau)\rangle$, or its rate per iteration $\langle\Sigma(\tau)\rangle/\tau$, is a suitable indicator of the thermodynamic costs of the computation. On the other hand, when considering the stopping-times average $\langle\Sigma(\mathcal{T})\rangle$, we observe a sublinear scaling at moderate values of the limit time, i.e. $\langle\Sigma(\mathcal{T})\rangle\sim \tau^{\alpha}$, with $\vert\alpha \vert <1$, reaching a plateau at large $\tau$ (see filled symbols in Fig.~\ref{fig:6}). 

We also find that $\langle\Sigma(\mathcal{T})\rangle$ is more sensitive than $\langle\Sigma(\tau)\rangle/\tau$ to the value of $p_0$ for all values of $\tau$ explored in our simulations. This reveals that the intrinsic mismatch cost at stopping times $\langle\Sigma(\mathcal{T})\rangle$ is a suitable quantity to quantify the average performance of a computation accomplished at a stochastic time. For example, the sensitivity of $\langle\Sigma(\mathcal{T})\rangle$ to string statistics could be fruitfully exploited as a probe of the performance of a DFA in processing different regular languages, an exciting avenue that we leave for future work.

\bibliography{refs}

\end{document}